\documentclass[ 10pt,journal, onecolumn]{IEEEtran}

\usepackage{refcheck}
\usepackage{graphicx}
\usepackage{amsmath}
\usepackage{amssymb}
\usepackage{graphics}
\usepackage{latexsym}
\usepackage[dvips]{epsfig}
\usepackage[nospace]{cite}
\usepackage{tabularx}
\usepackage{subfigure}
\usepackage{epstopdf}
\usepackage{mathdots}

\newtheorem{Definition}{Definition}
\newtheorem{Lemma}{Lemma}
\newtheorem{Corollary}[Lemma]{\bf \emph{Corollary}}
\newtheorem{Proposition}[Lemma]{Proposition}
\newtheorem{Theorem}{\bf \emph{Theorem}}

\newtheorem{Remark}{Remark}

\def\Pr{{\rm \bold {Pr}}}
\def\E{{\rm \bold  E}}
\def\rank{{\rm {rank}}}

\def\det{{\rm \text {det}}}

\begin{document}

\title{Two-Hop Interference Channels: \\ Impact of Linear Schemes}

\author{
   \IEEEauthorblockN{Ibrahim Issa, Silas L. Fong, and A. Salman Avestimehr}
   \thanks{I. Issa, S. L. Fong, and A. S. Avestimehr are with the School of Electrical and Computer Engineering, Cornell University, Ithaca, NY (email: ii47@cornell.edu, lf338@cornell.edu, and avestimehr@ece.cornell.edu).}
   \thanks{Preliminary parts of this work were presented at the 2013 International Symposium on Information Theory (ISIT)\cite{IFAISIT2013}.}

 }
%
%

\maketitle
\begin{abstract}
We consider the two-hop interference channel (IC), which consists of two source-destination pairs communicating with each other via two relays. We analyze the degrees of freedom (DoF) of this network when the relays are restricted to perform \emph{linear schemes}, and the channel gains are constant (i.e., slow fading). We show that, somewhat surprisingly, by using vector-linear strategies at the relays, it is possible to achieve 4/3 sum-DoF when the channel gains are real. The key achievability idea is to alternate relaying coefficients across time, to create different end-to-end interference structures (or topologies) at different times. Although each of these topologies has only 1 sum-DoF, we manage to achieve 4/3 by \emph{coding} across them. Furthermore, we develop a novel outer bound that matches our achievability, hence characterizing the sum-DoF of two-hop interference channels with linear schemes. As for the case of complex channel gains, we characterize the sum-DoF with linear schemes to be 5/3. We also generalize the results to the multi-antenna setting, characterizing the sum-DoF with linear schemes to be $2M-1/3$ (for complex channel gains), where $M$ is the number of antennas at each node.
\end{abstract}

\section{Introduction}

Multi-hopping is typically viewed as an effective approach to extend the coverage range of wireless networks, by bridging the gap between the sources and destinations via relays. However, it has also the potential to significantly impact network capacity by enabling new interference management techniques. For example, from the degrees of freedom (DoF) perspective, authors in~\cite{Jafar} considered a two-hop complex interference channel (IC) consisting of two sources, two relays, and two destinations, and they showed by introducing a new scheme called \emph{aligned-interference-neutralization} that the sum-DoF of this network is 2 (i.e., twice the sum-DoF of a single-hop IC). More recently, authors in~\cite{IlanKKK} have considered two-hop interference networks with $K$ sources, $K$ relays, and $K$ destinations (i.e., $K \times K \times K$ network), and they showed by developing a new scheme named \emph{aligned-network-diagonalization} that relays have the potential to asymptotically cancel the interference between all source-destination pairs, hence the cut-set bound is achievable (i.e., sum-DoF=$K$).

While the aforementioned results essentially demonstrate that significant DoF gains can be achieved by carefully designing the interference management strategies in multi-hop interference networks, they often require complicated relaying strategies. For instance, for the case of time-varying channels, relays need to code over many \emph{independent} channel realizations (i.e. requiring large channel diversity), and for the case of constant channels, they need to employ non-linear schemes and utilize large rational dimensions in order to align and neutralize interference. In this paper, we take a complementary approach and ask how much of these DoF gains can be realized if we limit the operation of relays to simple \emph{linear} strategies?

We first consider the two-hop interference channel with constant and real channel gains (i.e., slow fading and baseband), and assume that the relays are allowed to perform only linear operations. It is easy to see that if we consider \emph{scalar}-linear schemes with \emph{fixed} amplify-forward (AF) coefficients at the relays, then the end-to-end channel is equivalent to a single-hop IC that has only 1 sum-DoF. We show that, surprisingly, by only allowing the relay AF coefficients to be time-varying (allowing for \emph{vector}-linear schemes), we can exceed 1 sum-DoF and achieve 4/3. The key idea is as follows. With appropriate choice of amplify-forward coefficients at the relays, it is possible to create three specific end-to-end interference structures (or topologies), namely Z, S, or X. In short, the Z topology corresponds to the case that the end-to-end interference from source~1 to destination~2 is nulled, the S topology corresponds to the case that end-to-end interference from source 2 to destination~1 is nulled, and the X topology corresponds to the case that no end-to-end interference nulling has occurred. Although each of these topologies has only 1 sum-DoF, we show that it is possible to achieve 4/3 sum-DoF by creating different end-to-end topologies at different times, and employing an innovative coding strategy across them. 

We also develop a novel outer bound on the DoF of two-hop IC with arbitrary vector-linear strategies that matches our achievability,  thus characterizing the sum-DoF of two-hop IC using linear schemes to be 4/3. The main idea for the converse is the following. Consider a vector-linear scheme, where relays operate over blocks of $\ell$ transmit symbols. In each block, the effective end-to-end channel (i.e., between the sources and the destinations) can be viewed as a multi-antenna IC with $\ell$ antennas at each node, where we have some control over the channel realization through the choice of matrices at the relays. We prove that, regardless of the choice of relaying matrices, there exists a time-invariant linear relationship between each effective direct link and the effective interference links, which means that the end-to-end multi-antenna IC is ill-conditioned. This gives rise to a tension between decreasing the ranks of interference links and increasing the ranks of direct links. By carefully examining this tension, we prove that the sum-DoF is upper bounded by 4/3.

Next, we consider  the setting with multiple antennas, say $M$ antennas, at each node (i.e. MIMO two-hop IC). This setup has been considered in~\cite{CaireLattice}, in which the authors show that, using lattice schemes, a sum-DoF of $2M-1$ is achievable. Also, it is known that for this channel $2M$ sum-DoF is achievable (i.e., the cut-set bound), by simply neglecting the possible cooperation between the antennas and applying the result of~\cite{IlanKKK} for $K \times K \times K$ interference networks. Again, we ask what can we achieve if relays are restricted to linear strategies?

In the setting with $M$ antennas at each node, we characterize the sum-DoF of linear schemes to be $2M-2/3$. The main idea for the achievability scheme is as follows. As before, the choice of relaying matrices dictates the end-to-end topology. However, in this setup, many topologies can be created, which makes the task of designing the scheme more difficult. We propose a 3-phase scheme which codes across 3 topologies, which we call the MIMO-S, MIMO-Z, and MIMO-X topologies. In the MIMO-S topology, all the end-to-end interference from source 2 is neutralized at destination 1, and only one antenna from source 1 causes interference at destination 2. Similarly, for the MIMO-Z topology, only one antenna from source 2 causes interference at destination 1. In the MIMO-X topology, however, one antenna from each source is causing interference at the other destination. The relaying matrices that create the above topologies correspond to the solutions of specific Sylvester equations\footnote{The Sylvester equation is a matrix equation of the form $AX+XB=C$, where $A$, $B$, $C$, and $X$ are square matrices, and the problem is to find $X$ (for a given $A$, $B$, and $C$).}, with the constraint that the solutions are invertible. The conditions for the existence of such solutions have been studied extensively in the literature (e.g., \cite{Luenberger,sylvesterEquation}).  Using these results, we show that the aforementioned three topologies can be created for almost all values of the channel gains. Finally, we show that by \emph{coding} across these topologies we can achieve $2M-2/3$ sum-DoF.

As for the converse, the key ingredient is proving a relationship between the end-to-end direct links and end-to-end interference links. In particular, we show that if any of the direct links has full rank, then at least one of the interference links must be non-zero. This relationship, coupled with two genie-aided bounds, yields our result.

We also generalize the results to the case of complex channel gains. In the single-antenna case, it was shown in~\cite{Jafar} that 3/2 sum-DoF is achievable by using a linear scheme based on asymmetric complex signaling. Also, more recently in~\cite{CaireFiniteField}, a new scheme named PCoF-CIA (Precoded Compute and Forward with Channel Integer Alignment) has been proposed to achieve 3/2 sum-DoF.  We can evidently apply our scheme (designed for the case of real channel gains) and follow the same nulling and coding procedure to achieve $4/3$ sum-DoF. However, we propose a better approach. We allow coding over the in-phase and quadrature-phase components of the channel, and show that, from the DoF perspective, the network can be viewed as a two-hop IC with real channel gains and $2$-antennas at each node (corresponding to in-phase and quadrature-phase components). Then, by applying our scheme for the 2-antenna setting, we can achieve $4-2/3$ \emph{real} DoF in the equivalent network, which equates to $\frac{4-2/3}{2}=5/3$ sum-DoF in the original two-hop IC with complex channel gains. This improves over all previously known results with linear schemes. We also prove the optimality of our scheme, hence characterize the linear sum-DoF of two-hop IC with complex channel gains to be indeed $5/3$. The results are also extended to the $M$-antenna setting with complex channel gains, characterizing the linear sum-DoF to be $2M-1/3$.

Finally, we present a numerical analysis of our proposed schemes. Although the main focus of this paper is the characterization of degrees of freedom (i.e., capacity analysis at high signal-to-noise ratio (SNR) regime), the simplicity of our schemes also allows for the analytical computation of the achieved rates at any finite SNR. Therefore, we compare our linear scheme for the two-hop IC with several state-of-the-art schemes, and demonstrate the capacity gains at finite SNR.

\textbf{Other Related Works.}
Other than the aforementioned works that focus on the degrees of freedom of two-hop interference channels, there have also been several works on the capacity analysis of such networks. For example, authors in~\cite{Mohajer} approximate the capacity of networks of the form ZZ and ZS, using the deterministic approach~\cite{AvestimehrDeterministic}.
In~\cite{Simeone} and~\cite{Thejaswi}, authors adopt an approach that applies rate-splitting at the sources based on the Han-Kobayashi scheme~\cite{HanKobayashi}, and decode-and-forward at the relays to cooperatively deliver the messages. However, this approach essentially treats the two-hop IC as a cascade of two interference channels, and thus cannot achieve more than 1 sum-DoF.

The rest of this paper is organized as follows. In Section~\ref{model&results}, we define the model and state our three main results. In Section~\ref{twoHopICSingleAntenna}, we prove our first result for the single-antenna two-hop IC with real channel gains. In Section~\ref{twoHopICMultipleAntennas}, we consider the MIMO two-hop IC with real channel gains, and we extend the result to the case of complex channel gains in Section~\ref{twoHopICMIMOComplex}. Finally, in Section~\ref{numericalResults}, we discuss our numerical results.

\section{Network Model \& Statement of Main Results} \label{model&results}

\begin{figure}[htp]
\centering
\includegraphics[width=2.7 in, height=1 in, bb = 96 249 584 435, clip=true]{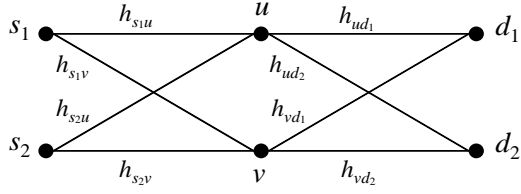}
\caption{Two-hop IC.}\label{twoHopIC}
\end{figure}

The two-hop IC, illustrated in Figure~\ref{twoHopIC}, consists of two sources, two relays and two destinations. The two sources are indexed by $s_1$ and $s_2$, the two relays are indexed by $u$ and $v$ and the two destinations are indexed by $d_1$ and $d_2$. Each node is equipped with a single antenna\footnote{The case for multi-antenna nodes will be discussed in Section~\ref{twoHopICMultipleAntennas}.}. The channel gains of the first hop are denoted by
\begin{equation*}
\mathbf{H_1}=\left[\begin{array}{cc}h_{s_1u} & h_{s_2u}\\ h_{s_1v} & h_{s_2v}\end{array}\right] \in \mathbb{R}^{2 \times 2},
 \end{equation*}
 and the channel gains of the second hop are denoted by
 \begin{equation*}
 \mathbf{H_2}=\left[\begin{array}{cc}h_{ud_1} & h_{vd_1}\\ h_{ud_2} & h_{vd_2}\end{array}\right] \in \mathbb{R}^{2 \times 2}.
  \end{equation*}
We assume that the channel gains are real-valued and drawn from some continuous distribution, and fixed during the course of communication. All the nodes have the knowledge of $\mathbf{H_1}$ and $\mathbf{H_2}$. For each $i\in\{1,2\}$, $s_i$ chooses a message $W_i\in\{1, 2, \ldots, 2^{n R_i}\}$ which is intended for $d_i$ only. Each message $W_i$ is uniformly distributed over $\{1, 2, \ldots, 2^{n R_i}\}$, and $W_1$ and $W_2$ are independent. The two sources transmit their messages to the destinations in~$n$ time slots. For each $i\in\{1, 2\}$, let $X_{i,k}\in \mathbb{R}$ denote the symbol transmitted by source~$s_i$ in the $k^{\text{th}}$ time slot. Then, the symbols received in the same time slot by $u$ and $v$, denoted by $Y_{u,k}\in \mathbb{R}$ and $Y_{v,k}\in \mathbb{R}$ respectively, satisfy
\begin{equation}
\left[\begin{array}{c}Y_{u,k}\\Y_{v,k} \end{array}\right] = \mathbf{H_1}\left[\begin{array}{c}X_{1,k}\\X_{2,k} \end{array}\right] + \left[\begin{array}{c}Z_{u,k}\\Z_{v,k} \end{array}\right] , \label{firstChannelAddition}
\end{equation}
where $Z_{u,k}\sim \mathcal{N}(0,1)$ and $Z_{v,k}\sim \mathcal{N}(0,1)$. In addition, for each $r\in\{u,v\}$,  let $X_{r,k}\in \mathbb{R}$ denote the symbol transmitted by relay~$r$ in the $k^{\text{th}}$ time slot. Then, the symbols received in the same time slot by $d_1$ and $d_2$, denoted by $Y_{1,k}\in \mathbb{R}$ and $Y_{2,k}\in \mathbb{R}$ respectively, satisfy
\begin{equation}
\left[\begin{array}{c}Y_{1,k}\\Y_{2,k} \end{array}\right] = \mathbf{H_2}\left[\begin{array}{c}X_{u,k}\\X_{v,k} \end{array}\right] + \left[\begin{array}{c}Z_{1,k}\\Z_{2,k} \end{array}\right] , \label{secondChannelAddition}
\end{equation}
where $Z_{1,k}\sim \mathcal{N}(0,1)$ and $Z_{2,k}\sim \mathcal{N}(0,1)$. We assume that $(X_1^n, X_2^n)$, $Z_u^n$, $Z_v^n$, $Z_1^n$, and $Z_2^n$ are independent, and we assume that $\{(Z_{u,k}, Z_{v,k}, Z_{1,k},Z_{2,k})\}_{k=1}^n$ are independent. After~$n$ time slots, node~$d_i$ declares~$\hat W_i$ to be the transmitted~$W_i$ based on $Y_i^n$ for each $i\in\{1,2\}$.
For each $i\in\{1, 2, u, v\}$, any codeword $[x_{i,1} \,  x_{i,2} \, \ldots \, x_{i,n}]^T$ that is transmitted over the network should satisfy
$||x_i^n||^2 \le n P$,
where~$P$ represents the power constraint for all the nodes.
\pagebreak
\begin{Definition} \label{defCode}
An $(n, R_1, R_2)$-code on the two-hop IC consists of the following:
\begin{enumerate}
\item A message set
$
\mathcal{W}_i=\{1, 2, \ldots, 2^{nR_i}\}
$
at~$s_i$ for each $i\in\{1,2\}$.
\item  An encoding function
$f_i : \mathcal{W}_i \rightarrow \mathbb{R}^n$
 at~$s_i$ for each $i\in\{1,2\}$
such that
$
X_i^n=f_i(W_i)$.
In addition, every codeword $x_i^n$ must satisfy the power constraint
$
||x_i^n||^2 \le nP
$.
\item An encoding function
$
f_{r, k} : \mathbb{R}^{k-1}\rightarrow \mathbb{R}
$
 at each relay $r\in\{u,v\}$ and each $k\in\{1, 2, \ldots, n\}$
such that \linebreak $X_{r,k}=f_{r,k}(Y_r^{k-1})$.
In addition, every codeword $x_r^n$ must satisfy the power constraint
$
||x_r^n||^2\le nP 
$.
\item A decoding function
$
g_i  : \mathbb{R}^n \rightarrow \mathcal{W}_i
$
  at~$d_i$ for each $i\in\{1,2\}$ such that
$
 \hat W_i= g_i(Y_i^n)$.
 \end{enumerate}
 \end{Definition}
 \bigskip
 Without loss of generality, we assume
\begin{equation}
X_{1,k^-}= X_{2,k^-}=Z_{u,k^-}= Z_{v,k^-}= Y_{u,k^-}=Y_{v,k^-} =0 \label{zeroConvention}
\end{equation}
for each $k^- \le 0$ in the rest of the paper.
\bigskip
\begin{Definition} \label{defErrorProbability}
For an $(n, R_1, R_2)$-code, the average probability of decoding error of $W_i$ is defined as
$P_{e,i}^n = \Pr\{\hat W_i \ne W_i\}$ for each $i\in\{1,2\}$.
\end{Definition}
\bigskip
\begin{Definition} \label{defVectorLinear}
Let $\ell$ be a natural number and $\mathcal{U}$ be a finite set of real numbers. An $(\ell n, R_1, R_2)$-code on the two-hop IC is said to be \textit{$\ell$-linear on $\mathcal{U}$} if there exist $\{\mathbf{A}_k\in \mathcal{U}^{\ell\times \ell} \}_{k=1}^{n}$ and $\{\mathbf{B}_k \in \mathcal{U}^{\ell\times \ell} \}_{k=1}^{n}$ such that for each $k\in\{1, 2, \ldots, n\}$,
\[
\left[\begin{array}{cccc}X_{u,\ell(k-1)+1} & X_{u,\ell(k-1)+2} & \ldots & X_{u,\ell k}\end{array}\right]^T =
\mathbf{A}_k\left[\begin{array}{cccc} Y_{u,\ell(k-2)+1} & Y_{u,\ell(k-2)+2} & \ldots & Y_{u,\ell(k-1)} \end{array}\right]^T
\]
and
\[
\left[\begin{array}{cccc}X_{v,\ell(k-1)+1}& X_{v,\ell(k-1)+2}& \ldots & X_{v,\ell k}\end{array}\right]^T =
\mathbf{B}_k \left[\begin{array}{cccc} Y_{v,\ell(k-2)+1} & Y_{v,\ell(k-2)+2} & \ldots & Y_{v,\ell(k-1)} \end{array}\right]^T.
\]
In other words, the relays are operating over blocks of length $\ell$, and the symbols in each transmitted block are linear combinations of the $\ell$ symbols received in the previous block. We call $(\{\mathbf{A}_k\in \mathcal{U}^{\ell \times \ell} \}_{k=1}^{n},\{\mathbf{B}_k\in \mathcal{U}^{\ell \times \ell} \}_{k=1}^{n})$ a \textit{relaying kernel} of the code.
\end{Definition}
\bigskip
\begin{Definition} \label{defAchievable}
A rate pair $(R_1, R_2)$ is
 \textit{$\ell$-linear achievable on $\mathcal{U}$} if there exists a sequence of $(\ell n, R_1, R_2)$-codes that are $\ell$-linear on $\mathcal{U}$ such that $\lim\limits_{n\rightarrow \infty} P_{e,i}^{\ell n}= 0$ for each $i\in\{1,2\}$.
\end{Definition}
\bigskip
\begin{Definition} \label{defDoF}
The \emph{linear sum-DoF} of the two-hop IC, denoted by $\mathcal{D}$, is defined by
\[
\mathcal{D} = \sup_{\ell, \mathcal{U}}\lim_{P\rightarrow\infty}\! \sup\left\{\left. \frac{R_1+R_2}{\frac{1}{2}\log_2 P} \:\right|(R_1, R_2) \text{ is $\ell$-linear achievable on }\mathcal{U}\right\}.
\]
\end{Definition}
\bigskip
The first main result of this paper is the characterization of $\mathcal{D}$ as follows:
\bigskip
\begin{Theorem} \label{thmDoF}
The linear sum-DoF of the two-hop IC is 4/3 for almost all values of real channel gains. In particular, $\mathcal{D} = 4/3$ if the channel gains satisfy the following conditions:
\begin{enumerate}
\item[](c-1) All the channel gains are non-zero.
\item[](c-2) $\det\left(\mathbf{H}_i\right) \ne 0$ for each $i\in\{1,2\}$.
\vspace{0.04 in}
\item[](c-3) $\det\left(\left[ \begin{array}{cc}h_{s_2u}h_{ud_1} & h_{s_1u}h_{ud_2} \\ h_{s_2v}h_{vd_1} & h_{s_1v}h_{vd_2} \end{array}  \right]\right) \ne 0$.
\end{enumerate}
\end{Theorem}
\bigskip
\begin{Remark}
If we do not restrict the operations of the relays to be linear (i.e. we consider codes that satisfy Definition~\ref{defCode}, but not necessarily Definition~\ref{defVectorLinear}), then it is shown in~\cite{Jafar} that 2 sum-DoF (i.e. the cut-set bound) is achievable using \emph{aligned interference neutralization}. Theorem~\ref{thmDoF} shows that if we restrict the relays to simpler linear operations, then we can achieve 4/3 sum-DoF. Furthermore, unlike the channel conditions needed for~\cite{Jafar} to achieve 2 sum-DoF , the above three conditions are insensitive to the rationality or irrationality of channel parameters.

\end{Remark}
\bigskip
In Section~\ref{twoHopICMultipleAntennas}, we extend the result for the two-hop MIMO interference channel with real channel gains, where each node is equipped with $M$ antennas. We similarly define a \emph{linear} code for this network in Section~\ref{twoHopICMultipleAntennas} (cf.\ Definition~\ref{defMIMOLinear}), and then prove the second main result of our paper as follows.
\bigskip
\begin{Theorem} \label{thmDoFMIMO}
The linear sum-DoF of the two-hop MIMO IC with $M$ antennas at each node is $2M - 2/3$ for almost all values of real channel gains.
\end{Theorem}
\bigskip

We finally extend Theorem~\ref{thmDoFMIMO} to complex channel gains and obtain the following corollary.
\bigskip
\begin{Corollary} \label{ComplexMIMO}
The linear sum-DoF of the two-hop MIMO IC with $M$ antennas at each node is $2M - 1/3$ for almost all values of complex channel gains.
\end{Corollary}
\bigskip
\begin{Remark}
Note that the same achievability scheme used for the MIMO IC with real channel gains can be used for the one with complex channel gains, thus achieving $2M-2/3$ sum-DoF for the complex case. However, we can get an additional gain by separating and coding over the in-phase and quadrature-phase components of the channel.
\end{Remark}
\bigskip
\begin{Remark}
For the single-antenna two-hop IC with complex gains, authors in~\cite{Jafar} and~\cite{CaireFiniteField} propose schemes that achieve 3/2 sum-DoF without using rational dimensions, where the former relies on linear coding and the latter utilizes lattice coding and nulls the end-to-end interference by linear precoding/decoding over the finite field. Corollary~\ref{ComplexMIMO} shows that we can actually exceed 3/2. In particular, it states that 5/3 sum-DoF is achievable using linear schemes.
\end{Remark}
\bigskip
\begin{Remark}
In a recent result, authors in~\cite{CaireLattice} show that for the two-hop MIMO IC with $M$ antennas at each node, $2M-1$ sum-DoF is achievable using Precoded Compute and Forward (PCoF) with Channel Integer Alignment (CIA). Corollary~\ref{ComplexMIMO} shows that linear schemes can also outperform PCoF with CIA in terms of sum-DoF.
\end{Remark}
\bigskip
\begin{Remark}
Although the theorems above focus only on degrees of freedom, we can analytically compute the rates achieved by our proposed schemes at any SNR. We demonstrate the capacity gains at finite SNR as compared to state-of-the-art schemes in Section~\ref{numericalResults}.
\end{Remark}
\bigskip
We will now proceed to prove Theorem~\ref{thmDoF}, Theorem~\ref{thmDoFMIMO}, and Corollary~\ref{ComplexMIMO} in Sections~\ref{twoHopICSingleAntenna},~\ref{twoHopICMultipleAntennas}, and~\ref{twoHopICMIMOComplex} respectively.

\section{Two-Hop IC with Single-Antenna Nodes} \label{twoHopICSingleAntenna}
In this section, we will prove Theorem~\ref{thmDoF}. First, we describe a linear code that achieves 4/3 sum-DoF. Then, we will prove that 4/3 is an upper bound on the sum-DoF for any linear code.

\subsection{Achievability Proof of Theorem~\ref{thmDoF}} \label{achieve}

We will show that $\ell$-linear schemes can achieve 4/3 sum-DoF if conditions (c-1)--(c-3) are satisfied. In particular, somewhat surprisingly, this can be done for $\ell=1$ only. In this case, the matrices chosen at the relays are just real scalars, however they can be time-varying.

The achievability scheme consists of three phases, during which each source sends two distinct symbols, and at the end of the three phases each receiver is able to reconstruct an interference-free (but noisy) version of its desired symbols.

For simplicity of notation, for $\ell=1$, set $\mathbf{A}_k = \alpha_k$ and $\mathbf{B}_k = \beta_k$. Then the received signals at the destinations at each time $k$ can be written as
\begin{equation*}
\begin{split}
\begin{bmatrix} Y_{1,k} \\ Y_{2,k} \end{bmatrix} &  = \mathbf{H_2} \begin{bmatrix} \alpha_k & 0 \\ 0 & \beta_k \end{bmatrix} \mathbf{H_1} \begin{bmatrix} X_{1,k-1} \\ X_{2,k-1} \end{bmatrix} + \begin{bmatrix} \tilde{Z}_{1,k} \\ \tilde{Z}_{2,k} \end{bmatrix} \\
& = \mathbf{G}_k \begin{bmatrix} X_{1,k-1} \\ X_{2,k-1} \end{bmatrix} + \begin{bmatrix} \tilde{Z}_{1,k} \\ \tilde{Z}_{2,k} \end{bmatrix},
\end{split}
\end{equation*}
where
\begin{equation*}
\tilde{Z}_{i,k} = h_{ud_i}\alpha_k Z_{u,k-1}+h_{vd_i}\beta_k Z_{v,k-1}+Z_{i,k}
\end{equation*} is the effective noise at destination $d_i$, $i \in \{1,2\}$,
and $\mathbf{G}_k =  \mathbf{H_2} \begin{bmatrix} \alpha_k & 0 \\ 0 & \beta_k \end{bmatrix} \mathbf{H_1}$ is the equivalent end-to-end channel matrix given by
\begin{equation*} \label{Gk}
\mathbf{G}_k =
\begin{bmatrix} h_{ud_1} h_{s_1u}\alpha_k +h_{vd_1} h_{s_1v}  \beta_k  & h_{ud_1} h_{s_2u} \alpha_k  + h_{vd_1} h_{s_2v}  \beta_k  \\  h_{ud_2} h_{s_1u} \alpha_k + h_{vd_2} h_{s_1v}  \beta_k  & h_{ud_2} h_{s_2u} \alpha_k  +h_{vd_2} h_{s_2v}  \beta_k \end{bmatrix}.
\end{equation*}

\noindent For notational convenience, let $\mathbf{G}_k = \begin{bmatrix} g_{11,k} & g_{12,k} \\ g_{21,k} & g_{22,k} \end{bmatrix}$. Then, the received signal at destination $d_i$, $i \in \{1,2\}$, at time $k$ is
\begin{equation} \label{RSignal}
Y_{i,k}=g_{i1,k}X_{1,k-1}+g_{i2,k}X_{2,k-1}+\tilde{Z}_{i,k},  \text{\quad $k \in \{1,2,\dots,n\}$}.
\end{equation}
\noindent Note that the variance of $\tilde{Z}_{i,k}$ depends only on channel gains and relay coefficients (chosen from $\mathcal{U}$), therefore it does not scale with $P$.

We will now describe the three phases of our linear achievability scheme in detail. Set
\begin{equation*}
\mathcal{U}=c\{0,1, -h_{ud_1} h_{s_2u}/h_{vd_1} h_{s_2v}, - h_{ud_2} h_{s_1u}/h_{vd_2} h_{s_1v}\},
\end{equation*} where the constant $c \in \mathbb{R}^+$ is chosen to satisfy the power constraint $P$ at the relays. More specifically,
\begin{equation*}
c= \min\left\{\sqrt{1/(h_{s_1u}^2 \hspace{-1mm} +h_{s_2u}^2 \hspace{-1mm}+1)}, l\sqrt{1/(h_{s_1v}^2+h_{s_2v}^2+1)} \right\},
\end{equation*}
where $l  =  \min\{ {|h_{vd_1} h_{s_2v}/h_{ud_1} h_{s_2u}| , |h_{vd_2} h_{s_1v}/h_{ud_2} h_{s_1u}|} \}.$ Note that the denominators are non-zero by condition (c-1).\\

\begin{figure}[!hbt]
\centering
\subfigure[Phase 1]{\includegraphics[scale=1]{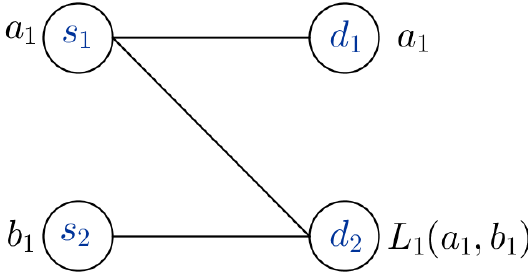}} \hspace{0.2in}
\subfigure[Phase 2]{\includegraphics[scale=1]{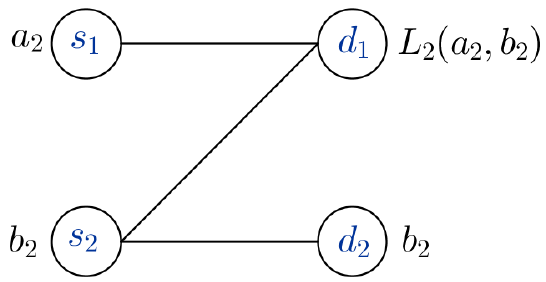}} \hspace{0.2in}
\subfigure[Phase 3]{\includegraphics[scale=1]{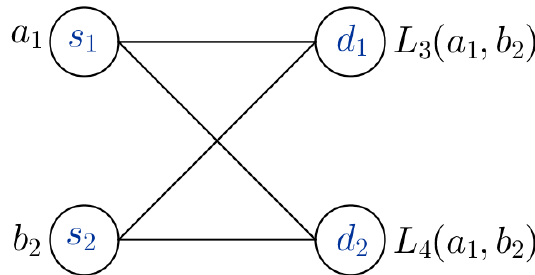}}
\caption{Illustration of achievability scheme. At each phase, the transmit symbols by sources are shown on the left. The  received signals at destinations are given on the right, where the noise is dropped and $L(x,y)$ denotes a linear combination of $x$ and $y$. }
\label{scheme}
\end{figure}

\noindent \textbf{Phase 1.} In this phase, $s_1$ and $s_2$ send two symbols $a_1$ and $b_1$ respectively $(a_1^2,b_1^2 \leq P)$. We choose the relay coefficients such that the interference from $s_2$ is canceled at $d_1$. More specifically, we set $\alpha_1=c$ and $\beta_1=-c h_{ud_1} h_{s_2u}/h_{vd_1} h_{s_2v}$. By inserting this choice of $\beta_1$ and $\alpha_1$ in (\ref{RSignal}), $d_1$ and $d_2$ will respectively receive
\begin{equation} \label{timeslot1}
y_{1,1}=g_{11,1}a_1+z_{1,1}, \text{ and } y_{2,1}= \underbrace{g_{21,1}a_1+g_{22,1}b_1}_{L_1(a_1,b_1)}+z_{2,1},
\end{equation}
where $g_{11,1}\ne 0$ and $g_{22,1}\ne 0$ (due to conditions (c-1), (c-2), and (c-3)), and $L_1(a_1,b_1)$ indicates a linear equation in $a_1$ and $b_1$. Thus, as shown in Figure \ref{scheme}(a), $d_1$ and $d_2$ now respectively have noisy versions of $a_1$ and $L_1(a_1,b_1)$.\\

\noindent \textbf{Phase 2.} In this phase, $s_1$ and $s_2$ send two new symbols $a_2$ and $b_2$ $(a_2^2,b_2^2 \leq P)$. However, this time, we cancel the effect of $s_1$ at $d_2$, by letting $\alpha_2 = c $ and $\beta_2 =-c h_{ud_2} h_{s_1u}/h_{vd_2} h_{s_1v}$. Then $d_1$ and $d_2$ will respectively receive
\begin{equation} \label{timeslot2}
y_{1,2}=\underbrace{g_{11,2}a_2+g_{12,2}b_2}_{L_2(a_2,b_2)}+z_{1,2}, \text{ and } y_{2,1}=g_{22,2}b_2+z_{2,2},
\end{equation}
where $g_{11,2} \ne 0$ and $g_{22,2} \ne 0$ (due to conditions (c-1), (c-2), and (c-3)), and $L_2(a_2,b_2)$ indicates a linear equation in $a_2$ and $b_2$. Thus, as shown in Figure \ref{scheme}(b), $d_1$ and $d_2$ now respectively have noisy versions of $L_2(a_2,b_2)$ and $b_2$.\\

\noindent \textbf{Phase 3.} Now notice that if, at phase 3, destination $d_1$ receives a linear combination of $a_1$ and $b_2$ ($L_3(a_1,b_2)$), then it can solve for (a noisy version of) $a_2$ given equations (\ref{timeslot1}) and (\ref{timeslot2}). Similarly, if $d_2$ receives $L_4(a_1,b_2)$ then it can also solve for (a noisy version of) $b_1$ given equations (\ref{timeslot1}) and (\ref{timeslot2}). Thus, as shown in Figure \ref{scheme}(c), in phase 3, $s_1$ sends $a_1$, $s_2$ sends $b_2$, and we choose $\alpha_3=c$ and $\beta_3=0$, so that $d_1$ and $d_2$ receive
\begin{equation} \label{timeslot3}
y_{1,3}=\underbrace{g_{11,3}a_1+g_{12,3}b_2}_{L_3(a_1,b_2)}+z_{1,3},
\text{ and }
y_{2,3}=\underbrace{g_{21,3}a_1+g_{22,3}b_2}_{L_4(a_1,b_2)}+z_{2,3},
\end{equation}
where $g_{12,3} \ne 0$, and $g_{21,3} \ne 0$ (due to condition (c-1)). Therefore, after the three phases, $d_1$ can construct
  \begin{equation}
 y_1^{a_1} = a_1+z_{1,1}/g_{11,1}, \label{firstDataStream}
   \end{equation}
   and
   \begin{equation}
 y_1^{a_2} = a_2+\frac{1}{g_{11,2}}z_{1,2}-\frac{g_{12,2}}{g_{11,2}g_{12,3}}z_{1,3}+\frac{g_{11,3}g_{12,2}}{g_{11,1}g_{11,2}g_{12,3}}z_{1,1}.
   \label{secondDataStream}
    \end{equation}
    from $(y_{1,1}, y_{1,2},y_{1,3})$. Let $\sigma_1^2$ and $\sigma_2^2$ be the variances of the noise terms in equations (\ref{firstDataStream}) and (\ref{secondDataStream}). Note that they depend only on channel gains and relay coefficients. Hence, they are constants that do not scale with $P$. Then, by using a proper outercode, we can achieve a rate of
\begin{align}
R_1  & = \frac{1}{6} \left( \log \left(1+\frac{P}{\sigma_1^2}\right)+\log \left(1+\frac{P}{\sigma_2^2} \right)  \right)  \label{singleAntennaRealAchRate} \\
&   \geq \frac{1}{3} \log \frac{P}{\sigma_1 \sigma_2}. \notag
\end{align}
So $d_1$ can achieve $2/3$ DoF. Similarly, $d_2$ can also achieve $2/3$ DoF, hence achieving a total of $4/3$ sum-DoF.
\bigskip
\begin{Remark}
Note that the described scheme can also be viewed as a one-phase linear code with $\ell=3$, where $s_1$ sends $\begin{bmatrix}
a_1 & a_2 & a_1
\end{bmatrix}^T$, $s_2$ sends $\begin{bmatrix}
b_1 & b_2 & b_2
\end{bmatrix}^T$, and relays $u$ and $v$ set their amplifying matrices to be
\begin{equation*}
\mathbf{A}_1 = \begin{bmatrix}
c & 0 & 0 \\
0 & c & 0 \\
0 & 0 & c
\end{bmatrix} \text{ and }
\mathbf{B}_1= \begin{bmatrix}
-ch_{ud_1} h_{s_2u}/h_{vd_1} h_{s_2v} & 0 & 0 \\
0 & - ch_{ud_2} h_{s_1u}/h_{vd_2} h_{s_1v} & 0 \\
0 & 0 & 0
\end{bmatrix}\text{ respectively.}
\end{equation*}
However, our initial description better illustrates the ``spirit'' of the scheme, in terms of understanding the choice of the amplifying factors at the relays and highlighting the opportunity of coding over different topologies. In fact, each individual topology shown in Figure~\ref{scheme} has a sum-DoF of 1, whilst we managed to achieve 4/3 sum-DoF by coding across them.
\end{Remark}
\bigskip
\begin{Remark}
The coding strategy that is used in our scheme was first used for a binary fading IC, in which the channel links are either ``on'' or ``off'' \cite[Appendix A]{AlirezaDelayed}. A similar coding idea was also shown to be useful and provided DoF gains in the context of two-user IC with alternating connectivity~\cite{JafarAlternating}.
\end{Remark}
\bigskip

\subsection{Converse Proof of Theorem~\ref{thmDoF}}\label{upperBound}


Assume $(R_1, R_2)$ is $\ell$-achievable on $\mathcal{U}$ for some $\ell \in \mathbb{N}$ and $\mathcal{U}\subset \mathbb{R}$. It then follows from Definition~\ref{defAchievable} that there exists a sequence of $(\ell n, R_1, R_2)$-codes that are $\ell$-linear on $\mathcal{U}$ such that
 \begin{equation}
\lim_{n\rightarrow \infty} P_{e,i}^{\ell n} = 0 \label{errorProbability}
\end{equation}
for each $i\in\{1,2\}$. We now fix this sequence of $(\ell n, R_1, R_2)$-codes and their corresponding relaying kernels  $(\{\mathbf{A}_k\in \mathcal{U}^{\ell \times \ell} \}_{k=1}^{n},\{\mathbf{B}_k\in \mathcal{U}^{\ell \times \ell} \}_{k=1}^{n})$. Let
\[
\ell_k=\{\ell(k-1)+1, \ell(k-1)+2, \ldots, \ell k\},
\]
where $\ell_k$ represents the time slots of block~$k$.
Then, at each block $k\in\{1, 2, \ldots, n\}$, we have the following relationship between the received signals at the destinations and the transmit signals at the sources:
\begin{align}
\left[\begin{array}{c}Y_{1,\ell_k} \\ Y_{2,\ell_k} \end{array}\right] &\stackrel{\eqref{secondChannelAddition}}{=} \left[\begin{array}{c}  h_{ud_1} X_{u,\ell_k}  \\  h_{ud_2} X_{u,\ell_k}  \end{array}\right]+\left[\begin{array}{c}  h_{vd_1} X_{v,\ell_k} \\  h_{vd_2} X_{v,\ell_k} \end{array}\right] + \left[\begin{array}{c}Z_{1,\ell_k} \\ Z_{2,\ell_k} \end{array}\right] \notag \\
& \stackrel{\text{(a)}}{=} \left[\begin{array}{c}  h_{ud_1} \mathbf{A}_k Y_{u,\ell_{k-1}}  \\  h_{ud_2}  \mathbf{A}_k Y_{u,\ell_{k-1}}  \end{array}\right]+\left[\begin{array}{c}  h_{vd_1}  \mathbf{B}_k Y_{v,\ell_{k-1}} \\  h_{vd_2}  \mathbf{B}_k Y_{v,\ell_{k-1}} \end{array}\right] + \left[\begin{array}{c}Z_{1,\ell_k} \\ Z_{2,\ell_k} \end{array}\right] \notag \\
& \stackrel{\text{(b)}}{=} \left[\begin{array}{cc}\mathbf{G}_{\mathbf{11},k} & \mathbf{G}_{\mathbf{12},k} \\ \mathbf{G}_{\mathbf{21},k} & \mathbf{G}_{\mathbf{22},k}
\end{array}\right]
\left[\begin{array}{c}X_{1,\ell_{k-1}} \\ X_{2,\ell_{k-1}}\end{array}\right] + \left[ \begin{array}{cc}  h_{ud_1} \mathbf{A}_k &  h_{vd_1} \mathbf{B}_k \\  h_{ud_2} \mathbf{A}_k &  h_{vd_2} \mathbf{B}_k \end{array}\right] \left[\begin{array}{c}Z_{u,\ell_{k-1}} \\Z_{v,\ell_{k-1}} \end{array}\right] + \left[\begin{array}{c}Z_{1,\ell_k}\\ Z_{2,\ell_k}\end{array}\right], \label{YkProof}
\end{align}
where
\begin{enumerate}
\item[(a)] follows from Definition~\ref{defVectorLinear}.
\item[(b)] follows from substituting $Y_{u,\ell_{k-1}}$ and $Y_{v,\ell_{k-1}}$ by \eqref{firstChannelAddition} and defining the end-to-end matrix from $s_j$ to $d_i$, denoted by $\mathbf{G}_{\mathbf{ij},k}$, for each $k\in\{1, 2, \ldots, n\}$ as follows:
 \begin{equation}
\mathbf{G}_{\mathbf{ij},k}= h_{s_j u}h_{ud_i} \mathbf{A}_k   +   h_{s_j v}h_{v d_i}\mathbf{B}_k. \label{Gij}
\end{equation}
\end{enumerate}
In addition, for each $i\in\{1,2\}$, let
\begin{equation}
P_k=\E[||X_{1,\ell_k}||^2+||X_{2,\ell_k}||^2] \label{averagePowerPi}
\end{equation}
be the average sum-power of block~$k$ transmitted by the sources, which is averaged over the codebooks of the sources. We state the following key lemma which implies $\mathcal{D}\le 4/3$.
\bigskip
\begin{Lemma} \label{lemmaMain}
For any sequence of $(\ell n, R_1, R_2)$-codes and their corresponding $\mathbf{G}_{\mathbf{ij},k}$ as defined above, we have
for sufficiently large~$n$
\begin{equation*}
R_1 + R_2 \le \tau_1 + \frac{1}{2\ell n}\sum_{k=1}^n (\rank(\mathbf{G}_{\mathbf{21},k})+\rank(\mathbf{G}_{\mathbf{21},k})) \log_2 (1+P_{k-1}/\ell), \tag*{Bound (i)}
\end{equation*}
\begin{equation*}
R_1 + R_2 \le \tau_2 + \frac{1}{2\ell n} \sum_{k=1}^n (2\ell -\rank(\mathbf{G}_{\mathbf{12},k}))\log_2 (1+P_{k-1}/\ell]) \tag*{Bound (ii)}
\end{equation*}
and
\begin{equation*}
R_1 + R_2 \le \tau_3 + \frac{1}{2\ell n} \sum_{k=1}^n (2\ell -\rank(\mathbf{G}_{\mathbf{21},k}))\log_2 (1+P_{k-1}/\ell), \tag*{Bound (iii)}
\end{equation*}
where $P_{k-1}$ is defined in \eqref{averagePowerPi}, and $\tau_1$, $\tau_2$ and $\tau_3$ are some constants that do not depend on $n$ and $P$.
\end{Lemma}
\bigskip
Before proving Lemma~\ref{lemmaMain}, we demonstrate how it implies $\mathcal{D} \le 4/3$ and hence Theorem~\ref{thmDoF}. Summing Bound (i), Bound (ii) and Bound (iii) in Lemma~\ref{lemmaMain} and dividing $3$ on both sides of the resultant inequality, we have for sufficiently large~$n$
 \begin{align}
R_1+R_2 & \le \frac{\tau_1 + \tau_2 + \tau_3}{3}+ \frac{2}{3n}\sum_{k=1}^n\log_2(1+P_{k-1}/\ell) \notag \\
&\stackrel{\eqref{zeroConvention}}{\le} \frac{\tau_1 + \tau_2 + \tau_3}{3}+ \frac{2}{3n}\sum_{k=1}^n\log_2(1+P_k/\ell) \notag \\
&\stackrel{\text{(a)}}{\le} \frac{\tau_1 + \tau_2 + \tau_3}{3}+ \frac{2}{3}\log_2\left(1+\frac{ \sum_{k=1}^n P_k}{\ell n}\right)  \notag \\
&\stackrel{\text{(b)}}{\le} \frac{\tau_1 + \tau_2 + \tau_3}{3}+ \frac{2}{3}\log_2(1+2P),
\label{fromLemmaToTheorem}
 \end{align}
 where
\begin{enumerate}
\item[(a)] follows from applying Jensen's inequality to the concave function $\log_2(1+x)$.
\item[(b)] follows from Definition~\ref{defCode} that $||X_i^{\ell n}||^2 \le \ell n P$ for each $i\in\{1,2\}$.
\end{enumerate}
It then follows from \eqref{fromLemmaToTheorem} and Definition~\ref{defDoF} that $\mathcal{D} \le 4/3$.
We now proceed to prove Lemma~\ref{lemmaMain}.
\bigskip\\
\subsubsection{Proof for Bound (i) in Lemma~\ref{lemmaMain}}
Fix a sequence of $(\ell n, R_1, R_2)$-codes and their corresponding $\mathbf{G}_{\mathbf{ij},k}$. Let
 \begin{equation}
 \tilde Y_{1,\ell_k} = \mathbf{G}_{\mathbf{11},k} X_{1, \ell_{k-1}} + \mathbf{G}_{\mathbf{12},k} X_{2, \ell_{k-1}} + Z_{1, \ell_k} \label{tildeY1}
 \end{equation}
 and
  \begin{equation}
 \tilde Y_{2,\ell_k} = \mathbf{G}_{\mathbf{21},k} X_{1, \ell_{k-1}} + \mathbf{G}_{\mathbf{22},k} X_{2, \ell_{k-1}} + Z_{2, \ell_k}  \label{tildeY2}
 \end{equation}
 be less noisy versions of $Y_{1,\ell_k}$ and $Y_{2,\ell_k}$ respectively for each $k\in\{1, 2, \ldots, n\}$ (i.e., removing the impact of $Z_{u, \ell_{k-1}}$ and $Z_{v, \ell_{k-1}}$ in \eqref{YkProof}).
Since $W_i$ is uniformly distributed over $\{1, 2, \ldots, 2^{\ell n R_i}\}$ for each $i\in\{1,2\}$, it follows that
\begin{align}
& \ell n (R_1 + R_2) \notag\\
&= H(W_1) + H(W_2) \notag\\
& = I(W_1; \tilde Y_1^{\ell n})+ I(W_2; \tilde Y_2^{\ell n}) + H(W_1| \tilde Y_1^{\ell n})+H(W_2|\tilde Y_2^{\ell n}) \notag\\
& \stackrel{\text{(a)}}{\le} I(X_1^{\ell n}; \tilde Y_1^{\ell n})+ I(X_2^{\ell n}; \tilde Y_2^{\ell n}) + H(W_1|\tilde Y_1^{\ell n})+H(W_2|\tilde Y_2^{\ell n}) \notag \\
& \stackrel{\text{(b)}}{=} I(X_1^{\ell n}; \tilde Y_1^{\ell n})+ I(X_2^{\ell n}; \tilde Y_2^{\ell n}) + H(W_1|\tilde Y_1^{\ell n}, Z_u^{\ell n}, Z_v^{\ell n})+H(W_2|\tilde Y_2^{\ell n},Z_u^{\ell n}, Z_v^{\ell n}) \notag \\
& \stackrel{\text{(c)}}{\le} I(X_1^{\ell n}; \tilde Y_1^{\ell n})+ I(X_2^{\ell n}; \tilde Y_2^{\ell n}) + H(W_1|Y_1^{\ell n})+H(W_2|Y_2^{\ell n}) \notag \\
& \stackrel{\text{(d)}}{\le} I(X_1^{\ell n}; \tilde Y_1^{\ell n})+ I(X_2^{\ell n}; \tilde Y_2^{\ell n}) + 2 + P_{e,1}^{\ell n}\ell n R_1 + P_{e,2}^{\ell n}\ell n R_2\label{thmProofTemp1}
\end{align}
where
\begin{enumerate}
\item[(a)] follows from Definition~\ref{defCode}, \eqref{tildeY1} and \eqref{tildeY2} that $W_i\rightarrow X_i^{\ell n} \rightarrow \tilde Y_i^{\ell n}$ forms a Markov Chain for each $i\in\{1,2\}$.
\item[(b)] follows from the fact that $(W_1, W_2,\tilde Y_1^{\ell n}, \tilde Y_2^{\ell n})$ and $(Z_u^{\ell n}, Z_v^{\ell n})$ are independent.
\item[(c)] follows from the fact that for each $i\in\{1,2\}$, $Y_i^{\ell n}$ is a function of $(\tilde Y_i^{\ell n}, Z_u^{\ell n}, Z_v^{\ell n})$ (cf. \eqref{YkProof}, \eqref{tildeY1} and \eqref{tildeY2}).
\item[(d)] follows from Fano's inequality.
\end{enumerate}
We now state the following lemma, proved in Appendix~\ref{Lemma3}, to upper bound $I(X_1^{\ell n}; \tilde Y_1^{\ell n})+ I(X_2^{\ell n}; \tilde Y_2^{\ell n})$ in \eqref{thmProofTemp1}.
\bigskip
\begin{Lemma}\label{lemmaLinearDecomposition}
For any sequence of $(\ell n, R_1, R_2)$-codes with their corresponding $\mathbf{G}_{\mathbf{ij},k}$, there exist four real numbers denoted by $\lambda_1$, $\lambda_2$, $\mu_1$ and $\mu_2$. which are only functions of $(\mathbf{H_1},\mathbf{H_2})$, such that
\begin{equation}
 \mathbf{G}_{\mathbf{11},k} = \lambda_1\mathbf{G}_{\mathbf{12},k} + \lambda_2\mathbf{G}_{\mathbf{21},k} \label{G11ThmStatementSumRank}
 \end{equation}
 and
\begin{equation}
\mathbf{G}_{\mathbf{22},k} = \mu_1\mathbf{G}_{\mathbf{12},k} + \mu_2\mathbf{G}_{\mathbf{21},k}\label{G22ThmStatementSumRank}
 \end{equation}
  for all $k\in\{1, 2, \ldots, n\}$.
\end{Lemma}
\bigskip
The importance of Lemma~\ref{lemmaLinearDecomposition} is that it captures the relationship between the direct links and the interference links. More specifically, it expresses the direct links as explicit functions of the interference links. This means that the corresponding MIMO channel, described by equation~\eqref{YkProof}, is ill-conditioned. Using \eqref{G11ThmStatementSumRank} and \eqref{G22ThmStatementSumRank} in Lemma~\ref{lemmaLinearDecomposition}, we obtain from \eqref{tildeY1} and \eqref{tildeY2} that
\begin{equation}
\tilde Y_{1, \ell_k}=  (\lambda_1\mathbf{G}_{\mathbf{12},k} + \lambda_2\mathbf{G}_{\mathbf{21},k})X_{1, \ell_{k-1}} +  \mathbf{G}_{\mathbf{12},k}  X_{2,\ell_{k-1}} + Z_{1, \ell_k} \label{Y1Lk}
\end{equation}
and
\begin{equation}
\tilde Y_{2, \ell_k} = \mathbf{G}_{\mathbf{21},k}X_{1, \ell_{k-1}} +  (\mu_1\mathbf{G}_{\mathbf{12},k} + \mu_2\mathbf{G}_{\mathbf{21},k})  X_{2,\ell_{k-1}} + Z_{2, \ell_k} \label{Y2Lk}
\end{equation}
 for each $k\in\{1, 2, \ldots, n\}$.
Following \eqref{thmProofTemp1}, we consider
\begin{align}
&I(X_1^{\ell n}; \tilde Y_1^{\ell n})+ I(X_2^{\ell n}; \tilde Y_2^{\ell n}) \notag \\
& = h(\tilde Y_1^{\ell n}) - h(\tilde Y_2^{\ell n}|X_2^{\ell n}) + h(\tilde Y_2^{\ell n}) - h(\tilde Y_1^{\ell n}|X_1^{\ell n}) \notag \\
& \stackrel{\text{(a)}}{=} \underbrace{h( \{(\lambda_1\mathbf{G}_{\mathbf{12},k} + \lambda_2\mathbf{G}_{\mathbf{21},k})X_{1, \ell_{k-1}} +  \mathbf{G}_{\mathbf{12},k}  X_{2,\ell_{k-1}} + Z_{1, \ell_k}\}_{k=1}^n) -h(\{\mathbf{G}_{\mathbf{21},k}X_{1, \ell_{k-1}}+Z_{2, \ell_k}\}_{k=1}^n)}_{\triangleq I_1} \notag \\
&\qquad + \underbrace{h( \{\mathbf{G}_{\mathbf{21},k}X_{1, \ell_{k-1}} +  (\mu_1\mathbf{G}_{\mathbf{12},k} + \mu_2\mathbf{G}_{\mathbf{21},k})  X_{2,\ell_{k-1}} + Z_{2, \ell_k}\}_{k=1}^n)-h(\{\mathbf{G}_{\mathbf{12},k}X_{2, \ell_{k-1}}+Z_{1, \ell_k}\}_{k=1}^n)}_{\triangleq I_2} \label{thmProofTemp1***}
\end{align}
where (a) follows from \eqref{Y1Lk}, \eqref{Y2Lk} and the fact that $X_1^{\ell n}$, $X_2^{\ell n}$ and $(Z_1^{\ell n}, Z_2^{\ell n})$ are independent.
We then state the following lemma, proved in Appendix~\ref{Lemma4}, to bound $I_1$ and $I_2$ as defined in \eqref{thmProofTemp1***}.
\bigskip
\begin{Lemma} \label{lemmaDifferenceEntropy}
Let $Z_1^n$ and $Z_2^n$ be two continuous random vectors and let $X^n$ and $Y^n$ be two general random vectors such that $Z_1^n$, $Z_2^n$ and $(X^n,Y^n)$ are independent.
Then, for any $n\times n$ matrix $\mathbf{L}$,
\[
h(X^n+Z_1^n) - h(Y^n+Z_2^n) \le h(X^n - \mathbf{L} Y^n + Z_1^n- \mathbf{L} Z_2^n|Y^n+Z_2^n)-h(Z_2^n).
\]
\end{Lemma}
\bigskip
In order to bound $I_1$, we apply Lemma~\ref{lemmaDifferenceEntropy} by setting
\begin{equation*}
\begin{cases}
X^n=\{(\lambda_1\mathbf{G}_{\mathbf{12},k} + \lambda_2\mathbf{G}_{\mathbf{21},k})X_{1, \ell_{k-1}} +  \mathbf{G}_{\mathbf{12},k}  X_{2,\ell_{k-1}}\}_{k=1}^n, \\
Y^n=\{\mathbf{G}_{\mathbf{21},k}X_{1, \ell_{k-1}}\}_{k=1}^n, \\
Z_1^n = \{Z_{1,\ell_k}\}_{k=1}^n, \\
 Z_2^n = \{Z_{2,\ell_k}\}_{k=1}^n, \\
 \mathbf{L}=\lambda_2 \mathbf{I}_{\ell n},
\end{cases}
\end{equation*}
and obtain
\begin{equation}
I_1 \le  h(\{\mathbf{G}_{\mathbf{12},k}(\lambda_1 X_{1, \ell_{k-1}} +  X_{2,\ell_{k-1}}) + Z_{1, \ell_k} - \lambda_2 Z_{2, \ell_k}\}_{k=1}^n)- h(Z_2^{\ell n}). \label{I1temp}
\end{equation}
Following similar procedures for proving \eqref{I1temp}, we obtain
\begin{equation}
I_2 \le  h(\{\mathbf{G}_{\mathbf{21},k}(X_{1, \ell_{k-1}} + \mu_2X_{2,\ell_{k-1}})  + Z_{2, \ell_k}-\mu_1Z_{1, \ell_k}\}_{k=1}^n) - h(Z_1^{\ell n}). \label{I2temp}
\end{equation}
Since $\{Z_{1,m}\}_{m=1}^{\ell n}$ are independent, $\{Z_{2,m}\}_{m=1}^{\ell n}$ are independent and the differential entropy of $\mathcal{N}(0,1)$ is positive,
it then follows from \eqref{thmProofTemp1***}, \eqref{I1temp} and \eqref{I2temp} that
\begin{align}
&I(X_1^{\ell n}; \tilde Y_1^{\ell n})+ I(X_2^{\ell n}; \tilde Y_2^{\ell n}) \notag \\ &\quad \le \sum_{k=1}^n (h(\mathbf{G}_{\mathbf{12},k}(\lambda_1 X_{1, \ell_{k-1}} +  X_{2,\ell_{k-1}}) + Z_{1, \ell_k} - \lambda_2 Z_{2, \ell_k})
 +h(\mathbf{G}_{\mathbf{21},k}(X_{1, \ell_{k-1}} + \mu_2X_{2,\ell_{k-1}})  + Z_{2, \ell_k}-\mu_1Z_{1, \ell_k})).\label{thmFirstIneq}
\end{align}
We finally need the following lemma, proved in Appendix~\ref{Lemma5}, to bound the terms in \eqref{thmFirstIneq}.
\bigskip
\begin{Lemma} \label{lemmaDifferentialEntropyBound1}
There exist two real numbers, denoted by $\kappa$ and $\kappa^\prime$, that do not depend on $n$ and $P$ such that for each $k\in\{1, 2, \ldots, n\}$,
\[
 h(\mathbf{G}_{\mathbf{12},k}(\lambda_1 X_{1, \ell_{k-1}} +  X_{2,\ell_{k-1}}) + Z_{1, \ell_k} - \lambda_2 Z_{2, \ell_k}) \le \rank(\mathbf{G}_{\mathbf{12},k}) \log_2 \sqrt{1+P_{k-1}/\ell} + \kappa
\]
and
\[
 h(\mathbf{G}_{\mathbf{21},k}(X_{1, \ell_{k-1}} + \mu_2X_{2,\ell_{k-1}})  + Z_{2, \ell_k}-\mu_1Z_{1, \ell_k})  \le \rank(\mathbf{G}_{\mathbf{21},k}) \log_2 \sqrt{1+P_{k-1}/\ell} + \kappa^\prime,
\]
where $P_{k-1}$ is defined in \eqref{averagePowerPi}.
\end{Lemma}
\bigskip
Using \eqref{thmProofTemp1}, \eqref{thmFirstIneq} and Lemma~\ref{lemmaDifferentialEntropyBound1}, we obtain
 \begin{equation}
\sum_{i=1}^2(1-P_{e,i}^{\ell n})R_i \le \tau_1+ \frac{1}{2\ell n}\sum_{k=1}^n (\rank(\mathbf{G}_{\mathbf{21},k})+\rank(\mathbf{G}_{\mathbf{21},k})) \log_2 (1+P_{k-1}/\ell) \label{firstBoundFinalstatement}
 \end{equation}
for some $\tau_1$ that does not depend on $n$ and $P$. It then follows from \eqref{firstBoundFinalstatement} and \eqref{errorProbability} that Bound (i) holds for sufficiently large~$n$. \bigskip\\
\subsubsection{Proof for Bound (ii) in Lemma~\ref{lemmaMain}} \label{proofBound2}
Following \eqref{thmProofTemp1}, we consider
\begin{align}
& I(X_1^{\ell n}; \tilde Y_1^{\ell n})+ I(X_2^{\ell n}; \tilde Y_2^{\ell n})\notag \\
& \stackrel{\text{(a)}}{\le} I(X_1^{\ell n}; \tilde Y_1^{\ell n})+ I(X_2^{\ell n}; \tilde Y_2^{\ell n}|X_1^{\ell n})\label{thmProofTemp2s} \\
& = h(\tilde Y_1^{\ell n}) - h(\tilde Y_2^{\ell n}|X_1^{\ell n}, X_2^{\ell n}) + h(\tilde Y_2^{\ell n}|X_1^{\ell n}) - h(\tilde Y_1^{\ell n}|X_1^{\ell n})\notag \\
& \stackrel{\text{(b)}}{=} h(\tilde Y_1^{\ell n})- h(Z_2^{\ell n})  + h(\{\mathbf{G}_{\mathbf{22},k}X_{2, \ell_{k-1}}+Z_{2, \ell_k}\}_{k=1}^n) - h(\{\mathbf{G}_{\mathbf{12},k}X_{2, \ell_{k-1}}+Z_{1, \ell_k}\}_{k=1}^n)\notag \\
&\stackrel{\text{(c)}}{\le}  h(\tilde Y_1^{\ell n})- h(Z_2^{\ell n})  + \notag \\
 &\quad   +h(\{\mathbf{G}_{\mathbf{22},k}X_{2, \ell_{k-1}} + Z_{2, \ell_k}\}_{k=1}^n | \{   \mathbf{G}_{\mathbf{12},k} X_{2, \ell_{k-1}}+Z_{1, \ell_k}   \}_{k=1}^n ) - h(Z_1^{\ell n})\notag \\
  &\stackrel{\text{(d)}}{\le} \sum_{k=1}^n (h(\tilde Y_{1, \ell_k})    +h(\mathbf{G}_{\mathbf{22},k}X_{2, \ell_{k-1}} + Z_{2, \ell_k}|\mathbf{G}_{\mathbf{12},k} X_{2, \ell_{k-1}}+Z_{1, \ell_k})), \label{thmSecondInequality2}
\end{align}
where
\begin{enumerate}
\item[(a)] follows from the fact that $X_1^{\ell n}$ and $X_2^{\ell n}$ are independent.
\item[(b)] follows from \eqref{tildeY1}, \eqref{tildeY2} and the fact that $X_1^{\ell n}$, $X_2^{\ell n}$ and $(Z_1^{\ell n}, Z_2^{\ell n})$ are independent.
\item[(c)] follows from Lemma~\ref{lemmaDifferenceEntropy}.
\item[(d)] follows from the facts that $\{Z_{1,m}\}_{m=1}^{\ell n}$ are independent, $\{Z_{2,m}\}_{m=1}^{\ell n}$ are independent and the differential entropy of $\mathcal{N}(0,1)$ is positive.
\end{enumerate}
We need the following lemma, proved in Appendix~\ref{Lemma6} by following the genie-aided bound approach, to bound the terms in~\eqref{thmSecondInequality2}.
\bigskip
\begin{Lemma} \label{lemmaDifferentialEntropyBound2}
There exist two real numbers, denoted by $\kappa$ and $\kappa^\prime$, that do not depend on $n$ and $P$ such that for each $k\in\{1, 2, \ldots, n\}$,
\[
 h(\tilde Y_{1, \ell_k}) \le \ell  \log_2 \sqrt{1+P_{k-1}/\ell} + \kappa
\]
and
\[
 h(\mathbf{G}_{\mathbf{22},k}X_{2, \ell_{k-1}} + Z_{2, \ell_k}|\mathbf{G}_{\mathbf{12},k} X_{2, \ell_{k-1}}+Z_{1, \ell_k})  \le (\ell-\rank(\mathbf{G}_{\mathbf{12},k})) \log_2 \sqrt{1+P_{k-1}/\ell} + \kappa^\prime,
\]
where $P_{k-1}$ is defined in \eqref{averagePowerPi}.
\end{Lemma}
\bigskip
Using \eqref{thmProofTemp1}, \eqref{thmSecondInequality2} and Lemma~\ref{lemmaDifferentialEntropyBound2}, we obtain
\begin{equation}
\sum_{i=1}^2(1-P_{e,i}^{\ell n})R_i \le \tau_2 + \frac{1}{2\ell n} \sum_{k=1}^n (2\ell-\rank(\mathbf{G}_{\mathbf{12},k}))\log_2 (1+P_{k-1}/\ell) \label{bound2*s}
\end{equation}
for some $\tau_2$ that does not depend on $n$ and $P$. It then follows from \eqref{bound2*s} and \eqref{errorProbability} that Bound (ii) holds for sufficiently large~$n$. \\
\subsubsection{Proof for Bound (iii) in Lemma~\ref{lemmaMain}} \label{proofBound3}
Following similar procedures for proving \eqref{thmProofTemp2s} and then \eqref{thmSecondInequality2}, we obtain
 \begin{align}
&  I(X_1^{\ell n}; \tilde Y_1^{\ell n})+ I(X_2^{\ell n}; \tilde Y_2^{\ell n})\notag \\
&\, \le I(X_1^{\ell n}; \tilde Y_1^{\ell n}|X_2^{\ell n})+ I(X_2^{\ell n}; \tilde Y_2^{\ell n}) \notag \\
 &\, \le   \sum_{k=1}^n (h(\tilde Y_{2, \ell_k})    +h(\mathbf{G}_{\mathbf{11},k}X_{1, \ell_{k-1}} + Z_{1, \ell_k}|\mathbf{G}_{\mathbf{21},k} X_{1, \ell_{k-1}}+Z_{2, \ell_k})).
\label{thmProofBound3Temp1s}
\end{align}
In addition, following similar procedures for proving \eqref{bound2*s} from \eqref{thmSecondInequality2} and \eqref{thmProofTemp1}, we obtain from \eqref{thmProofBound3Temp1s} and \eqref{thmProofTemp1} that
\begin{equation}
\sum_{i=1}^2(1-P_{e,i}^{\ell n})R_i \le \tau_3 + \frac{1}{2\ell n} \sum_{k=1}^n (2\ell-\rank(\mathbf{G}_{\mathbf{21},k}))\log_2 (1+P_{k-1}/\ell) \label{bound3*s}
\end{equation}
for some $\tau_3$ that does not depend on $n$ and $P$. It then follows from \eqref{bound3*s} and \eqref{errorProbability} that Bound (iii) holds for sufficiently large~$n$. 

\section{Two-Hop IC with Multiple-Antenna Nodes} \label{twoHopICMultipleAntennas}
In this section, we consider a more general setting, in which each node has $M$ antennas. In particular, we prove Theorem~\ref{thmDoFMIMO}.
\subsection{Network Model}\label{formulationMIMO}

\begin{figure}
\centering
\includegraphics[scale=0.65]{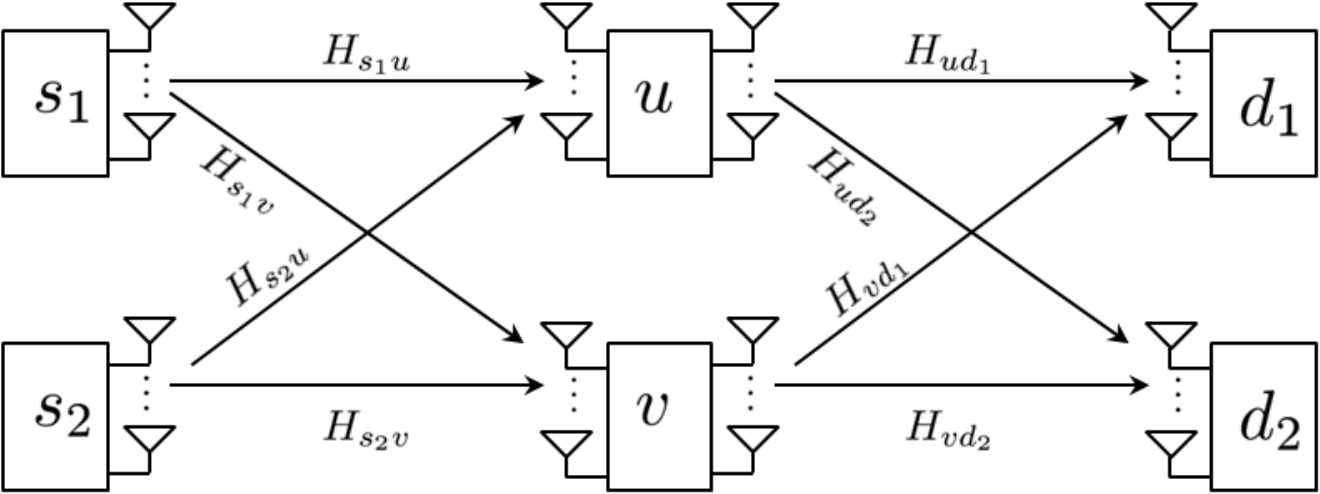}
\caption{Two-Hop MIMO IC.}\label{twoHopICMIMO}
\end{figure}

This section considers the two-hop MIMO IC, in which each node is equipped with $M$ antennas. The two-hop MIMO IC is illustrated in Figure~\ref{twoHopICMIMO}, where each $H_{ij}\in \mathbb{R}^{M \times M}$ characterizes the channels between node~$i$ and node~$j$. The channel gains of the first hop are denoted by
\begin{equation*}
\mathbf{H_1}=\left[\begin{array}{cc}H_{s_1 u} & H_{s_2 u}\\ H_{s_1 v} & H_{s_2 v}\end{array}\right] \in \mathbb{R}^{2M \times 2M},
 \end{equation*}
 and the channel gains of the second hop are denoted by
 \begin{equation*}
 \mathbf{H_2}=\left[\begin{array}{cc}H_{ud_1} & H_{vd_1}\\ H_{ud_2} & H_{vd_2}\end{array}\right] \in \mathbb{R}^{2M \times 2M}.
  \end{equation*}
All the nodes have the knowledge of $\mathbf{H_1}$ and $\mathbf{H_2}$.
To facilitate discussion, let $M_k = \{M(k-1)+1, M(k-1)+2, \ldots, Mk\}$ for each $k\in\{1, 2, \ldots, n\}$. For each $i\in\{1, 2\}$, let $X_{i, M_k}\in \mathbb{R}^M$ be the symbols transmitted by source~$s_i$ in the $k^{\text{th}}$ time slot, where $X_{i,M(k-1) + m}\in \mathbb{R}$ denotes the symbol transmitted by $s_i$ through its $m^{\text{th}}$ antenna. In addition, for each $r\in\{u,v\}$, let $Y_{r,M(k-1)+m}\in \mathbb{R}$ denote the symbol received by relay~$r$ through its $m^{\text{th}}$ antenna in the $k^{\text{th}}$ time slot. Then, the symbols received by $u$ and $v$ in the $k^{\text{th}}$ time slot time slot satisfy
\begin{equation}
\left[\begin{array}{c}Y_{u,M_k}\\Y_{v,M_k} \end{array}\right] = \mathbf{H_1}\left[\begin{array}{c}X_{1,M_k}\\X_{2,M_k} \end{array}\right] + \left[\begin{array}{c}Z_{u,M_k}\\Z_{v,M_k} \end{array}\right], \label{firstChannelAdditionMIMO}
\end{equation}
where $Z_{r, M(k-1)+m}\sim \mathcal{N}(0,1)$ denote the noise variable received by relay~$r$ through its $m^{\text{th}}$ antenna for each $r\in\{u,v\}$. Similarly, let $X_{r, M_k}\in \mathbb{R}^M$ be the symbols transmitted by relay~$r$ in the $k^{\text{th}}$ time slot for each $r\in\{u,v\}$. Then, the symbols received in the same time slot by $d_1$ and $d_2$, denoted by $Y_{1,M_k}\in \mathbb{R}^M$ and $Y_{2,M_k}\in \mathbb{R}^M$ respectively, satisfy
\begin{equation}
\left[\begin{array}{c}Y_{1,M_k}\\Y_{2, M_k} \end{array}\right] = \mathbf{H_2}\left[\begin{array}{c}X_{u,M_k}\\X_{v,M_k} \end{array}\right] + \left[\begin{array}{c}Z_{1,M_k}\\Z_{2,M_k} \end{array}\right], \label{secondChannelAdditionMIMO}
\end{equation}
where $Z_{i, M(k-1)+m}\sim \mathcal{N}(0,1)$ denote the noise variable received by $d_i$ through its $m^{\text{th}}$ antenna for each $i\in\{1,2\}$.  We assume that $X_1^{Mn}$, $X_2^{Mn}$, $Z_u^{Mn}$, $Z_v^{Mn}$, $Z_1^{Mn}$ and $Z_2^{Mn}$ are independent, and we assume that $\{(Z_{u,k}, Z_{v,k}, Z_{1,k},Z_{2,k})\}_{k=1}^{Mn}$ are independent.
For each $i\in\{1, 2, u, v\}$, any codeword $[x_{i,1} \,  x_{i,2} \, \ldots \, x_{i,Mn}]^T$ that is transmitted over the network should satisfy
$||x_i^{Mn}||^2 \le nP$,
where~$P$ represents the power constraint for all the nodes.
The definitions of an $(n,R_1,R_2)$ code and error probability of a code are very similar to those for single-antenna case (cf.\ Definitions~\ref{defCode} and~\ref{defErrorProbability}), and are thus omitted. We now define linear schemes and linear sum-DoF for the MIMO case.
  \bigskip
\begin{Definition} \label{defMIMOLinear}
Let $\mathcal{U}$ be a finite set of real numbers. An $(n, R_1, R_2)$-code on the two-hop MIMO IC is said to be \textit{linear on $\mathcal{U}$} if there exist $\{\mathbf{A}_k\in \mathcal{U}^{M \times M} \}_{k=1}^{n}$ and $\{\mathbf{B}_k \in \mathcal{U}^{M\times M}\}_{k=1}^{n}$ such that for each $k\in\{1, 2, \ldots, n\}$,
\[
X_{u,M_k} = \mathbf{A}_k Y_{u,M_{k-1}}
\]
and
\[
X_{v,M_k} = \mathbf{B}_k Y_{v,M_{k-1}}.
\]
In other words, the relays are operating over blocks of length $M$, where the symbols in each block to be transmitted in the $k^{\text{th}}$ time slot are linear combinations of the $M$ symbols received in the $(k-1)^{\text{th}}$ time slot. We call $(\{\mathbf{A}_k\in \mathcal{U}^{M \times M} \}_{k=1}^{n},\{\mathbf{B}_k\in \mathcal{U}^{M \times M} \}_{k=1}^{n})$ a \textit{relaying kernel} of the code.
\end{Definition}
\bigskip
\begin{Definition} \label{defAchievableMIMO}
A rate pair $(R_1, R_2)$ is  \textit{linear achievable on $\mathcal{U}$} if there exists a sequence of $(n, R_1, R_2)$-codes that are linear on $\mathcal{U}$ such that $\lim\limits_{n\rightarrow \infty} P_{e,i}^{ n}= 0$ for each $i\in\{1,2\}$.
\end{Definition}
\bigskip
\begin{Definition} \label{defDoFMIMO}
The \emph{linear sum-DoF} of the two-hop MIMO IC with real channel gains, denoted by $\mathcal{D}_{\text{MIMO}}$, is defined by
\[
\mathcal{D}_{\text{MIMO}}= \sup_{\mathcal{U}}\lim_{P\rightarrow\infty}\! \sup\left\{\left. \frac{R_1+R_2}{\frac{1}{2}\log_2 P} \:\right|(R_1, R_2) \text{ is linear achievable on }\mathcal{U}\right\}.
\]
\end{Definition}
\bigskip
In the following two subsections, we will prove Theorem~\ref{thmDoFMIMO}, by first proving that we can achieve $2M-2/3$ sum-DoF, and then proving the converse statement.

\subsection{Achievability Proof of Theorem~\ref{thmDoFMIMO}} \label{achMIMO}

\begin{figure}[!hbt]
\centering
\subfigure[Phase 1]{\includegraphics[scale=0.7]{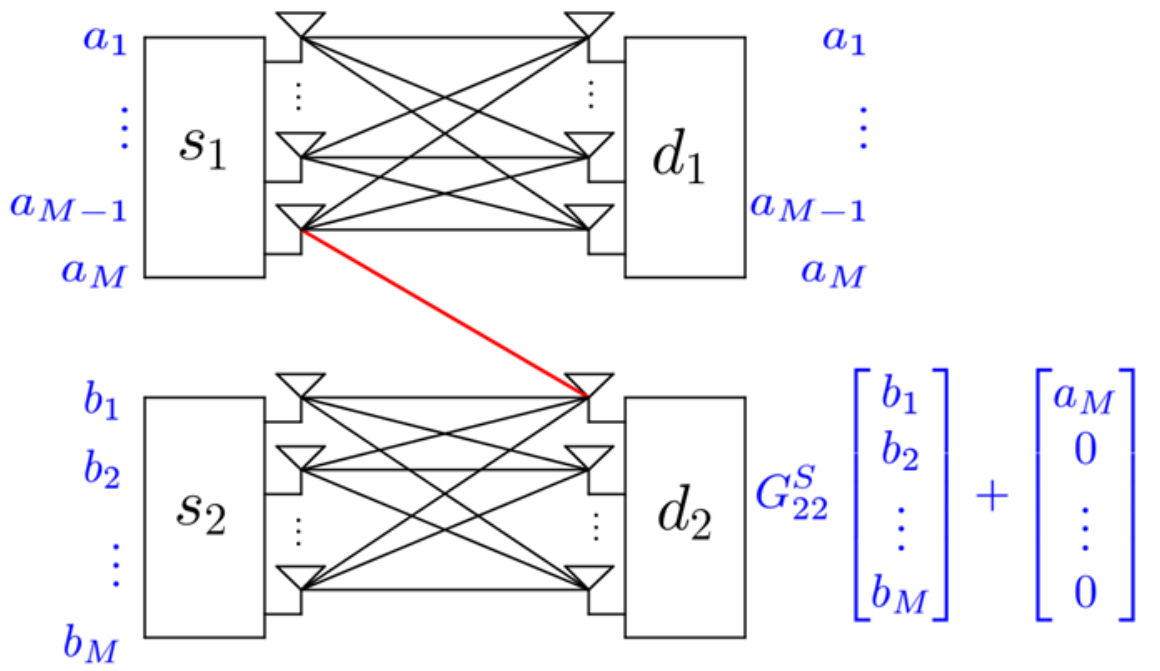}} \hspace{0.2in}
\subfigure[Phase 2]{\includegraphics[scale=0.7]{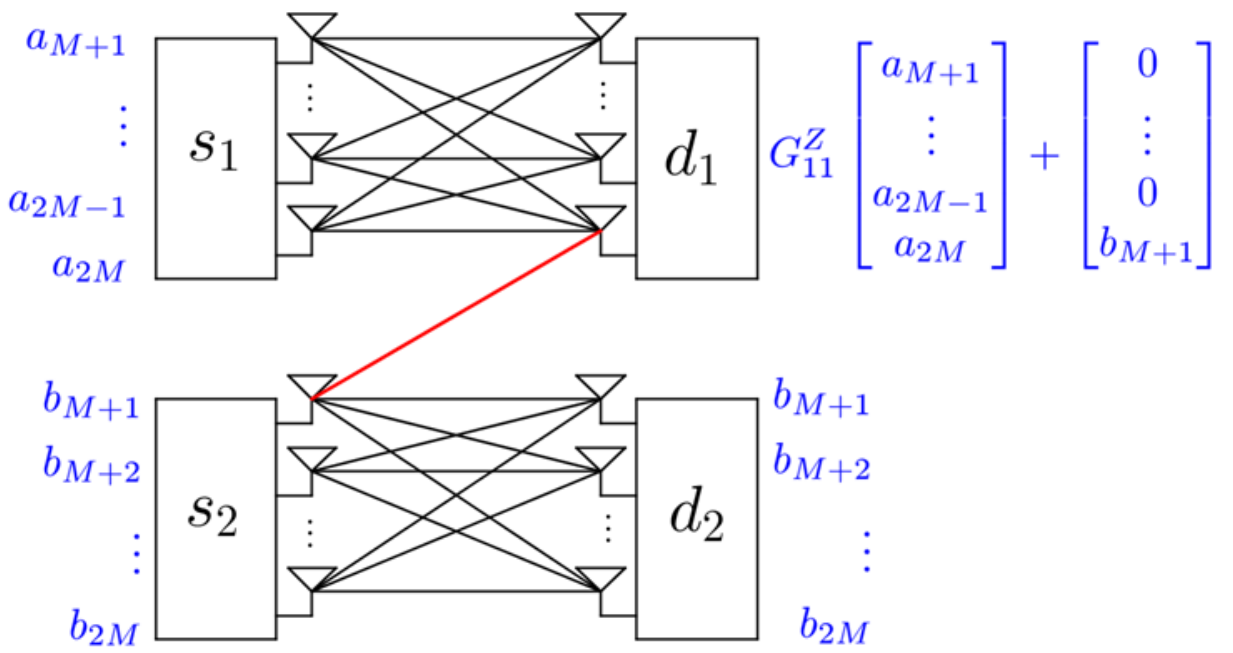}\vspace{0.1in}}
\subfigure[Phase 3]{\includegraphics[scale=0.7]{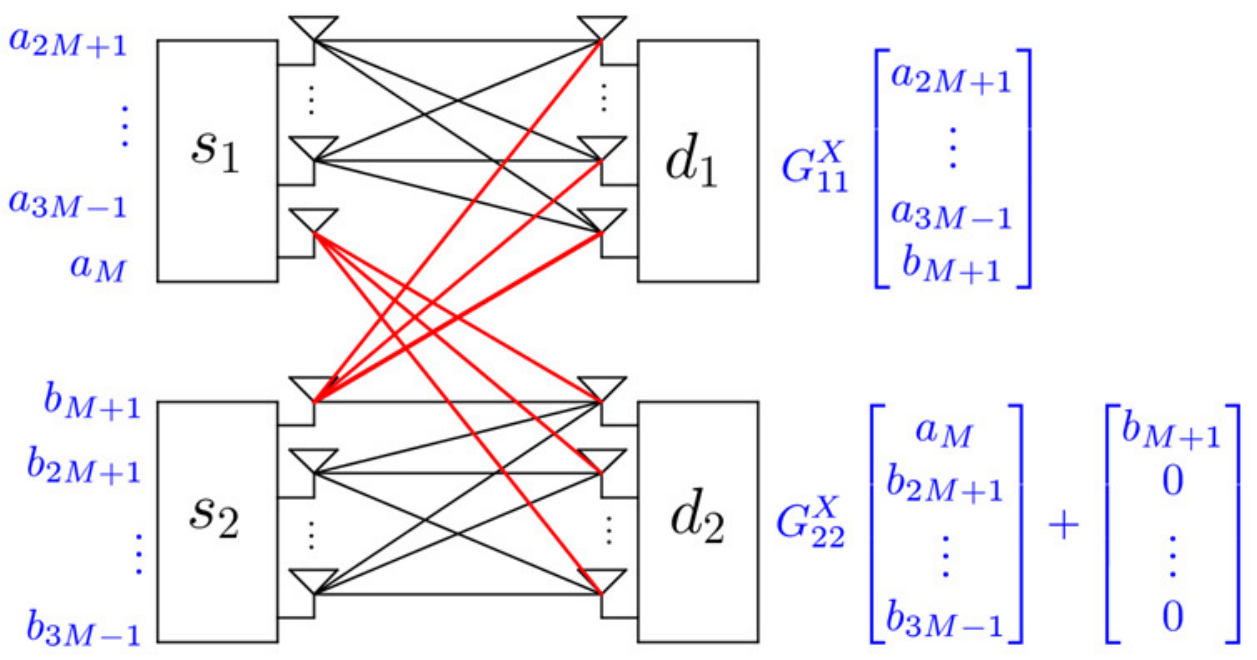}}
\caption{Illustration of achievability scheme for two-hop MIMO channel. At each phase, the transmit symbols by sources are shown on the left. The  received signals at destinations are given on the right, where the noise is dropped.}
\label{MIMOscheme}
\end{figure}

Before describing the achievability, we will first set up some notation and prove a main lemma. For any two $M \times M$ matrices, denoted by $\mathbf{A}$ and $\mathbf{B}$, define
\begin{align}
\mathbf{G}_{\mathbf{ij}}^{\mathbf{(A,B)}} & = H_{ud_i} \mathbf{A} H_{s_ju} +  H_{vd_i} \mathbf{B} H_{s_jv}. \label{GijM}
\end{align}
For $i,j \in \{1,2\}$, $\mathbf{G}_{\mathbf{ij}}^{\mathbf{(A,B)}}$ denotes the effective end-to-end link between source $s_j$ and destination $d_i$ when relays $u$ and $v$ set their amplifying matrices to be $\mathbf{A}$ and $\mathbf{B}$ respectively. In other words, for any time~$k$, if we define the matrices used by relays $u$ and $v$ by $\mathbf{A}_k$ and $\mathbf{B}_k$ respectively, then the $M\times 1$ vectors received at destinations $d_1$ and $d_2$ can be written as
\begin{align}
\left[\begin{array}{c}Y_{1,M_k} \\ Y_{2,M_k} \end{array}\right] &\stackrel{\eqref{secondChannelAdditionMIMO}}{=} \left[\begin{array}{c}  H_{ud_1} X_{u,M_k}  \\  H_{ud_2} X_{u,M_k}  \end{array}\right]+\left[\begin{array}{c}  H_{vd_1} X_{v,M_k} \\  H_{vd_2} X_{v,M_k} \end{array}\right] + \left[\begin{array}{c}Z_{1,M_k} \\ Z_{2,M_k} \end{array}\right] \notag \\
& \stackrel{\text{(a)}}{=} \left[\begin{array}{c}  H_{ud_1} \mathbf{A}_k Y_{u,M_{k-1}}  \\  H_{ud_2}  \mathbf{A}_k Y_{u,M_{k-1}}  \end{array}\right]+\left[\begin{array}{c}  H_{vd_1}  \mathbf{B}_k Y_{v,M_{k-1}} \\  H_{vd_2}  \mathbf{B}_k Y_{v,M_{k-1}} \end{array}\right] + \left[\begin{array}{c}Z_{1,M_k} \\ Z_{2,M_k} \end{array}\right] \notag \\
& \stackrel{\text{(b)}}{=} \left[\begin{array}{cc}\mathbf{G}_{\mathbf{11}}^{(\mathbf{A}_k,\mathbf{B}_k)} & \mathbf{G}_{\mathbf{12}}^{(\mathbf{A}_k,\mathbf{B}_k)} \\ \mathbf{G}_{\mathbf{21}}^{(\mathbf{A}_k,\mathbf{B}_k)} & \mathbf{G}_{\mathbf{22}}^{(\mathbf{A}_k,\mathbf{B}_k)}
\end{array}\right]
\left[\begin{array}{c}X_{1,M_{k-1}} \\ X_{2,M_{k-1}}\end{array}\right] + \left[ \begin{array}{cc}  H_{ud_1} \mathbf{A}_k &  H_{vd_1} \mathbf{B}_k \\  H_{ud_2} \mathbf{A}_k &  H_{vd_2} \mathbf{B}_k \end{array}\right] \left[\begin{array}{c}Z_{u,M_{k-1}} \\Z_{v,M_{k-1}} \end{array}\right] + \left[\begin{array}{c}Z_{1,M_k}\\ Z_{2,M_k}\end{array}\right], \label{YkProofMIMOe}
\end{align}
for each $k\in\{1,2,\ldots, n\}$, where
\begin{enumerate}
\item[(a)] follows from Definition~\ref{defMIMOLinear}.
\item[(b)] follows from follows from \eqref{firstChannelAdditionMIMO} and \eqref{GijM}.
\end{enumerate}
\bigskip
\noindent Now, let $c= \begin{bmatrix}
1 &  0^{1 \times M-1}
\end{bmatrix}^T$, and $c'= \begin{bmatrix}
 0^{1 \times M-1} & 1
\end{bmatrix}^T$. We state the following two lemmas, which are proved in Appendix~\ref{LemmaS}.
\bigskip
\begin{Lemma} \label{lemmaS}
For almost all values of channel gains, $\exists $ a pair of matrices $\mathbf{A^S}$ and $\mathbf{B^S}$, such that $\mathbf{G}_{\mathbf{12}}^{\mathbf{(A^S,B^S)}} = \mathbf{0}^{M\times M}$, $\text{rank}(\mathbf{G}_{\mathbf{11}}^{\mathbf{(A^S,B^S)}}) =\text{rank}(\mathbf{G}_{\mathbf{22}}^{\mathbf{(A^S,B^S)}})=M$, and $\mathbf{G}_{\mathbf{21}}^{\mathbf{(A^S,B^S)}} = \begin{bmatrix}
\mathbf{0}^{M \times M-1} & c
\end{bmatrix}$.
\end{Lemma}
\bigskip
\begin{Lemma} \label{lemmaZ}
For almost all values of channel gains, $\exists $ a pair of matrices $\mathbf{A^Z}$ and $\mathbf{B^Z}$, such that $\mathbf{G}_{\mathbf{21}}^{\mathbf{(A^Z,B^Z)}} = \mathbf{0}^{M\times M}$, $\text{rank}(\mathbf{G}_{\mathbf{11}}^{\mathbf{(A^Z,B^Z)}}) =\text{rank}(\mathbf{G}_{\mathbf{22}}^{\mathbf{(A^Z,B^Z)}})=M$, and $\mathbf{G}_{\mathbf{12}}^{\mathbf{(A^Z,B^Z)}} = \begin{bmatrix}
c' & \mathbf{0}^{M \times M-1}
\end{bmatrix}$.
\end{Lemma}
\bigskip
\begin{Remark}
In words, Lemma~\ref{lemmaS} states that there exists a choice of matrices at the relays such that in the equivalent end-to-end channel, only one antenna (namely the $M^{\text{th}}$) from source $s_1$ causes interference at one antenna (namely the first) at destination $d_2$, while source $s_2$ causes no interference and the direct links are invertible. The equivalent end-to-end topology, which we will call the MIMO-S topology, is shown in Figure~\ref{MIMOscheme}(a). Similarly, Lemma~\ref{lemmaZ} states that the symmetric topology, which we will call the MIMO-Z topology (shown in Figure~\ref{MIMOscheme}(b)), can also be created, where only one antenna (the first) from source $s_2$ causes interference at one antenna (the $M^{\text{th}}$) at destination $d_1$, while source $s_1$ causes no interference and the direct links are invertible.
\end{Remark}
\bigskip
\begin{Remark} \label{LemmaAspect}
Note that the important aspect of Lemmas~\ref{lemmaS} and~\ref{lemmaZ} is the fact that we can create an end-to-end channel where only \emph{one} antenna causes interference, and the direct links are invertible. This means that we reduced the interference to the minimum possible, whilst not affecting the rank of the direct links. Also note that it is irrelevant which specific antenna causes interference, and whether its signal is received at one or more antennas at the other destination. We merely chose the specific topologies above for ease of proof (and luckily, they make nicer figures).
\end{Remark}
\bigskip

\noindent Before stating the third lemma, we need to set up some notation. Let \begin{align}
\tilde{X}_{1,M_k} = [X_{1,M(k-1)+1}, X_{1,M(k-1)+2}, \dots, X_{1,M(k-1)+M-1}, X_{2,M(k-1)+1}]^T
\label{modifieds1} \\
\intertext{and} \tilde{X}_{2,M_k} = [X_{1,Mk},  X_{2,M(k-1)+2},  X_{2,M(k-1)+3},\dots, X_{2,Mk}]^T.
\label{modifieds2}
\end{align}
In words, $\tilde{X}_{1,M_k}$ denotes the symbols transmitted, at time $k$, by the first $M-1$ antennas of source $s_1$ and the first antenna of source $s_2$. We will call this modified ``source'' $\tilde{s}_1$. We will denote the channel submatrices between the modified source and relays $u$ and $v$ by $H_{\tilde{s}_1 u}$ and $H_{\tilde{s}_1 v}$ respectively. Similarly, $\tilde{X}_{2,M_k}$ denotes the symbols transmitted, at time $k$, by the $M^{\text{th}}$ antenna of source $s_1$  and the last $M-1$ antennas of source $s_2$. We will call this modified source $\tilde{s}_2$. We will denote the channel submatrices between the modified source and relays $u$ and $v$ by $H_{\tilde{s}_2 u}$ and $H_{\tilde{s}_2 v}$ respectively. We also define
\begin{equation*}
\mathbf{\tilde{G}}_{\mathbf{ij}}^{\mathbf{(A,B)}} = H_{ud_i} \mathbf{A} H_{\tilde{s}_ju} +  H_{vd_i} \mathbf{B} H_{\tilde{s}_jv}. \label{Gijt}
\end{equation*}

\bigskip
\begin{Remark}
For $i,j \in \{1,2\}$, $\mathbf{\tilde{G}}_{\mathbf{ij}}^{\mathbf{(A,B)}}$ denotes the effective end-to-end link between source $\tilde{s}_j$ and destination $d_i$ when relays $u$ and $v$ set their amplifying matrices to be $\mathbf{A}$ and $\mathbf{B}$ respectively. In other words, if we define the matrices used by relays $u$ and $v$ by $\mathbf{A}_k$ and $\mathbf{B}_k$ respectively and follow similar procedures for deriving \eqref{YkProofMIMOe}, then we can write the $M\times 1$ vectors received at destinations $d_1$ and $d_2$ as
\begin{equation} \label{YkProofMIMOe2}
\begin{bmatrix}
Y_{1,M_k} \\ Y_{2,M_k}
\end{bmatrix} = \left[\begin{array}{cc}\mathbf{\tilde{G}}_{\mathbf{11}}^{(\mathbf{A}_k,\mathbf{B}_k)} & \mathbf{\tilde{G}}_{\mathbf{12}}^{(\mathbf{A}_k,\mathbf{B}_k)} \\ \mathbf{\tilde{G}}_{\mathbf{21}}^{(\mathbf{A}_k,\mathbf{B}_k)} & \mathbf{\tilde{G}}_{\mathbf{22}}^{(\mathbf{A}_k,\mathbf{B}_k)}
\end{array}\right]
\left[\begin{array}{c}\tilde{X}_{1,M_{k-1}} \\ \tilde{X}_{2,M_{k-1}}\end{array}\right] + \left[ \begin{array}{cc}  H_{ud_1} \mathbf{A}_k &  H_{vd_1} \mathbf{B}_k \\  H_{ud_2} \mathbf{A}_k &  H_{vd_2} \mathbf{B}_k \end{array}\right] \left[\begin{array}{c}Z_{u,M_{k-1}} \\Z_{v,M_{k-1}} \end{array}\right] + \left[\begin{array}{c}Z_{1,M_k}\\ Z_{2,M_k}\end{array}\right],
\end{equation}
where $\tilde{X}_{1,M_k}$ and $\tilde{X}_{2,M_k}$ are defined in \eqref{modifieds1} and \eqref{modifieds2} respectively.
\end{Remark}
\bigskip
\begin{Lemma} \label{lemmaX}
For almost all values of channel gains, $\exists$ a unique pair of matrices $\mathbf{A^X}$ and $\mathbf{B^X}$, such that $\mathbf{\tilde{G}}_{\mathbf{12}}^{\mathbf{(A^X,B^X)}} = \mathbf{0}^{M\times M}$, $\text{rank}(\mathbf{\tilde{G}}_{\mathbf{11}}^{\mathbf{(A^X,B^X)}}) =\text{rank}(\mathbf{\tilde{G}}_{\mathbf{22}}^{\mathbf{(A^X,B^X)}})=M$, and $\mathbf{\tilde{G}}_{\mathbf{21}}^{\mathbf{(A^X,B^X)}} = \begin{bmatrix}
 \mathbf{0}^{M \times M-1} & c
\end{bmatrix}$.
\end{Lemma}
\bigskip
\begin{IEEEproof}
Note that in the proof of Lemma~\ref{lemmaS}, we did not need any precoding. Therefore, the proof is the same up to replacing $H_{s_1u}$, $H_{s_1v}$, $H_{s_2u}$, and $H_{s_2v}$ by $H_{\tilde{s}_1u}$, $H_{\tilde{s}_1v}$, $H_{\tilde{s}_2u}$, and $H_{\tilde{s}_2v}$ respectively.
\end{IEEEproof}
\bigskip
The topology created by Lemma~\ref{lemmaX} is shown in Figure~\ref{MIMOscheme}(c). Finally, for any time $k$, we will denote by $\mathbf{A}_k$ and $\mathbf{B}_k$ the matrices used by relays $u$ and $v$ respectively.
%
For shorter notation, we will let
\begin{equation*}
\tilde{Z}_{i,M_k}=H_{ud_i} \mathbf{A}_k Z_{u,M_{k-1}} + H_{vd_i} \mathbf{B}_k Z_{v,M_{k-1}} + Z_{i,M_k}
\end{equation*}
 be the effective noise at destination $d_i$ at any time $k$.
We can now describe the achievability scheme, which consists of 3 phases. Set $\mathcal{U}$ such that $d\{\mathbf{A^S},\mathbf{A^Z},\mathbf{A^X},\mathbf{B^S},\mathbf{B^Z},\mathbf{B^X} \} \subset \mathcal{U}^{M\times M}$, where the constant $d \in \mathbb{R}$ is chosen to satisfy the power constraint $P$ at the relays (cf.\ Definition~\ref{defMIMOLinear}). More specifically,
\begin{align*}
d & = \min \left\{ (1/l) \sqrt{1/ \left( \|H_{s_1u}\|^2+\|H_{s_2u}\|^2)+M  \right) }, (1/l) \sqrt{1/ \left( \|H_{s_1v}\|^2+\|H_{s_2v}\|^2)+M\right) }   \right\}, \\ \intertext{where}
l &  = \max \{\|\mathbf{A^S}\|,\|\mathbf{A^Z}\|,\|\mathbf{A^X}\|,|\mathbf{B^S}\|,\|\mathbf{B^Z}\|,\|\mathbf{B^X}\|\},
\end{align*}
where for any matrix $\mathbf{L}$, $\|\mathbf{L}\|$ denotes the $L_2$-induced norm of the matrix.\\

\noindent \textbf{Phase 1.} In this phase, $s_1$ sends $M$ symbols $a_1$ through $a_M$. Similarly, $s_2$ sends $M$ symbols $b_1$ through $b_M$.
We choose the amplifying matrices at the relays such that the interference from $s_2$ is canceled at $d_1$, and the interference from the first $M-1$ antennas of $s_1$ is canceled at $d_2$. More specifically, we set $\mathbf{A}_1=d\mathbf{A^S}$ and $\mathbf{B}_1=d\mathbf{B^S}$. Then by Lemma~\ref{lemmaS} and (\ref{YkProofMIMOe}), $d_1$ and $d_2$ will respectively receive
\begin{align} \label{phase1}
y_{1,M_1} & =\mathbf{G}_{\mathbf{11}}^{(\mathbf{A^S},\mathbf{B^S})}\begin{bmatrix} a_1 & a_2 & \dots & a_M \end{bmatrix}^T +\tilde{z}_{1,M_1}, \notag \\
 y_{2,M_1} & =d \begin{bmatrix}
a_M & \mathbf{0}^{1 \times M-1}
\end{bmatrix}^T +\mathbf{G}_{\mathbf{22}}^{(\mathbf{A^S},\mathbf{B^S})}\begin{bmatrix} b_1 & b_2 & \dots & b_M \end{bmatrix}^T+\tilde{z}_{2,M_1},
\end{align}
where $\text{rank}(\mathbf{G}_{\mathbf{11}}^{(\mathbf{A^S},\mathbf{B^S})})=\text{rank}(\mathbf{G}_{\mathbf{22}}^{(\mathbf{A^S},\mathbf{B^S})})=M$. Therefore, as can be seen in Figure~\ref{MIMOscheme}(a), $d_1$ can now compute a noisy version of the symbols $a_1$ through $a_M$, and $d_2$ can compute a version of the symbols $b_1$ through $b_M$ that is corrupted by \emph{one} interference symbol $a_M$ and noise.\\

\noindent \textbf{Phase 2.} In this phase, $s_1$ sends $M$ new symbols $a_{M+1}$ through $a_{2M}$. Similarly, $s_2$ sends $M$ new symbols $b_{M+1}$ through $b_{2M}$.
We choose the amplifying matrices at the relays such that the interference from $s_1$ is canceled at $d_2$, and the interference from the last $M-1$ antennas of $s_2$ is canceled at $d_1$. More specifically, we set $\mathbf{A}_2=d\mathbf{A^Z}$ and $\mathbf{B}_2=d\mathbf{B^Z}$. Then by Lemma~\ref{lemmaZ} and (\ref{YkProofMIMOe}), $d_1$ and $d_2$ will respectively receive
\begin{align} \label{phase2}
y_{1,M_2} & =\mathbf{G}_{\mathbf{11}}^{(\mathbf{A^Z},\mathbf{B^Z})} \begin{bmatrix} a_{M+1} & a_{M+2} & \dots & a_{2M} \end{bmatrix}^T +d \begin{bmatrix}
\mathbf{0}^{1 \times M-1} & b_{M+1}
\end{bmatrix}^T +\tilde{z}_{1,M_2}, \notag \\
 y_{2,M_2} & =\mathbf{G}_{\mathbf{22}}^{(\mathbf{A^Z},\mathbf{B^Z})}\begin{bmatrix} b_{M+1} & b_{M+2} & \dots & b_{2M} \end{bmatrix}^T+\tilde{z}_{2,M_2},
\end{align}
where $\text{rank}(\mathbf{G}_{\mathbf{11}}^{(\mathbf{A^Z},\mathbf{B^Z})})=\text{rank}(\mathbf{G}_{\mathbf{22}}^{(\mathbf{A^Z},\mathbf{B^Z})})=M$. Therefore, as can be seen in Figure~\ref{MIMOscheme}(b), $d_2$ can now compute a noisy version of the symbols $b_{M+1}$ through $b_{2M}$, and $d_1$ can compute a version of the symbols $a_{M+1}$ through $a_{2M}$ that is corrupted by \emph{one} interference symbol $b_{M+1}$ and noise.\\

\noindent \textbf{Phase 3.} Now notice that if, at phase 3, $d_1$ can solve for $b_{M+1}$,  then it can solve for the symbols $a_{M+1}$ through $a_{2M}$ from equation (\ref{phase2}). Similarly, $d_2$ can solve for the symbols $b_1$ through $b_{M}$ from equation (\ref{phase1}), if it can solve for $a_M$. Therefore, $s_1$ repeats symbol $a_M$ through its last antenna, and sends $M-1$ new symbols $a_{2M+1}$ to $a_{3M-1}$ through its first $M-1$ antennas. $s_2$ repeats symbol $b_{M+1}$ through its first antenna, and sends $M-1$ new symbols $b_{2M+1}$ to $b_{3M-1}$ through its last $M-1$ antennas. In other words, at phase 3, \\
 $\tilde{x}_{1,M_3} = \begin{bmatrix}
a_{2M+1} & a_{2M+2} & \dots & a_{3M-1} & b_{M+1}
\end{bmatrix}^T$, and $\tilde{x}_{2,M_3} = \begin{bmatrix}
a_{M} & b_{2M+1} &  b_{2M+2} &\dots & b_{3M-1}
\end{bmatrix}^T$. \\
Now, by letting $\mathbf{A}_3 = d \mathbf{A^X}$ and $\mathbf{B}_3 = d \mathbf{B^X}$, Lemma~\ref{lemmaX} and (\ref{YkProofMIMOe2}) guarantee that $d_1$ and $d_2$ will receive
\begin{align} \label{phase3}
y_{1,M_3} & = \mathbf{\tilde{G}}_{\mathbf{11}}^{(\mathbf{A^X},\mathbf{B^X})} \begin{bmatrix}
a_{2M+1} & a_{2M+2} & \dots & a_{3M-1} & b_{M+1} \end{bmatrix}^T + \tilde{z}_{1,M_3}, \notag \\
y_{2,M_3} & = \mathbf{\tilde{G}}_{\mathbf{22}}^{(\mathbf{A^Z},\mathbf{B^Z})}  \begin{bmatrix}
a_{M} & b_{2M+1} &  b_{2M+2} &\dots & b_{3M-1}
\end{bmatrix}^T + d \begin{bmatrix}
b_{M+1} & \mathbf{0}^{1 \times M-1}
\end{bmatrix}^T + \tilde{z}_{2,M_3},
\end{align}
where $\text{rank}(\mathbf{\tilde{G}}_{\mathbf{11}}^{(\mathbf{A^X},\mathbf{B^X})})=\text{rank}(\mathbf{\tilde{G}}_{\mathbf{22}}^{(\mathbf{A^X},\mathbf{B^X})})=M$. Therefore, $d_1$ can now compute a noisy version of the symbol $b_{M+1}$ in addition to the symbols $a_{2M+1}$ through $a_{3M-1}$. Then, $d_1$ can subtract the effect of $b_{M+1}$ in (\ref{phase2}) to solve for $a_{M+1}$ through $a_{2M}$. Finally, we see that in 3 phases, $d_1$ can decode $3M-1$ symbols, thus achieving a DoF of $M-1/3$. As for destination $d_2$, it can cancel $b_{M+1}$ from (\ref{phase3}), since it already has it from phase 2. So now it can compute a noisy version of the symbol $a_M$ in addition to the symbols $b_{2M+1}$ through $b_{3M-1}$. Then, $d_2$ can subtract the effect of $a_M$ in (\ref{phase1}) to solve for $b_1$ through $b_{M}$. Therefore, $d_2$ also achieves $M-1/3$ DoF, which yields a sum-DoF of $2M-2/3$.
\bigskip
\begin{Remark}
It is worth noting the following about the achievability scheme: the ``neighboring'' antennas, the ones causing interference, actually follow the same scheme described in Section~\ref{achieve}, while the remaining antennas are oblivious to the interference and always send fresh symbols.
\end{Remark}
\bigskip

\subsection{Converse Proof of Theorem~\ref{thmDoFMIMO}}\label{upperBoundMIMO}

Assume $(R_1, R_2)$ is linear achievable on $\mathcal{U}$ for some $\mathcal{U}\subset \mathbb{R}$. It then follows from Definition~\ref{defAchievableMIMO} that there exists a sequence of $(n, R_1, R_2)$-codes that are linear on $\mathcal{U}$ such that
 \begin{equation}
\lim_{n\rightarrow \infty} P_{e,i}^{n} = 0 \label{errorProbabilityMIMO}
\end{equation}
for each $i\in\{1,2\}$. We now fix this sequence of $(n, R_1, R_2)$-codes and their corresponding relaying kernels  $(\{\mathbf{A}_k\in \mathcal{U}^{M \times M} \}_{k=1}^{n},\{\mathbf{B}_k\in \mathcal{U}^{M \times M} \}_{k=1}^{n})$. Equation \eqref{YkProofMIMOe} specifies the relationship between the $M$ received signals at the destinations and the $M$ transmit signals at the sources in the $k^{\text{th}}$ time slot. Since the relaying kernels (i.e., $\mathbf{A}_k$'s and $\mathbf{B}_k$'s) are fixed, for notational simplicity we define
 \begin{equation*}
\mathbf{G}_{\mathbf{ij},k}=  \mathbf{G}_{\mathbf{ij}}^{(\mathbf{A}_k, \mathbf{B}_k)}. \label{GijMIMO}
\end{equation*}
Hence, the symbols received at destinations at destinations $d_1$ and $d_2$ become
\begin{equation}
\left[\begin{array}{c}Y_{1,M_k} \\ Y_{2,M_k} \end{array}\right] = \left[\begin{array}{cc}\mathbf{G}_{\mathbf{11},k} & \mathbf{G}_{\mathbf{12},k} \\ \mathbf{G}_{\mathbf{21},k} & \mathbf{G}_{\mathbf{22},k}
\end{array}\right]
\left[\begin{array}{c}X_{1,M_{k-1}} \\ X_{2,M_{k-1}}\end{array}\right] + \left[ \begin{array}{cc}  H_{ud_1} \mathbf{A}_k &  H_{vd_1} \mathbf{B}_k \\  H_{ud_2} \mathbf{A}_k &  H_{vd_2} \mathbf{B}_k \end{array}\right] \left[\begin{array}{c}Z_{u,M_{k-1}} \\Z_{v,M_{k-1}} \end{array}\right] + \left[\begin{array}{c}Z_{1,M_k}\\ Z_{2,M_k}\end{array}\right] \label{YkProofMIMO}
\end{equation}
by \eqref{YkProofMIMOe}. In addition, for each $i\in\{1,2\}$, let
\begin{equation}
P_k=\E[||X_{1,M_k}||^2+||X_{2,M_k}||^2] \label{averagePowerPiMIMO}
\end{equation}
be the average sum-power transmitted by the sources in the $k^{\text{th}}$ time slot, averaged over the codebooks of the sources. We now state the following lemma which implies $\mathcal{D}_{\text{MIMO}}\le 2M-2/3$.
\bigskip
\begin{Lemma} \label{lemmaMainMIMO}
For any $(R_1, R_2)$ that is linear achievable on $\mathcal{U}$, we have for sufficiently large~$n$
\begin{equation*}
R_1 + R_2 \le \tau_1 + \frac{1}{2 n}\sum_{k=1}^n (2M-2 + \rank(\mathbf{G}_{\mathbf{21},k})+\rank(\mathbf{G}_{\mathbf{21},k})) \log_2 (1+P_{k-1}), \tag*{Bound (I)}
\end{equation*}
\begin{equation*}
R_1 + R_2 \le \tau_2 + \frac{1}{2 n} \sum_{k=1}^n (2M  - \rank(\mathbf{G}_{\mathbf{12},k}))\log_2 (1+P_{k-1}) \tag*{Bound (II)}
\end{equation*}
and
\begin{equation*}
R_1 + R_2 \le \tau_3 + \frac{1}{2 n} \sum_{k=1}^n (2M  - \rank(\mathbf{G}_{\mathbf{21},k}))\log_2 (1+P_{k-1}), \tag*{Bound (III)}
\end{equation*}
where $P_{k-1}$ is defined in \eqref{averagePowerPiMIMO} and $\tau_1$, $\tau_2$ and $\tau_3$ are some constants that do not depend on $n$ and $P$.
\end{Lemma}
\bigskip
Before proving Lemma~\ref{lemmaMainMIMO}, we demonstrate how it implies $\mathcal{D} \le 2M-2/3$ and hence Theorem~\ref{thmDoFMIMO}. Summing Bound (I), Bound (II) and Bound (III) in Lemma~\ref{lemmaMainMIMO} and dividing $3$ on both sides of the resultant inequality, we have for sufficiently large~$n$
 \begin{align}
R_1+R_2 & \le \frac{\tau_1 + \tau_2 + \tau_3}{3}+ \frac{M-1/3}{ n}\sum_{k=1}^n\log_2(1+P_{k-1}) \notag \\
&\stackrel{\eqref{zeroConvention}}{\le} \frac{\tau_1 + \tau_2 + \tau_3}{3}+ \frac{M-1/3}{n}\sum_{k=1}^n\log_2(1+P_k) \notag \\
&\stackrel{\text{(a)}}{\le} \frac{\tau_1 + \tau_2 + \tau_3}{3}+ (M-1/3)\log_2 \left(1+\sum_{k=1}^n P_k/n \right)  \notag \\
&\stackrel{\text{(b)}}{\le} \frac{\tau_1 + \tau_2 + \tau_3}{3}+ (M-1/3)\log_2(1+2P),
\label{fromLemmaToTheoremMIMO}
 \end{align}
 where
\begin{enumerate}
\item[(a)] follows from applying Jensen's inequality to the concave function $\log_2(1+x)$.
\item[(b)] follows from the fact that $||X_i^{M n}||^2 \le nP$ for each $i\in\{1,2\}$.
\end{enumerate}
It then follows from \eqref{fromLemmaToTheoremMIMO} and Definition~\ref{defDoFMIMO} that $\mathcal{D} \le 2M-2/3$. We now proceed to prove Lemma~\ref{lemmaMainMIMO}.
\bigskip\\
\subsubsection{Proof for Bound (I) in Lemma~\ref{lemmaMainMIMO}}
Fix a sequence of $(n, R_1, R_2)$-codes and their corresponding $\mathbf{G}_{\mathbf{ij},k}$.
Let
 \begin{equation}
 \tilde Y_{1,M_k} = \mathbf{G}_{\mathbf{11},k} X_{1, M_{k-1}} + \mathbf{G}_{\mathbf{12},k} X_{2, M_{k-1}} + Z_{1, M_k} \label{tildeY1MIMO}
 \end{equation}
 and
  \begin{equation}
 \tilde Y_{2,M_k} = \mathbf{G}_{\mathbf{21},k} X_{1, M_{k-1}} + \mathbf{G}_{\mathbf{22},k} X_{2, M_{k-1}} + Z_{2, M_k}  \label{tildeY2MIMO}
 \end{equation}
 be less noisy versions of $Y_{1,M_k}$ and $Y_{2,M_k}$ respectively for each $k\in\{1, 2, \ldots, n\}$ (i.e., removing the impact of $Z_{u, M_{k-1}}$ and $Z_{v, M_{k-1}}$ in \eqref{YkProofMIMO}). Since $W_i$ is uniformly distributed over $\{1, 2, \ldots, 2^{n R_i}\}$ for each $i\in\{1,2\}$, it follows that
\begin{align}
&  n (R_1 + R_2) \notag\\
&= H(W_1) + H(W_2) \notag\\
& = I(W_1; \tilde Y_1^{M n})+ I(W_2; \tilde Y_2^{M n}) + H(W_1| \tilde Y_1^{M n})+H(W_2|\tilde Y_2^{M n}) \notag\\
& \stackrel{\text{(a)}}{\le} I(X_1^{M n}; \tilde Y_1^{M n})+ I(X_2^{M n}; \tilde Y_2^{M n}) + H(W_1|\tilde Y_1^{M n})+H(W_2|\tilde Y_2^{M n}) \notag \\
& \stackrel{\text{(b)}}{=} I(X_1^{M n}; \tilde Y_1^{M n})+ I(X_2^{M n}; \tilde Y_2^{M n}) + H(W_1|\tilde Y_1^{M n}, Z_u^{M n}, Z_v^{M n})+H(W_2|\tilde Y_2^{M n},Z_u^{M n}, Z_v^{M n}) \notag \\
& \stackrel{\text{(c)}}{\le} I(X_1^{M n}; \tilde Y_1^{M n})+ I(X_2^{M n}; \tilde Y_2^{M n}) + H(W_1|Y_1^{M n})+H(W_2|Y_2^{M n}) \notag \\
& \stackrel{\text{(d)}}{\le} I(X_1^{M n}; \tilde Y_1^{M n})+ I(X_2^{M n}; \tilde Y_2^{M n}) + 2 + P_{e,1}^{ n} n R_1 + P_{e,2}^{ n} n R_2\label{thmProofTemp1MIMO}
\end{align}
where
\begin{enumerate}
\item[(a)] follows from \eqref{tildeY1MIMO} and \eqref{tildeY2MIMO} that $W_i\rightarrow X_i^{M n} \rightarrow \tilde Y_i^{M n}$ forms a Markov Chain for each $i\in\{1,2\}$.
\item[(b)] follows from the fact that $(W_1, W_2,\tilde Y_1^{M n}, \tilde Y_2^{M n})$ and $(Z_u^{M n}, Z_v^{M n})$ are independent.
\item[(c)] follows from \eqref{YkProofMIMO}, \eqref{tildeY1MIMO} and \eqref{tildeY2MIMO} that for each $i\in\{1,2\}$, $Y_i^{M n}$ is a function of $(\tilde Y_i^{M n}, Z_u^{M n}, Z_v^{M n})$.
\item[(d)] follows from Fano's inequality.
\end{enumerate}
We now state the following lemma, proved in Appendix~\ref{proofOfLemmaLinearDecompositionMIMO}, to upper bound $I(X_1^{M n}; \tilde Y_1^{M n})+ I(X_2^{M n}; \tilde Y_2^{M n})$ in \eqref{thmProofTemp1MIMO}.
\bigskip
\begin{Lemma}\label{lemmaLinearDecompositionMIMO}
There exists a $K_{M,\mathcal{U}}>0$ that does not depend on $n$ and $P$ such that the following holds for any sequence of $(n, R_1, R_2)$-codes with their corresponding $\mathbf{G}_{\mathbf{ij},k}$: For each $k\in\{1, 2, \ldots, n\}$, there exist six matrices in $\mathbb{R}^{M \times M}$, denoted by $\boldsymbol{\Lambda}_{\mathbf{1},k}$, $\boldsymbol{\Lambda}_{\mathbf{2},k}$, $\boldsymbol{\Lambda}_{\mathbf{3},k}$, $\boldsymbol{\Omega}_{\mathbf{1},k}$, $\boldsymbol{\Omega}_{\mathbf{2},k}$ and $\boldsymbol{\Omega}_{\mathbf{3},k}$ respectively, and two matrices in $\mathbb{R}^{(M-1) \times M}$, denoted by $\boldsymbol{\Gamma}_{\mathbf{1},k}$ and $\boldsymbol{\Gamma}_{\mathbf{2},k}$ respectively, such that the magnitudes of the entries in each of the eight matrices are upper bounded by $K_{M,\mathcal{U}}$, and
\begin{equation}
 \mathbf{G}_{\mathbf{11},k} = \mathbf{G}_{\mathbf{12},k}\boldsymbol{\Lambda}_{\mathbf{1},k} + \boldsymbol{\Lambda}_{\mathbf{2},k}\mathbf{G}_{\mathbf{21},k} + \boldsymbol{\Lambda}_{\mathbf{3},k}\left[ \begin{array}{c} \boldsymbol{\Gamma}_{\mathbf{1},k}  \\ \mathbf{0}^{1\times M}\end{array}\right]\label{G11ThmStatementSumRankMIMO}
 \end{equation}
 and
\begin{equation}
\mathbf{G}_{\mathbf{22},k} = \boldsymbol{\Omega}_{\mathbf{1},k}\mathbf{G}_{\mathbf{12},k} + \mathbf{G}_{\mathbf{21},k}\boldsymbol{\Omega}_{\mathbf{2},k} + \boldsymbol{\Omega}_{\mathbf{3},k} \left[\begin{array}{c} \boldsymbol{\Gamma}_{\mathbf{2},k}  \\ \mathbf{0}^{1\times M}\end{array}\right]. \label{G22ThmStatementSumRankMIMO}
 \end{equation}
\end{Lemma}
\bigskip
In the rest of this section, fix $K_{M,\mathcal{U}}$ and the corresponding $\{\boldsymbol{\Lambda}_{\mathbf{1},k}, \boldsymbol{\Lambda}_{\mathbf{2},k}, \boldsymbol{\Lambda}_{\mathbf{3},k}, \boldsymbol{\Omega}_{\mathbf{1},k}, \boldsymbol{\Omega}_{\mathbf{2},k}, \boldsymbol{\Omega}_{\mathbf{3},k}, \boldsymbol{\Gamma}_{\mathbf{1},k}, \boldsymbol{\Gamma}_{\mathbf{2},k}\}_{k=1}^n$ described in Lemma~\ref{lemmaLinearDecompositionMIMO} such that for each $k\in\{1, 2, \ldots, n\}$, \eqref{G11ThmStatementSumRankMIMO} and \eqref{G22ThmStatementSumRankMIMO} are satisfied by $\boldsymbol{\Lambda}_{\mathbf{1},k}$, $\boldsymbol{\Lambda}_{\mathbf{2},k}$, $\boldsymbol{\Lambda}_{\mathbf{3},k}$, $\boldsymbol{\Omega}_{\mathbf{1},k}$, $\boldsymbol{\Omega}_{\mathbf{2},k}$, $\boldsymbol{\Omega}_{\mathbf{3},k}$, $\boldsymbol{\Gamma}_{\mathbf{1},k}$ and $\boldsymbol{\Gamma}_{\mathbf{2},k}$, in each of which $K_{M,\mathcal{U}}$ is an upper bound on the magnitudes of the entries. Using \eqref{G11ThmStatementSumRankMIMO} and \eqref{G22ThmStatementSumRankMIMO} in Lemma~\ref{lemmaLinearDecompositionMIMO}, we obtain from \eqref{tildeY1MIMO} and \eqref{tildeY2MIMO} that
\begin{equation}
\tilde Y_{1, M_k}=  \left(\mathbf{G}_{\mathbf{12},k}\boldsymbol{\Lambda}_{\mathbf{1},k} + \boldsymbol{\Lambda}_{\mathbf{2},k}\mathbf{G}_{\mathbf{21},k} + \boldsymbol{\Lambda}_{\mathbf{3},k} \left[\begin{array}{c} \boldsymbol{\Gamma}_{\mathbf{1},k}  \\ \mathbf{0}^{1\times M}\end{array}\right]\right)X_{1, M_{k-1}} +  \mathbf{G}_{\mathbf{12},k}  X_{2,M_{k-1}} + Z_{1, M_k} \label{Y1MkMIMOTemp*}
\end{equation}
and
\begin{equation}
\tilde Y_{2, M_k} = \mathbf{G}_{\mathbf{21},k}X_{1, M_{k-1}} +  \left(\boldsymbol{\Omega}_{\mathbf{1},k}\mathbf{G}_{\mathbf{12},k} + \mathbf{G}_{\mathbf{21},k} \boldsymbol{\Omega}_{\mathbf{2},k} + \boldsymbol{\Omega}_{\mathbf{3},k} \left[\begin{array}{c} \boldsymbol{\Gamma}_{\mathbf{2},k}  \\ \mathbf{0}^{1\times M}\end{array}\right]\right) X_{2,M_{k-1}} + Z_{2, M_k} \label{Y2MkMIMOTemp*}
\end{equation}
for each $k\in\{1, 2, \ldots, n\}$. Let $\hat Z_i^{Mn}$ be $Mn$ copies of $\mathcal{N}(0,1)$ for each $i\in\{1, 2\}$ such that $\hat Z_i^{Mn}$ and $(\tilde Y_i^{Mn}, X_i^{Mn})$ are independent. Then, letting
\begin{equation}
\vec Y_{1,k}^* = \left[\begin{array}{c} \boldsymbol{\Gamma}_{\mathbf{1},k}  \\ \mathbf{0}^{1\times M}\end{array}\right] X_{1, M_{k-1}} +  \hat Z_{1,M_k} \label{Y1*MIMO}
\end{equation}
and
\begin{equation}
\vec Y_{2,k}^* =\left[\begin{array}{c} \boldsymbol{\Gamma}_{\mathbf{2},k}  \\ \mathbf{0}^{1\times M}\end{array}\right] X_{2, M_{k-1}} +  \hat Z_{2,M_k}, \label{Y2*MIMO}
\end{equation}
we obtain from \eqref{Y1MkMIMOTemp*} and \eqref{Y2MkMIMOTemp*} that
\begin{equation}
\tilde Y_{1, M_k}=  \left(\mathbf{G}_{\mathbf{12},k}\boldsymbol{\Lambda}_{\mathbf{1},k} + \boldsymbol{\Lambda}_{\mathbf{2},k}\mathbf{G}_{\mathbf{21},k}\right)X_{1, M_{k-1}} + \boldsymbol{\Lambda}_{\mathbf{3},k} (\vec Y_{1,k}^* - \hat Z_{1,M_k}) + \mathbf{G}_{\mathbf{12},k}  X_{2,M_{k-1}} + Z_{1, M_k}  \label{Y1MkMIMO}
\end{equation}
and
\begin{equation}
\tilde Y_{2, M_k} = \mathbf{G}_{\mathbf{21},k}X_{1, M_{k-1}} +  \left(\boldsymbol{\Omega}_{\mathbf{1},k}\mathbf{G}_{\mathbf{12},k} + \mathbf{G}_{\mathbf{21},k} \boldsymbol{\Omega}_{\mathbf{2},k} \right) X_{2,M_{k-1}} + \boldsymbol{\Omega}_{\mathbf{3},k}(\vec Y_{2,k}^*-\hat Z_{2,M_k} )+ Z_{2, M_k} \label{Y2MkMIMO}
\end{equation}
for each $k\in\{1, 2, \ldots, n\}$. Following \eqref{thmProofTemp1MIMO}, we consider
\begin{align}
& I(X_1^{Mn}; \tilde Y_1^{Mn})+ I(X_2^{Mn}; \tilde Y_2^{Mn})\notag \\
& = h(\tilde Y_1^{Mn}) - h(\tilde Y_2^{Mn}|X_2^{Mn}) + h(\tilde Y_2^{Mn}) - h(\tilde Y_1^{Mn}|X_1^{Mn})\notag \\
& = h(\tilde Y_1^{Mn}|\{\vec Y_{1,k}^*\}_{k=1}^n) + I(\{\vec Y_{1,k}^*\}_{k=1}^n;\tilde Y_1^{Mn}) - h(\tilde Y_2^{Mn}|X_2^{Mn}) + h(\tilde Y_2^{Mn}|\{\vec Y_{2,k}^*\}_{k=1}^n) + I(\{\vec Y_{2,k}^*\}_{k=1}^n;\tilde Y_2^{Mn}) - h(\tilde Y_1^{Mn}|X_1^{Mn})\notag \\
&\le  h(\tilde Y_1^{Mn}|\{\vec Y_{1,k}^*\}_{k=1}^n) + h(\{\vec Y_{1,k}^*\}_{k=1}^n)-h(\{\vec Y_{1,k}^*\}_{k=1}^n|\tilde Y_1^{Mn}, X_1^{Mn}) - h(\tilde Y_2^{Mn}|X_2^{Mn}) \notag \\
 &\qquad + h(\tilde Y_2^{Mn}|\{\vec Y_{2,k}^*\}_{k=1}^n) + h(\{\vec Y_{2,k}^*\}_{k=1}^n)-h(\{\vec Y_{2,k}^*\}_{k=1}^n|\tilde Y_2^{Mn},  X_2^{Mn})  - h(\tilde Y_1^{Mn}|X_1^{Mn})\notag \\
 & \stackrel{\text{(a)}}{=}  h(\tilde Y_1^{Mn}|\{\vec Y_{1,k}^*\}_{k=1}^n) + h(\{\vec Y_{1,k}^*\}_{k=1}^n)-h(\hat Z_1^{Mn}) - h(\tilde Y_2^{Mn}|X_2^{Mn}) \notag \\
 &\qquad + h(\tilde Y_2^{Mn}|\{\vec Y_{2,k}^*\}_{k=1}^n) + h(\{\vec Y_{2,k}^*\}_{k=1}^n) -h(\hat Z_2^{Mn}) - h(\tilde Y_1^{Mn}|X_1^{Mn})\notag \\
& \stackrel{\text{(b)}}{\le}  h(\tilde Y_1^{Mn}|\{\vec Y_{1,k}^*\}_{k=1}^n) + h(\{\vec Y_{1,k}^*\}_{k=1}^n) - h(\tilde Y_2^{Mn}|X_2^{Mn}) + h(\tilde Y_2^{Mn}|\{\vec Y_{2,k}^*\}_{k=1}^n) + h(\{\vec Y_{2,k}^*\}_{k=1}^n) - h(\tilde Y_1^{Mn}|X_1^{Mn}), \label{thmFirstIneqMIMOTemp****}
\end{align}
 where
\begin{enumerate}
\item[(a)] follows from \eqref{Y1*MIMO}, \eqref{Y2*MIMO} and the fact that $\hat Z_i^{Mn}$ and $(\tilde Y_i^{Mn}, X_i^{Mn})$ are independent for each $i\in\{1,2\}$.
\item[(b)] follows from the facts that $\{\hat Z_{1,m}\}_{m=1}^{Mn}$ are independent, $\{\hat Z_{2,m}\}_{m=1}^{Mn}$ are independent and the differential entropy of $\mathcal{N}(0,1)$ is positive.
\end{enumerate}
Substituting $\tilde Y_{i,M_k}$ and $\vec Y_{i,k}^*$ in \eqref{thmFirstIneqMIMOTemp****} by \eqref{Y1*MIMO}, \eqref{Y2*MIMO}, \eqref{Y1MkMIMO}, \eqref{Y2MkMIMO} and using the fact that $X_1^{Mn}$, $X_2^{Mn}$ and $(Z_1^{Mn}, Z_2^{Mn})$ are independent, we obtain
\begin{align}
& I(X_1^{Mn}; \tilde Y_1^{Mn})+ I(X_2^{Mn}; \tilde Y_2^{Mn})\notag \\
& \le \underbrace{h( \{(\mathbf{G}_{\mathbf{12},k}\boldsymbol{\Lambda}_{\mathbf{1},k} \! + \!\boldsymbol{\Lambda}_{\mathbf{2},k}\mathbf{G}_{\mathbf{21},k})X_{1, M_{k-1}} \! + \! \mathbf{G}_{\mathbf{12},k}  X_{2,M_{k-1}}\! +\! Z_{1, M_k} \! -\! \boldsymbol{\Lambda}_{\mathbf{3},k}\hat Z_{1, M_k}\}_{k=1}^n)\! -\! h(\{\mathbf{G}_{\mathbf{21},k}X_{1, M_{k-1}}\!+\!Z_{2, M_k}\}_{k=1}^n)}_{\triangleq I_1}  \notag \\
 &\quad  + \underbrace{h(\{\mathbf{G}_{\mathbf{21},k}X_{1, M_{k-1}}\! + \! (\boldsymbol{\Omega}_{\mathbf{1},k}\mathbf{G}_{\mathbf{12},k}\! + \! \mathbf{G}_{\mathbf{21},k}\boldsymbol{\Omega}_{\mathbf{2},k})X_{2, M_{k-1}}\! + \! Z_{2, M_k}\! -\! \boldsymbol{\Omega}_{\mathbf{3},k}\hat Z_{2, M_k}\}_{k=1}^n) \!- \! h(\{\mathbf{G}_{\mathbf{12},k}X_{2, M_{k-1}}\!+\! Z_{1, M_k}\}_{k=1}^n)}_{\triangleq I_2} \notag \\
 &\quad + h(\{\vec Y_{1,k}^*\}_{k=1}^n) + h(\{\vec Y_{2,k}^*\}_{k=1}^n) \label{thmFirstIneqMIMOTemp*}
 \end{align}
In order to bound $I_1$ defined in \eqref{thmFirstIneqMIMOTemp*}, we apply Lemma~\ref{lemmaDifferenceEntropy} by setting
\begin{equation*}
\begin{cases}
X^n=\{(\mathbf{G}_{\mathbf{12},k}\boldsymbol{\Lambda}_{\mathbf{1},k} + \boldsymbol{\Lambda}_{\mathbf{2},k}\mathbf{G}_{\mathbf{21},k})X_{1, M_{k-1}} +  \mathbf{G}_{\mathbf{12},k}  X_{2,M_{k-1}} \}_{k=1}^n, \\
Y^n=\{\mathbf{G}_{\mathbf{21},k}X_{1, M_{k-1}}\}_{k=1}^n, \\
Z_1^n = \{Z_{1,M_k} - \boldsymbol{\Lambda}_{\mathbf{3},k}\hat Z_{1, M_k} \}_{k=1}^n, \\
 Z_2^n = \{Z_{2, M_k}\}_{k=1}^n, \\
  \mathbf{L}=\left[\begin{array}{cccc} \boldsymbol{\Lambda}_{\mathbf{2},1} &&& \\ &\boldsymbol{\Lambda}_{\mathbf{2},2} && \\ && \ddots & \\ &&& \boldsymbol{\Lambda}_{\mathbf{2},n}\end{array}\right], \, \text{$\mathbf{L}$ is block-diagonal,}
\end{cases}
\end{equation*}
and obtain
\begin{equation}
I_1 \le  h(\{\mathbf{G}_{\mathbf{12},k}(\boldsymbol{\Lambda}_{\mathbf{1},k} X_{1, M_{k-1}} +  X_{2,M_{k-1}}) + Z_{1, M_k}  -  \boldsymbol{\Lambda}_{\mathbf{3},k}\hat Z_{1, M_k} - \boldsymbol{\Lambda}_{\mathbf{2},k} Z_{2, M_k}\}_{k=1}^n)- h(Z_2^{Mn}). \label{I1tempMIMO}
\end{equation}
Following similar procedures for proving \eqref{I1tempMIMO}, we obtain
\begin{equation}
I_2 \le  h(\{\mathbf{G}_{\mathbf{21},k}(X_{1, M_{k-1}} + \boldsymbol{\Omega}_{\mathbf{2},k}X_{2,M_{k-1}})  + Z_{2, M_k}-  \boldsymbol{\Omega}_{\mathbf{3},k}\hat Z_{2, M_k}-\boldsymbol{\Omega}_{\mathbf{1},k}Z_{1, M_k}\}_{k=1}^n) - h(Z_1^{Mn}). \label{I2tempMIMO}
\end{equation}
Since $\{Z_{1,m}\}_{m=1}^{M n}$ are independent, $\{Z_{2,m}\}_{m=1}^{M n}$ are independent and the differential entropy of $\mathcal{N}(0,1)$ is positive,
it then follows from \eqref{thmFirstIneqMIMOTemp*}, \eqref{I1tempMIMO} and \eqref{I2tempMIMO} that
 \begin{align}
& I(X_1^{Mn}; \tilde Y_1^{Mn})+ I(X_2^{Mn}; \tilde Y_2^{Mn})\notag \\
&\le \sum_{k=1}^n (h(\mathbf{G}_{\mathbf{12},k}(\boldsymbol{\Lambda}_{\mathbf{1},k} X_{1, M_{k-1}} +  X_{2,M_{k-1}}) + Z_{1, M_k}  -  \boldsymbol{\Lambda}_{\mathbf{3},k}\hat Z_{1, M_k} - \boldsymbol{\Lambda}_{\mathbf{2},k} Z_{2, M_k})+h(\vec Y_{1,k}^*)\notag \\
 &\qquad   +h(\mathbf{G}_{\mathbf{21},k}(X_{1, M_{k-1}} + \boldsymbol{\Omega}_{\mathbf{2},k}X_{2,M_{k-1}})  + Z_{2, M_k}-  \boldsymbol{\Omega}_{\mathbf{3},k}\hat Z_{2, M_k}-\boldsymbol{\Omega}_{\mathbf{1},k}Z_{1, M_k}) +h(\vec Y_{2,k}^*)).\label{thmFirstIneqMIMO}
\end{align}
We now need the following lemma, proved in Appendix~\ref{appendixLemmaDifferentialEntropyBound1}, to bound the terms in \eqref{thmFirstIneqMIMO}.
\bigskip
\begin{Lemma} \label{lemmaDifferentialEntropyBound1MIMO}
There exist four real numbers, denoted by $\kappa_1$, $\kappa_2$, $\kappa_3$ and $\kappa_4$ respectively, that do not depend on $n$ and $P$ such that for each $k\in\{1, 2, \ldots, n\}$,
\begin{align*}
& h(\mathbf{G}_{\mathbf{12},k}(\boldsymbol{\Lambda}_{\mathbf{1},k} X_{1, M_{k-1}} +  X_{2,M_{k-1}}) + Z_{1, M_k}  -  \boldsymbol{\Lambda}_{\mathbf{3},k}\hat Z_{1, M_k} - \boldsymbol{\Lambda}_{\mathbf{2},k} Z_{2, M_k}) \notag \\
 &\quad \le \rank(\mathbf{G}_{\mathbf{12},k}) \log_2 \sqrt{1+P_{k-1}} + \kappa_1,
\end{align*}
\begin{align*}
& h(\mathbf{G}_{\mathbf{21},k}(X_{1, M_{k-1}} + \boldsymbol{\Omega}_{\mathbf{2},k}X_{2,M_{k-1}})  + Z_{2, M_k}-  \boldsymbol{\Omega}_{\mathbf{3},k}\hat Z_{2, M_k}-\boldsymbol{\Omega}_{\mathbf{1},k}Z_{1, M_k}) \notag \\
 &\quad \le \rank(\mathbf{G}_{\mathbf{21},k}) \log_2 \sqrt{1+P_{k-1}} + \kappa_2,
\end{align*}
\begin{equation*}
 h(\vec Y_{1,k}^*) \le (M-1) \log_2 \sqrt{1+P_{k-1}} + \kappa_3
\end{equation*}
and
\begin{equation*}
 h(\vec Y_{2,k}^*) \le (M-1) \log_2 \sqrt{1+P_{k-1}} + \kappa_4,
\end{equation*}
where $P_{k-1}$ is defined in \eqref{averagePowerPiMIMO}.
\end{Lemma}
\bigskip
Using \eqref{thmProofTemp1MIMO}, \eqref{thmFirstIneqMIMO} and Lemma~\ref{lemmaDifferentialEntropyBound1MIMO}, we obtain
 \begin{equation}
\sum_{i=1}^2(1-P_{e,i}^{n})R_i \le \tau_1+ \frac{1}{2n}\sum_{k=1}^n (2M-2 +\rank(\mathbf{G}_{\mathbf{21},k})+\rank(\mathbf{G}_{\mathbf{21},k})) \log_2 (1+P_{k-1}) \label{firstBoundFinalstatementMIMO}
 \end{equation}
 for some $\tau_1$ that does not depend on $n$ and $P$. It then follows from \eqref{firstBoundFinalstatementMIMO} and \eqref{errorProbabilityMIMO} that Bound (I) holds for sufficiently large~$n$.
  \bigskip\\
\subsubsection{Proof for Bounds (II) and (III) in Lemma~\ref{lemmaMainMIMO}}
Since the proofs for Bounds (II) and (III) in Lemma~\ref{lemmaMainMIMO} are almost identical to the proofs for Bounds (ii) and (iii) in Lemma~\ref{lemmaMain} and the proofs for Bound (ii) and Bound (iii) are similar (cf.\ Section~\ref{upperBound}), we only provide sketches of the proof for Bound (II). Following similar procedures for deriving \eqref{thmSecondInequality2}, we obtain
\begin{align}
& I(X_1^{M n}; \tilde Y_1^{M n})+ I(X_2^{M n}; \tilde Y_2^{M n})\notag \\
  &\le \sum_{k=1}^n (h(\tilde Y_{1, M_k})    +h(\mathbf{G}_{\mathbf{22},k}X_{2, M_{k-1}} + Z_{2, M_k}|\mathbf{G}_{\mathbf{12},k} X_{2, M_{k-1}}+Z_{1, M_k})). \label{thmSecondInequalityMIMO}
\end{align}
We bound the terms in \eqref{thmSecondInequalityMIMO} using the following lemma, whose proof is omitted because it is almost identical to the proof of Lemma~\ref{lemmaDifferentialEntropyBound2} in the Appendix~\ref{Lemma6}.
\bigskip
\begin{Lemma} \label{lemmaDifferentialEntropyBound2MIMO}
There exist two real numbers, denoted by $\kappa$ and $\kappa^\prime$, that do not depend on $n$ and $P$ such that for each $k\in\{1, 2, \ldots, n\}$,
\[
 h(\tilde Y_{1, M_k}) \le M  \log_2 \sqrt{1+P_{k-1}} + \kappa
\]
and
\[
 h(\mathbf{G}_{\mathbf{22},k}X_{2, M_{k-1}} + Z_{2, M_k}|\mathbf{G}_{\mathbf{12},k} X_{2, M_{k-1}}+Z_{1, M_k})  \le (M-\rank(\mathbf{G}_{\mathbf{12},k})) \log_2 \sqrt{1+P_{k-1}} + \kappa^\prime,
 \]
where $P_{k-1}$ is defined in \eqref{averagePowerPiMIMO}.
\end{Lemma}
\bigskip
Using \eqref{thmProofTemp1MIMO}, \eqref{thmSecondInequalityMIMO} and Lemma~\ref{lemmaDifferentialEntropyBound2MIMO}, we obtain
\begin{equation}
\sum_{i=1}^2(1-P_{e,i}^{M n})R_i \le \tau_2 + \frac{1}{2n} \sum_{k=1}^n (2M-\rank(\mathbf{G}_{\mathbf{12},k}))\log_2 (1+P_{k-1}) \label{bound2*MIMO}
\end{equation}
for some $\tau_2$ that do not depend on $n$ and $P$. It then follows from \eqref{bound2*MIMO} and \eqref{errorProbabilityMIMO} that Bound (II) holds for sufficiently large~$n$.

\section{Two-Hop MIMO IC with Complex Channel Gains}\label{twoHopICMIMOComplex}

This section extends the result in Section~\ref{twoHopICMultipleAntennas} for complex channel gains. We first define the two-hop MIMO IC with complex channel gains as well as linear schemes for the channel. Then, we prove the linear sum-DoF result stated in Corollary~\ref{ComplexMIMO}.
\subsection{Network Model}\label{formulationReal}
Let
\begin{equation}
\mathbf{H_1}=\left[\begin{array}{cc}H_{s_1 u} & H_{s_2 u}\\ H_{s_1 v} & H_{s_2 v}\end{array}\right] \in \mathbb{C}^{2M \times 2M} \label{H1Complex}
 \end{equation}
 and
 \begin{equation}
 \mathbf{H_2}=\left[\begin{array}{cc}H_{ud_1} & H_{vd_1}\\ H_{ud_2} & H_{vd_2}\end{array}\right] \in \mathbb{C}^{2M \times 2M} \label{H2Complex}
  \end{equation}
characterize the channels of the first and second hop respectively, where the channels between each node~$i$ and each node~$j$ are characterized by an $M\times M$ complex matrix. The model is the same as described in Section~\ref{formulationMIMO}, except that the quantities in~(\ref{firstChannelAdditionMIMO}) and~(\ref{secondChannelAdditionMIMO}) are now complex. Also, we will redefine a \emph{linear} code for the complex two-hop IC as follows.
\bigskip
\begin{Definition} \label{defMIMOLinearComplex}
Let $\mathcal{U}$ be a finite set of real numbers. An $(n, R_1, R_2)$-code on the two-hop MIMO IC is said to be \textit{linear on $\mathcal{U}$} if there exist $\{\mathbf{A}_k\in \mathcal{U}^{2M \times 2M} \}_{k=1}^{n}$ and $\{\mathbf{B}_k \in \mathcal{U}^{2M\times 2M}\}_{k=1}^{n}$ such that for each $k\in\{1, 2, \ldots, n\}$,
\[
\left[\begin{array}{c}\Re\{X_{u,M_k}\} \\ \Im\{X_{u,M_k}\}\end{array}\right] = \mathbf{A}_k  \left[\begin{array}{c} \Re\{Y_{u,M_{k-1}}\} \\ \Im\{Y_{u,M_{k-1}}\}\end{array}\right]
\]
and
\[
\left[\begin{array}{c}\Re\{X_{v,M_k}\} \\ \Im\{X_{v,M_k}\}\end{array}\right] = \mathbf{B}_k \left[\begin{array}{c} \Re\{Y_{v,M_{k-1}}\} \\ \Im\{Y_{v,M_{k-1}}\}\end{array}\right].
\]
In other words, the relays are operating over blocks of length $2M$ real symbols, where the real symbols in each block to be transmitted in the $k^{\text{th}}$ time slot are linear combinations of the $2M$ real symbols received in the $(k-1)^{\text{th}}$ time slot. We call $(\{\mathbf{A}_k\in \mathcal{U}^{2M \times 2M} \}_{k=1}^{n},\{\mathbf{B}_k\in \mathcal{U}^{2M \times 2M} \}_{k=1}^{n})$ a \textit{relaying kernel} of the code.
\end{Definition}
\bigskip
\begin{Definition} \label{defDoFMIMOComplex}
The linear sum-DoF of the two-hop MIMO IC with complex channel gains, denoted by $\bar{\mathcal{D}}_{\text{MIMO}}$, is defined by
\[
\bar{\mathcal{D}}_{\text{MIMO}}= \sup_{\mathcal{U}}\lim_{P\rightarrow\infty}\! \sup\left\{\left. \frac{R_1+R_2}{\log_2 P} \:\right|(R_1, R_2) \text{ is linear achievable on }\mathcal{U}\right\}.
\]
\end{Definition}

\subsection{Achievability Proof of Corollary~\ref{ComplexMIMO}} \label{ComplexAch}
First, we will show how we can view the two-hop MIMO IC with complex channel gains and $M$ antennas at each hop as a two-hop MIMO IC with appropriate real channel gains and $2M$ antennas at each node. Consider the following equation characterizing the input-output relationship for the first hop:
\begin{equation} \label{firstChannelAdditionMIMOComplex}
\begin{bmatrix}
Y_{u,M_k} \\ Y_{v,M_k}
\end{bmatrix} =
\begin{bmatrix}
H_{s_1u} & H_{s_2u} \\
H_{s_1v} & H_{s_2v}
\end{bmatrix}
\begin{bmatrix}
X_{1,M_k} \\ X_{2,M_k}
\end{bmatrix}
+ \begin{bmatrix}
Z_{u,M_k} \\ Z_{v,M_k}
\end{bmatrix}.
\end{equation}
So, the real and imaginary components of $Y_{u,M_k}$ can be written as:
\begin{equation} \label{realcomplexYu}
\begin{bmatrix}
\Re\{Y_{u,M_k}\} \\ \Im\{Y_{u,M_k}\}
\end{bmatrix} =
\begin{bmatrix}
\Re\{H_{s_1u}\} & -\Im\{H_{s_1u}\} \\
\Im\{H_{s_1u}\} & \Re\{H_{s_1u}\}
\end{bmatrix} \begin{bmatrix}
\Re\{X_{1,M_k}\} \\ \Im\{X_{1,M_k}\}
\end{bmatrix} + \begin{bmatrix}
\Re\{H_{s_2u}\} & -\Im\{H_{s_2u}\} \\
\Im\{H_{s_2u}\} & \Re\{H_{s_2u}\}
\end{bmatrix} \begin{bmatrix}
\Re\{X_{2,M_k}\} \\ \Im\{X_{2,M_k}\}
\end{bmatrix} + \begin{bmatrix}
\Re\{Z_{u,M_k}\} \\ \Im\{Z_{u,M_k}\}
\end{bmatrix}.
\end{equation}
Now, for each node $i$ and node $j$, define
\begin{equation}
\bar H_{ij}= \left[\begin{array}{cc} \Re\{H_{ij}\} & -\Im\{H_{ij}\} \\ \Im\{H_{ij}\} & \Re\{H_{ij}\}\end{array}\right]. \label{barHij}
 \end{equation}
So by writing a similar expression as \eqref{realcomplexYu} for $Y_{v,M_k}$, we can rewrite~\eqref{firstChannelAdditionMIMOComplex} as
\begin{equation} \label{firstChannelAdditionMIMOAugmented}
\begin{bmatrix}
\Re\{Y_{u,M_k}\} \\ \Im\{Y_{u,M_k}\} \\ \Re\{Y_{v,M_k}\} \\ \Im\{Y_{v,M_k}\}
\end{bmatrix} =
\begin{bmatrix}
\bar{H}_{s_1u} & \bar{H}_{s_2u} \\
\bar{H}_{s_1v} & \bar{H}_{s_2v}
\end{bmatrix} \begin{bmatrix}
\Re\{X_{1,M_k}\} \\ \Im\{X_{1,M_k}\} \\ \Re\{X_{2,M_k}\} \\ \Im\{X_{2,M_k}\}
\end{bmatrix} + \begin{bmatrix}
\Re\{Z_{u,M_k}\} \\ \Im\{Z_{u,M_k}\} \\ \Re\{Z_{v,M_k}\} \\ \Im\{Z_{v,M_k}\}
\end{bmatrix}.
\end{equation}
Similarly, we can write the equations characterizing the input-output relationship for the second hop as:
\begin{equation} \label{secondChannelAdditionMIMOAugmented}
\begin{bmatrix}
\Re\{Y_{1,M_k}\} \\ \Im\{Y_{1,M_k}\} \\ \Re\{Y_{2,M_k}\} \\ \Im\{Y_{2,M_k}\}
\end{bmatrix} =
\begin{bmatrix}
\bar{H}_{ud_1} & \bar{H}_{vd_1} \\
\bar{H}_{ud_2} & \bar{H}_{vd_2}
\end{bmatrix} \begin{bmatrix}
\Re\{X_{u,M_k}\} \\ \Im\{X_{u,M_k}\} \\ \Re\{X_{v,M_k}\} \\ \Im\{X_{v,M_k}\}
\end{bmatrix} + \begin{bmatrix}
\Re\{Z_{1,M_k}\} \\ \Im\{Z_{1,M_k}\} \\ \Re\{Z_{2,M_k}\} \\ \Im\{Z_{2,M_k}\}
\end{bmatrix}.
\end{equation}
Equations~\eqref{firstChannelAdditionMIMOAugmented} and~\eqref{secondChannelAdditionMIMOAugmented} show that the complex MIMO two-hop IC with $M$ antennas at each node is equivalent to a two-hop IC with appropriate real channel gains and $2M$ antennas at each node. This will allow us to use results derived for the case of real channel gains in Section \ref{twoHopICMultipleAntennas}.
We will call this equivalent view of the channel, i.e. the two-hop IC with $2M$ antennas at each node and channel matrices given by~(\ref{barHij}), the \emph{augmented} channel.

Now, define $\mathbf{G_{11}^{(A,B)}}$ as in (\ref{GijM}), but by replacing $H_{ij}$'s by $\bar{H}_{ij}$'s, and by letting $\mathbf{A}$ and $\mathbf{B}$ be $2M \times 2M$ real matrices. Similarly, define $\mathbf{G_{12}^{(A,B)}}$, $\mathbf{G_{21}^{(A,B)}}$, and $\mathbf{G_{22}^{(A,B)}}$.
Matrix $\mathbf{G_{ij}^{(A,B)}}$ now denotes the end-to-end link between source $s_j$ and destination $d_i$ in the augmented channel when relays $u$ and $v$ set their amplifying matrices to be $\mathbf{A}$ and $\mathbf{B}$ respectively. Now, notice that if Lemmas~\ref{lemmaS},~\ref{lemmaZ}, and~\ref{lemmaX} hold for almost values of the augment channel gains (or equivalently for almost all values of the complex channel gains), then we could apply the same scheme described in~\ref{achMIMO} to achieve $2(2M)-2/3$ sum-DoF. Since each real degree of freedom corresponds to half a complex degree of freedom (hence the normalization difference in Definitions~\ref{defDoFMIMOComplex} and~\ref{defDoFMIMO}), we get that $\mathcal{\bar{D}}_{\text{MIMO}} \geq 2M-1/3$ for almost all values of complex channel gains. It remains to show that the mentioned lemmas do hold, which is done in Appendix~\ref{CondComplexMIMO}.


\subsection{Converse Proof of Corollary~\ref{ComplexMIMO}}
Consider an $(n, R_1, R_2)$-code that is linear on $\mathcal{U}$, where $(\{\mathbf{A}_k\in \mathcal{U}^{2M \times 2M} \}_{k=1}^{n},\{\mathbf{B}_k\in \mathcal{U}^{2M\times 2M} \}_{k=1}^{n})$ is the relaying kernel of the code.
Theorem~\ref{thmDoFMIMO} implies that the linear sum-DoF of the augmented two-hop MIMO IC with real channel gains consisting of $2M$-antenna nodes is upper bounded by $4M - 2/3$ for almost all channel gains. However, we need first to make sure that the augmented channel satisfies some sufficient conditions for the converse of Theorem~\ref{thmDoFMIMO} to hold. This is done in Appendix~\ref{CondComplexMIMO}. Consequently, it follows from Definitions~\ref{defDoFMIMOComplex} and~\ref{defDoFMIMO} that $\bar{\mathcal{D}}_{\text{MIMO}}\le 2M-1/3$ for the two-hop MIMO IC with complex channel gains consisting of $M$-antenna nodes.

\section{Numerical Analysis} \label{numericalResults}
Although the main results of this paper, contained in Theorem~\ref{thmDoF}, Theorem~\ref{thmDoFMIMO} and Corollary~\ref{ComplexMIMO}, characterize only the DoF of the two-hop IC, the achievable rates of our linear schemes can also be numerically computed at any finite SNR.  In this section, we focus on the two-hop IC with single-antenna nodes and complex channel gains, and numerically evaluate the achievable sum-rate of our linear scheme. We consider two settings corresponding to moderate and high interference regimes, and we show that our vector-linear scheme performs well under both settings. In particular, our vector-linear scheme can outperform state-of-the-art schemes (described later) at 15dB for the moderate interference regime, and 25 dB for the high interference regime. The details of the simulations are described as follows.

\begin{figure}[htp]
\centering
\includegraphics[scale=0.4]{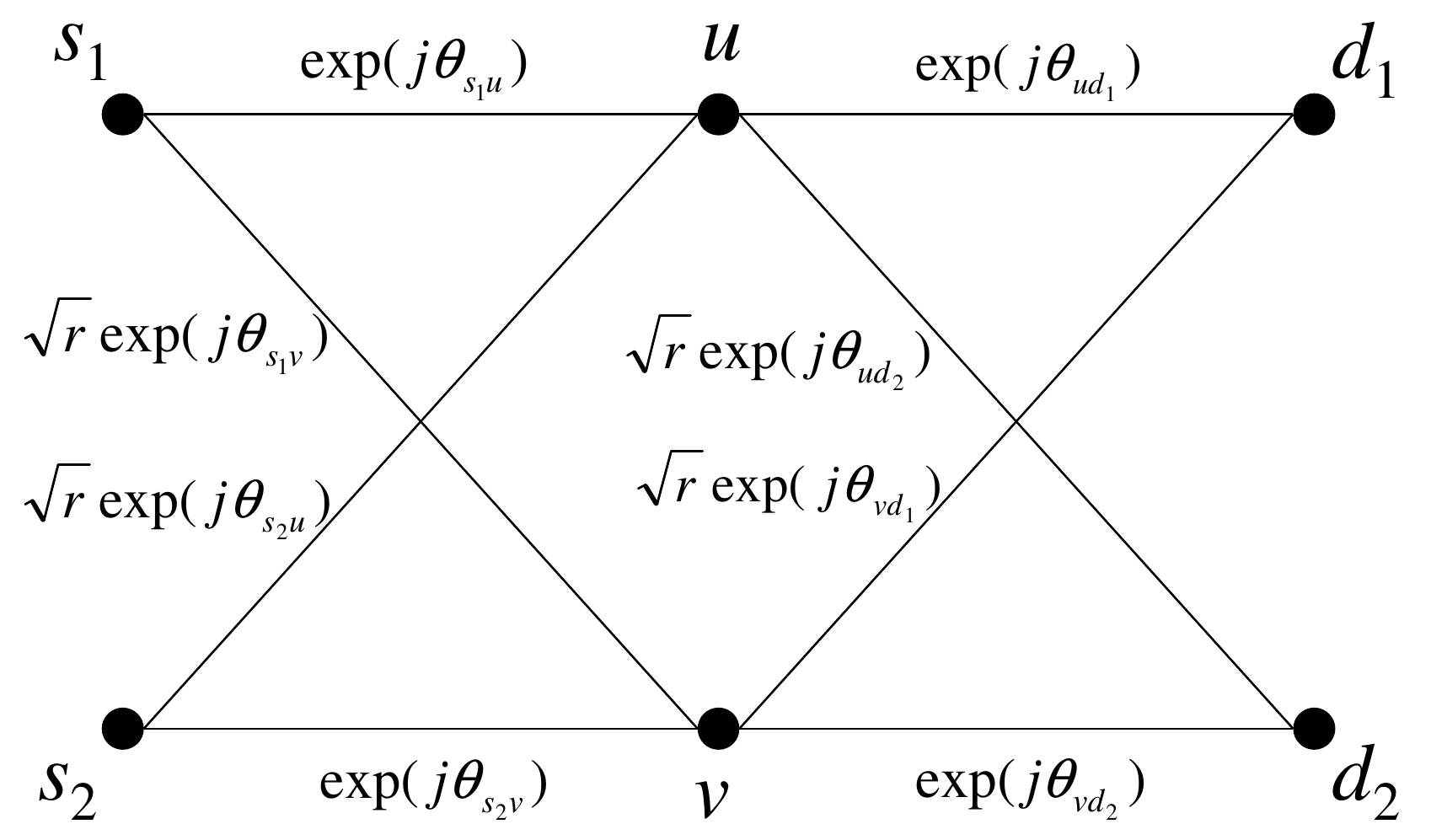}
\caption{A two-hop IC where each hop consists of direct channels with magnitude one and interference channels with magnitude $\sqrt{r}$.}  \label{twoHopICSimulationComplex}
\end{figure}
\begin{figure}[htp]
\centering
\subfigure[MIMO-S topology under $(\mathbf{A^S},\mathbf{B^S})$.]{\includegraphics[width=2.3 in, height=1.4 in]{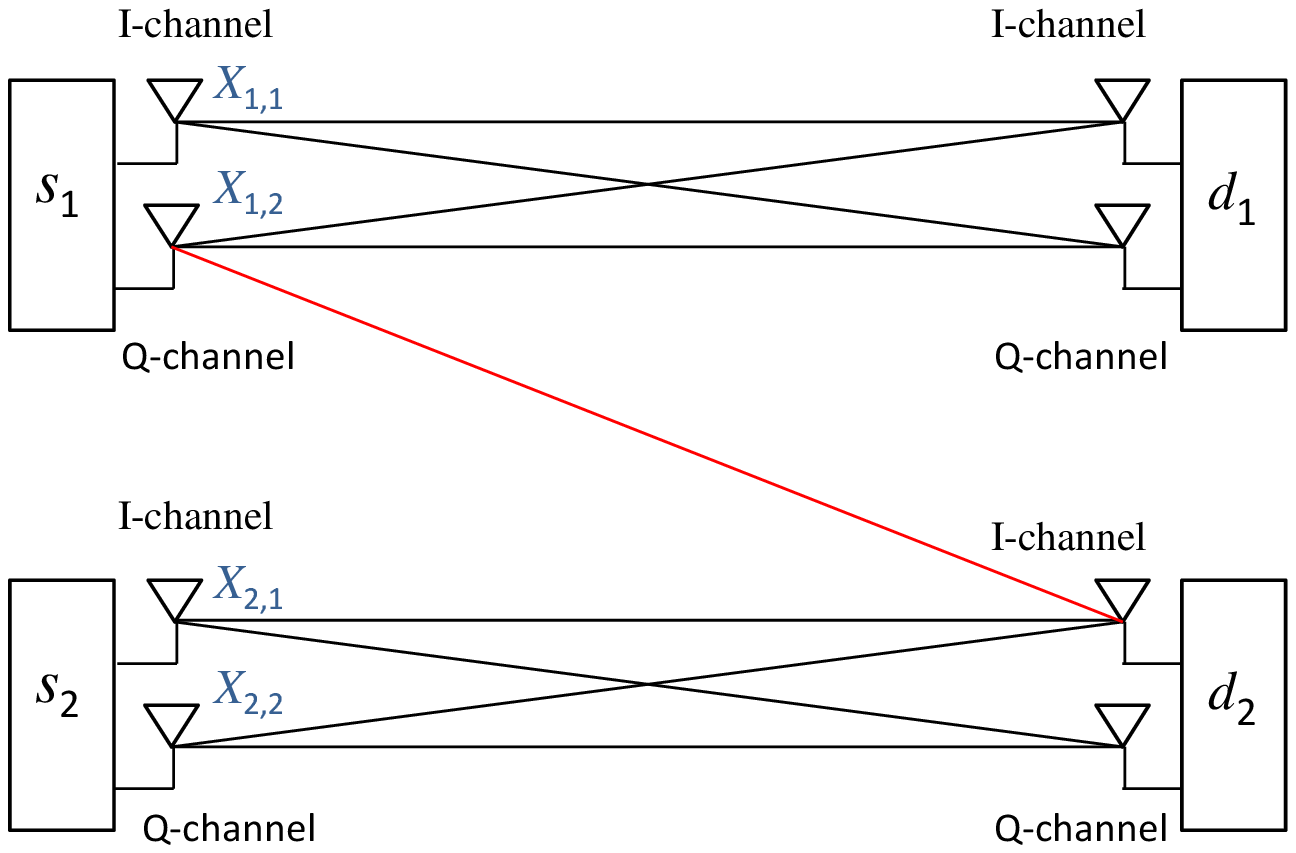}}
\subfigure[MIMO-Z topology under $(\mathbf{A^Z},\mathbf{B^Z})$.]{\includegraphics[width=2.3 in, height=1.4 in]{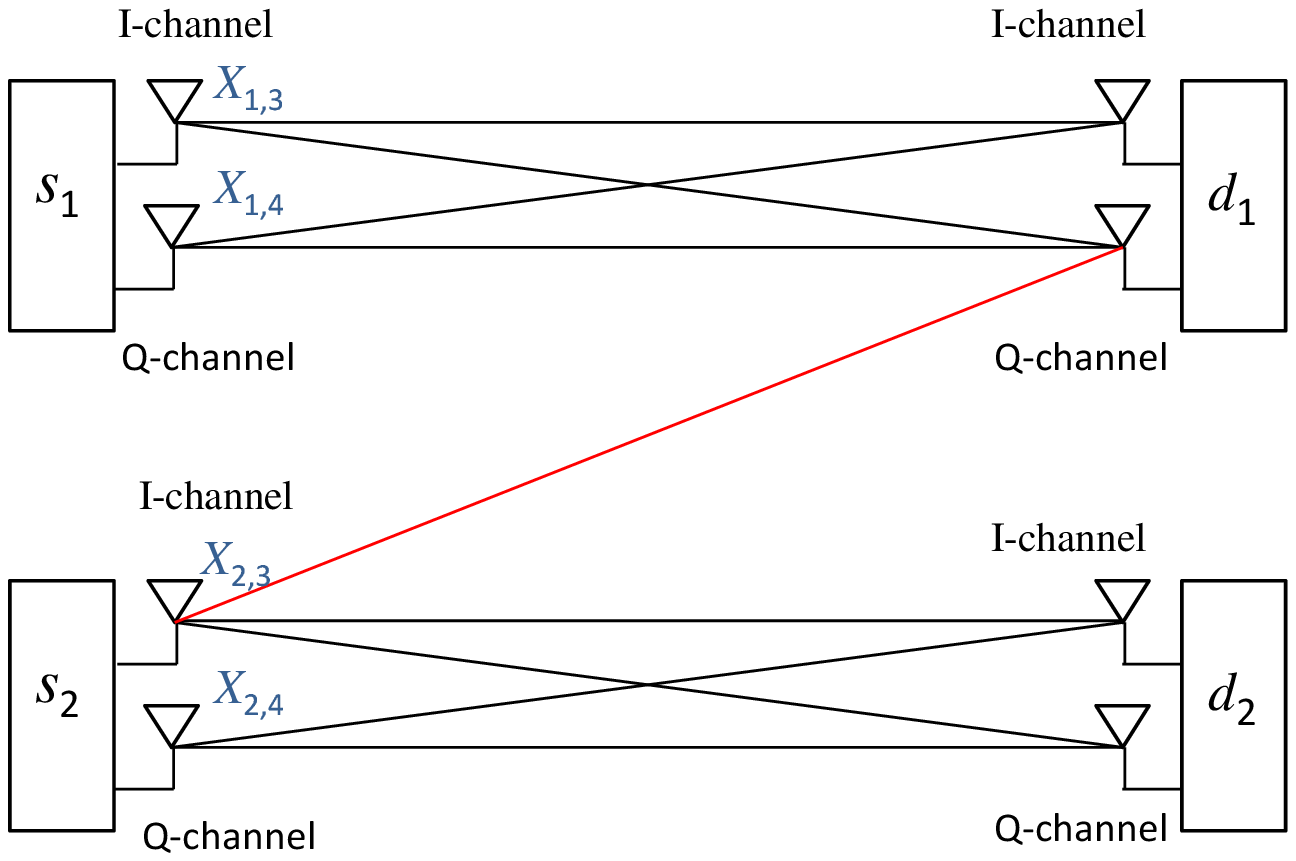}}
\subfigure[MIMO-X topology under $(\mathbf{A^X},\mathbf{B^X})$.]{\includegraphics[width=2.3 in, height=1.4 in]{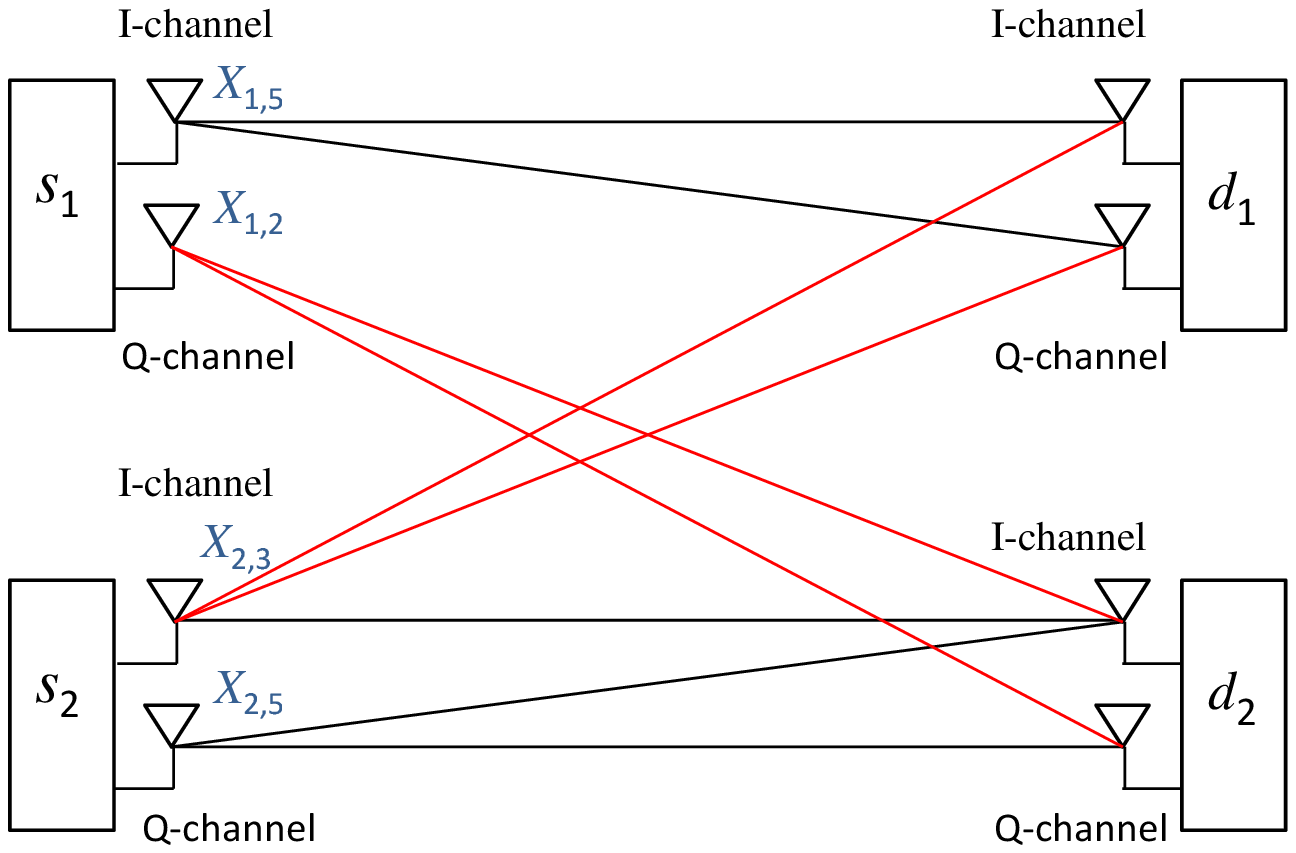}}
\caption{End-to-end interference structures.}  \label{twoHopICSimulationStructures}
\end{figure}

We consider a two-hop IC with single-antenna nodes illustrated in Figure~\ref{twoHopICSimulationComplex}, where $h_{ab}=e^{j\theta_{ab}} \in \mathbb{C}$ for each $(a,b)\in \{(s_1,u), (s_2,v), (u,d_1), (v,d_2)\}$ and $h_{cd}=\sqrt{r} e^{j\theta_{cd}} \in \mathbb{C}$ for each $\{(s_1,v), (s_2,u), (u,d_2), (v,d_1)\}$ for some $0<r \leq 1$. The parameter~$r$ represents the interference power of both the first-hop and second-hop ICs. As discussed in Section~\ref{twoHopICMIMOComplex}, we can view this two-hop IC with complex channel gains as a two-hop MIMO IC consisting of 2-antenna nodes with real channel gains and can therefore apply the 3-phase linear scheme described in Section~\ref{achMIMO} as follows: Source $s_1$ transmits five independent real symbols denoted by $X_{1,k}$'s and $s_2$ transmits five independent symbols denoted by $X_{2,k}$'s in the three phases as illustrated in Figure~\ref{twoHopICSimulationStructures}, where the actual complex symbols transmitted by $s_1$ in the first, second and third phases are $X_{1,1}+jX_{1,2}$, $X_{1,3}+jX_{1,4}$ and $X_{1,5}+jX_{1,2}$ respectively and the actual complex symbols transmitted by $s_2$ in the first, second and third phases are $X_{2,1} + jX_{2,2}$, $X_{2,3}+jX_{2,4}$ and $X_{2,1}+jX_{2,5}$ respectively. Then, the relays apply the relaying kernels $(\mathbf{A^S},\mathbf{B^S})$, $(\mathbf{A^Z},\mathbf{B^Z})$ and $(\mathbf{A^X},\mathbf{B^X})$ in the first, second and third phases to create the MIMO-S, MIMO-Z and MIMO-X topologies respectively. This is illustrated in Figure~\ref{twoHopICSimulationStructures}, where $(\mathbf{A^S},\mathbf{B^S})$, $(\mathbf{A^Z},\mathbf{B^Z})$ and $(\mathbf{A^X},\mathbf{B^X})$ are defined in Lemmas~\ref{lemmaS}, \ref{lemmaZ} and \ref{lemmaX} respectively. Following similar procedures for deriving two interference-free real-valued data streams specified by \eqref{firstDataStream} and \eqref{secondDataStream} from \eqref{timeslot1}, \eqref{timeslot2} and \eqref{timeslot3} for $s_1$ in three time slots, $s_1$ can derive an interference-free version of each of the $X_{1,k}$'s from \eqref{phase1}, \eqref{phase2} and \eqref{phase3}. Similarly, $s_2$ can derive an interference-free version of each of the $X_{2,k}$'s in three time slots. Consequently, we can compute the sum-rate achievable by our 3-phase scheme by summing the $\frac{1}{2} \log (1+\text{SNR})$ of the interference-free channels created for each symbol (similar to computing \eqref{singleAntennaRealAchRate} from \eqref{firstDataStream} and \eqref{secondDataStream}).

\begin{figure}[htp]
\centering
\subfigure[Symbols transmitted in first phase.]{\includegraphics[width=2.3 in, height=1.5 in]{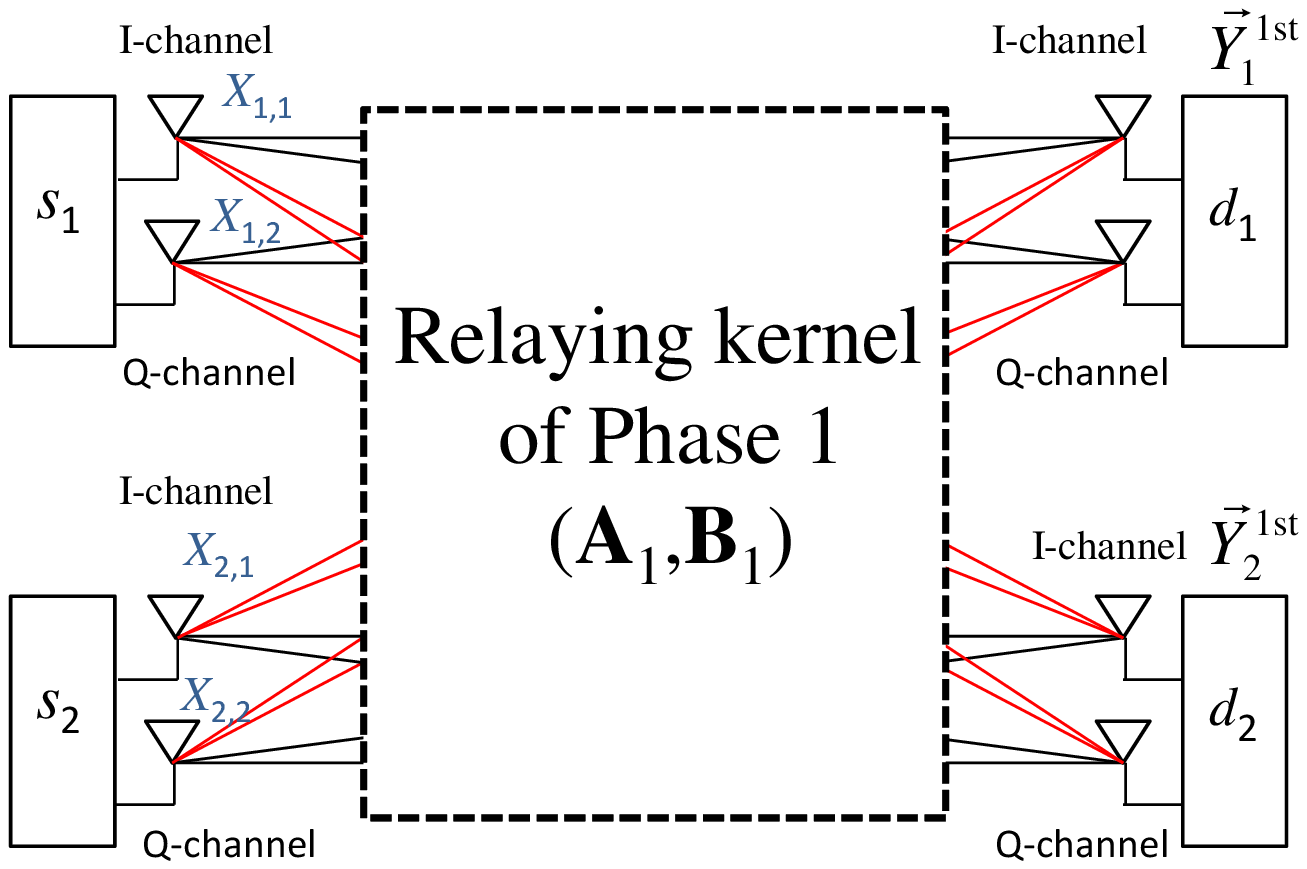}}
\subfigure[Symbols transmitted in second phase.]{\includegraphics[width=2.3 in, height=1.5 in]{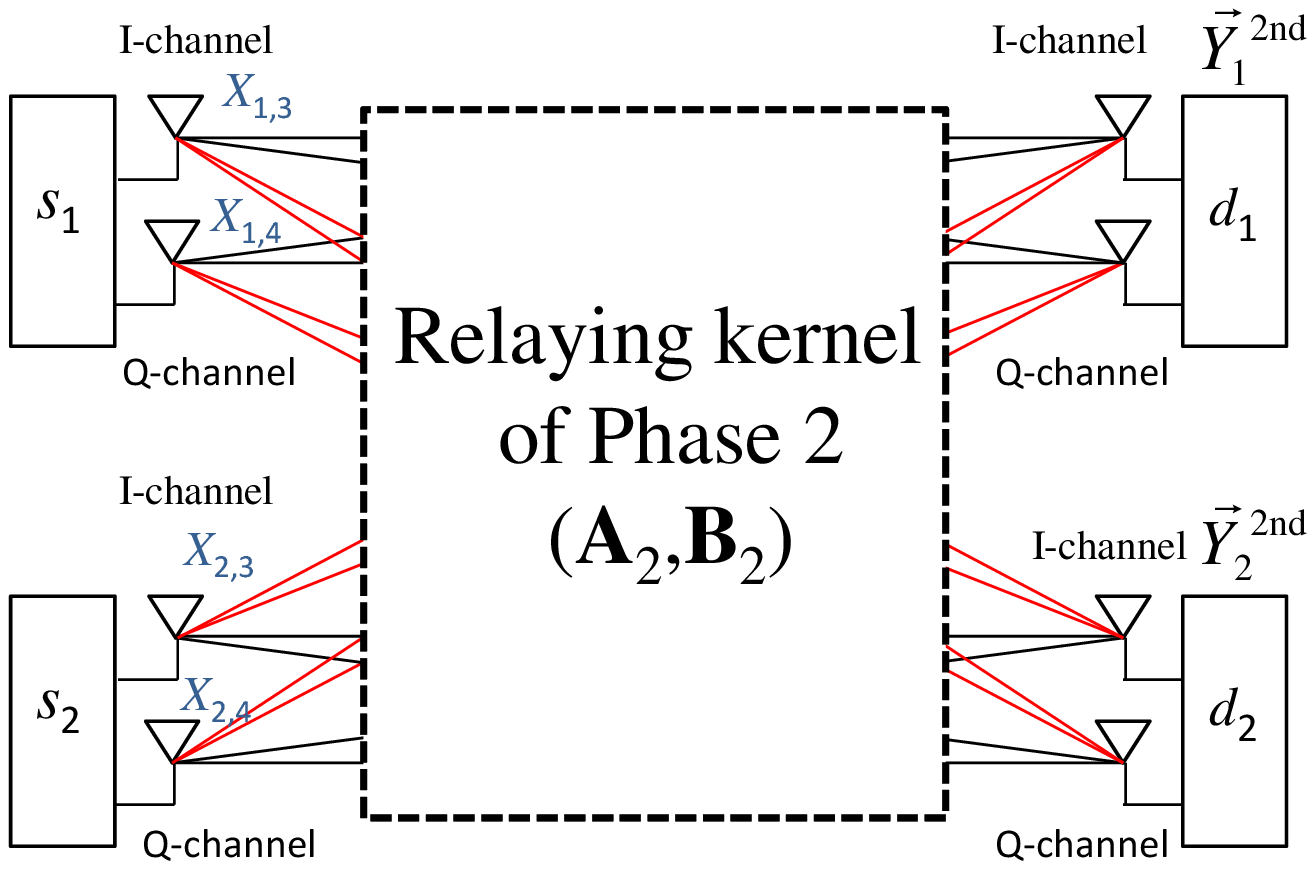}}
\subfigure[Symbols transmitted in third phase.]{\includegraphics[width=2.3 in, height=1.5 in]{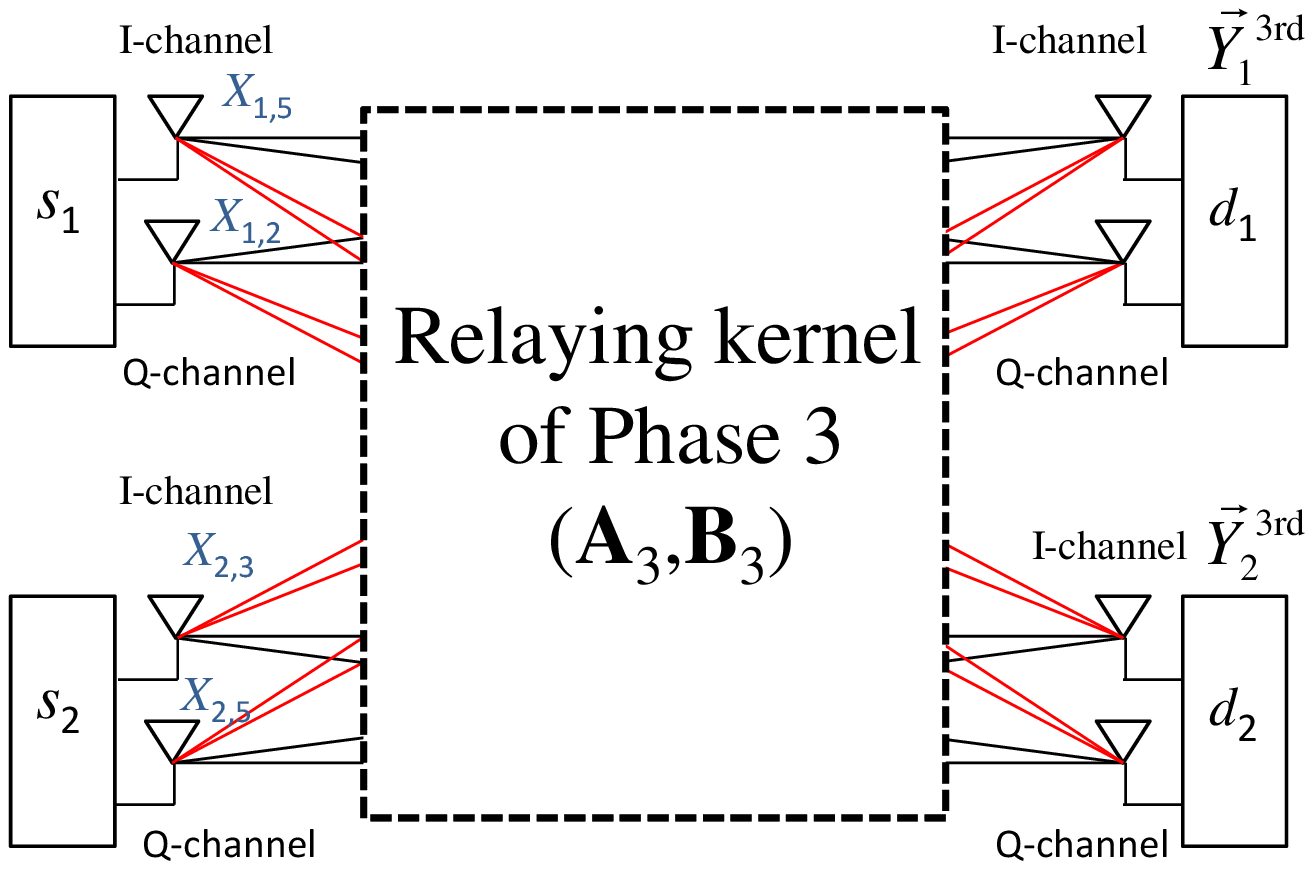}}
\caption{Patterns of transmit symbols in the three phases.}  \label{twoHopICSimulationStructuresImproved}
\end{figure}

Although the 3-phase scheme described above achieves the optimal sum-DoF, i.e., 5/3, its performance can be improved for finite SNR by not insisting on creating the MIMO-S, MIMO-Z and MIMO-X topologies in the three phases. Instead, if only the coding pattern of transmit symbols in the three phases are preserved as shown in Figure~\ref{twoHopICSimulationStructuresImproved}, then $s_1$ and $s_2$ can achieve $I(X_{1,1}, X_{1,2}, X_{1,3}, X_{1,4}, X_{1,5}; \vec Y_{1}^{\text{1st}},\vec Y_{1}^{\text{2nd}}, \vec Y_{1}^{\text{3rd}})$ and $I(X_{2,1}, X_{2,2}, X_{2,3}, X_{2,4}, X_{2,5}; \vec Y_{2}^{\text{1st}},\vec Y_{2}^{\text{2nd}}, \vec Y_{2}^{\text{3rd}})$ respectively, where $\vec Y_i^{\text{1st}}$, $\vec Y_i^{\text{2nd}}$ and $\vec Y_i^{\text{3rd}}$ are the symbols received by $d_i$ in the first, second and third phase respectively. More specifically, if we let $X_{1,k}$'s and $X_{2,k}$'s be independent random variables $\sim \mathcal{N}(0, \frac{P}{2})$ and let
\[
\mathcal{V}\! = \! \left\{ \! (\mathbf{A}, \mathbf{B})\in \mathbb{R}^{2\times 2} \!\!\times \! \mathbb{R}^{2\times 2} \! \left| ||\mathbf{A}|| \le \sqrt{\frac{P}{(||\bar H_{s_1,u}||^2 + ||\bar H_{s_2,u}||^2)P + 1}}\,,  ||\mathbf{B}|| \le \sqrt{\frac{P}{(||\bar H_{s_1,v}||^2 + ||\bar H_{s_2,v}||^2)P + 1}} \right.\right\}
 \]
 be the set of relaying kernels that respect the power constraint for the relays where $\bar H_{a,b}$ is defined in \eqref{barHij}, then the sum-rate achievable by the 3-phase scheme not insisting on creating the three topologies is
\begin{equation}
R_{\text{3-phase}} \triangleq \sup_{(\mathbf{A}_1, \mathbf{B}_1),(\mathbf{A}_2, \mathbf{B}_2),(\mathbf{A}_3, \mathbf{B}_3)\in \mathcal{V}}\left\{(R_1, R_2) \left| \:\parbox[c]{3.1 in}{$R_1\le  I(X_{1,1}, X_{1,2}, X_{1,3}, X_{1,4}, X_{1,5}; \vec Y_{1}^{\text{1st}},\vec Y_{1}^{\text{2nd}}, \vec Y_{1}^{\text{3rd}}),\\R_2\le I(X_{2,1}, X_{2,2}, X_{2,3}, X_{2,4}, X_{2,5}; \vec Y_{2}^{\text{1st}},\vec Y_{2}^{\text{2nd}}, \vec Y_{2}^{\text{3rd}})$}\right.\right \}, \label{R3Phase}
\end{equation}
where $(\mathbf{A}_k, \mathbf{B}_k)$ is the relaying kernel used in Phase~$k$. Although finding the optimal $((\mathbf{A}_1, \mathbf{B}_1),(\mathbf{A}_2, \mathbf{B}_2),(\mathbf{A}_3, \mathbf{B}_3))$ that maximizes $R_{\text{3-phase}}$ is a non-convex optimization problem as shown in \eqref{R3Phase}, we can still obtain in our simulation a heuristic sum-rate by first evaluating the closed form of the mutual information terms in \eqref{R3Phase} followed by conducting MATLAB constrained non-linear optimization initiated at the relaying kernels $((\mathbf{A^S},\mathbf{B^S}), (\mathbf{A^Z},\mathbf{B^Z}),(\mathbf{A^X},\mathbf{B^X}))$ (cf.\ Figure~\ref{twoHopICSimulationStructures}).

We now compare the achievable rate of our 3-phase scheme with other schemes in the literature. As for the benchmark, we consider two schemes: time-sharing (TDMA) and amplify-forward (AF) schemes. Under the TDMA scheme, the two sources transmit their messages in different time slots and the two relays forward the messages in different time slots in such a way that~$u$ forwards only the message of~$s_1$ and~$v$ forwards only the message of~$s_2$. Under the AF scheme, the sources transmit the codewords consisting of complex symbols simultaneously and each relay multiplies its received codeword with a time-invariant complex scalar followed by transmitting the resultant codeword to the destinations. Upon receiving the complex codewords from the relays, each destination under the AF scheme
decodes the message by treating interference as noise. For a finite~$P$, let
\begin{equation*}
R_{\text{TDMA}} = \log_2(1+P)
\end{equation*}
 be the sum-rate achievable by TDMA schemes, and let
 \begin{align*}
 R_{\text{AF}}&= \max_{\alpha, \beta\in \mathbb{C}: |\alpha|, |\beta| \le \sqrt{\frac{P}{P+1}}} \bigg\{ \log_2\left(1+\frac{|g_{11}|^2P}{|g_{12}|^2P+|\alpha|^2+|\beta|^2+1}\right) \notag \\
  &\qquad \hspace{1.6 in} + \log_2\left(1+\frac{|g_{22}|^2P}{|g_{21}|^2P+|\alpha|^2+|\beta|^2+1}\right) \bigg\}
 \end{align*}
 be the sum-rate achievable by AF schemes, where $\alpha$ and $\beta$ are the amplifying scalars chosen by $u$ and $v$ respectively and $g_{ij}\triangleq h_{s_ju}h_{ud_i}\alpha  + h_{s_jv}h_{vd_i} \beta$ is the end-to-end channel gain between~$s_j$ and~$d_i$.

 In addition, we also consider two recent schemes called \textit{compute-and-forward with aligned network diagonalization} (CoF-AND) and \textit{precoded compute-and-forward with channel integer alignment} (PCoF-CIA) \cite{CaireFiniteField} respectively. The main idea behind these schemes is to innovatively use lattice codes and transform the two-hop IC into a network defined on a finite field. Then, the relays cooperate to eliminate the end-to-end interference in the finite field domain by using aligned network diagonalization techniques \cite{IlanKKK} under CoF-AND and by using asymmetric complex signaling techniques \cite{asymmetricComplexSignaling} under PCoF-CIA. To facilitate discussion, let $R_{\text{CoF-AND}}$ and $R_{\text{PCoF-CIA}}$ be the sum-rates achievable by CoF-AND and PCoF-CIA respectively.

  \begin{figure}[htp]
\centering
\subfigure[$r=0.5$.]{\includegraphics[width=3 in, height=2.6 in]{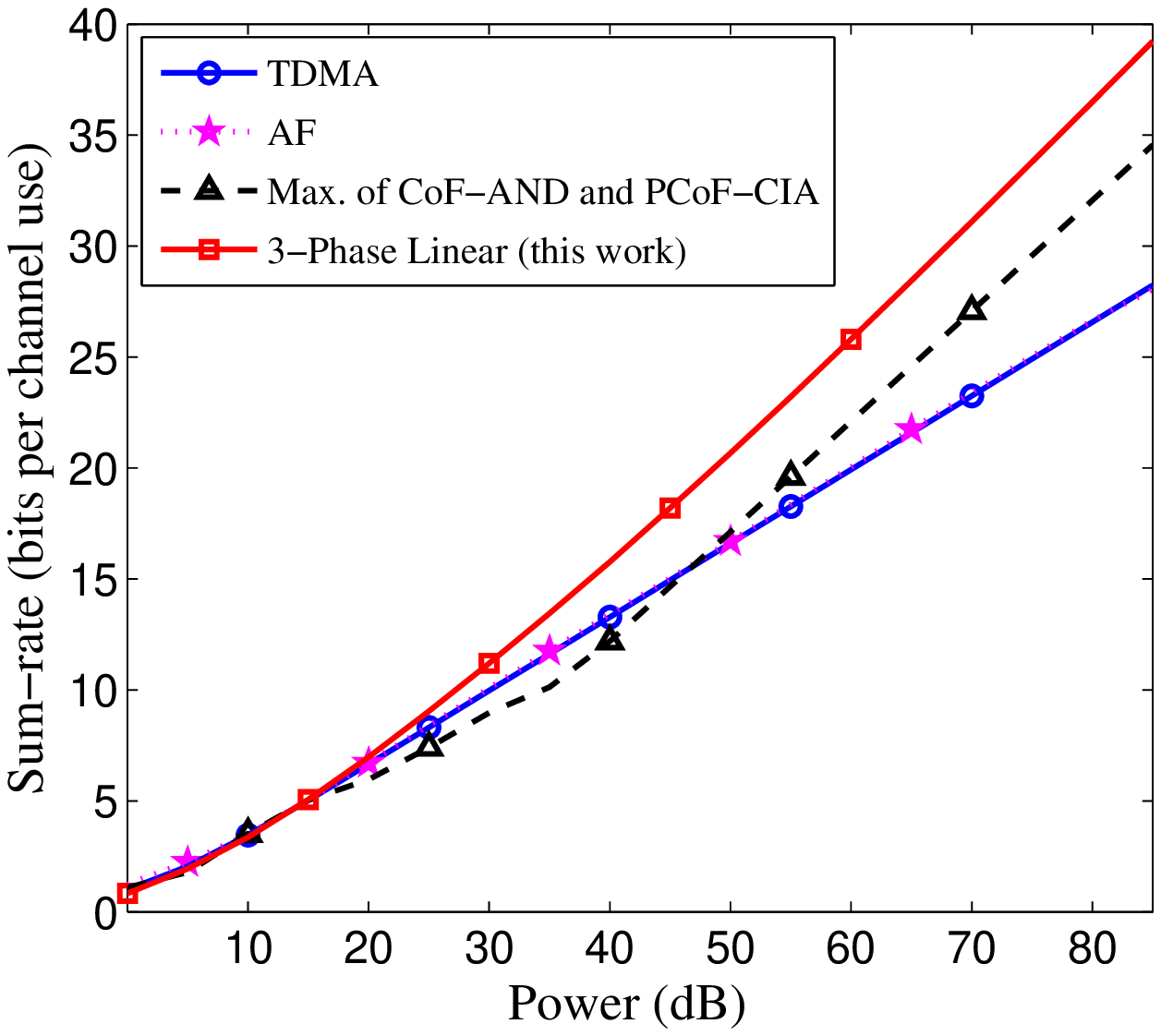}} \hspace{.4 in}
\subfigure[$r=1$.]{\includegraphics[width=3 in, height=2.6 in]{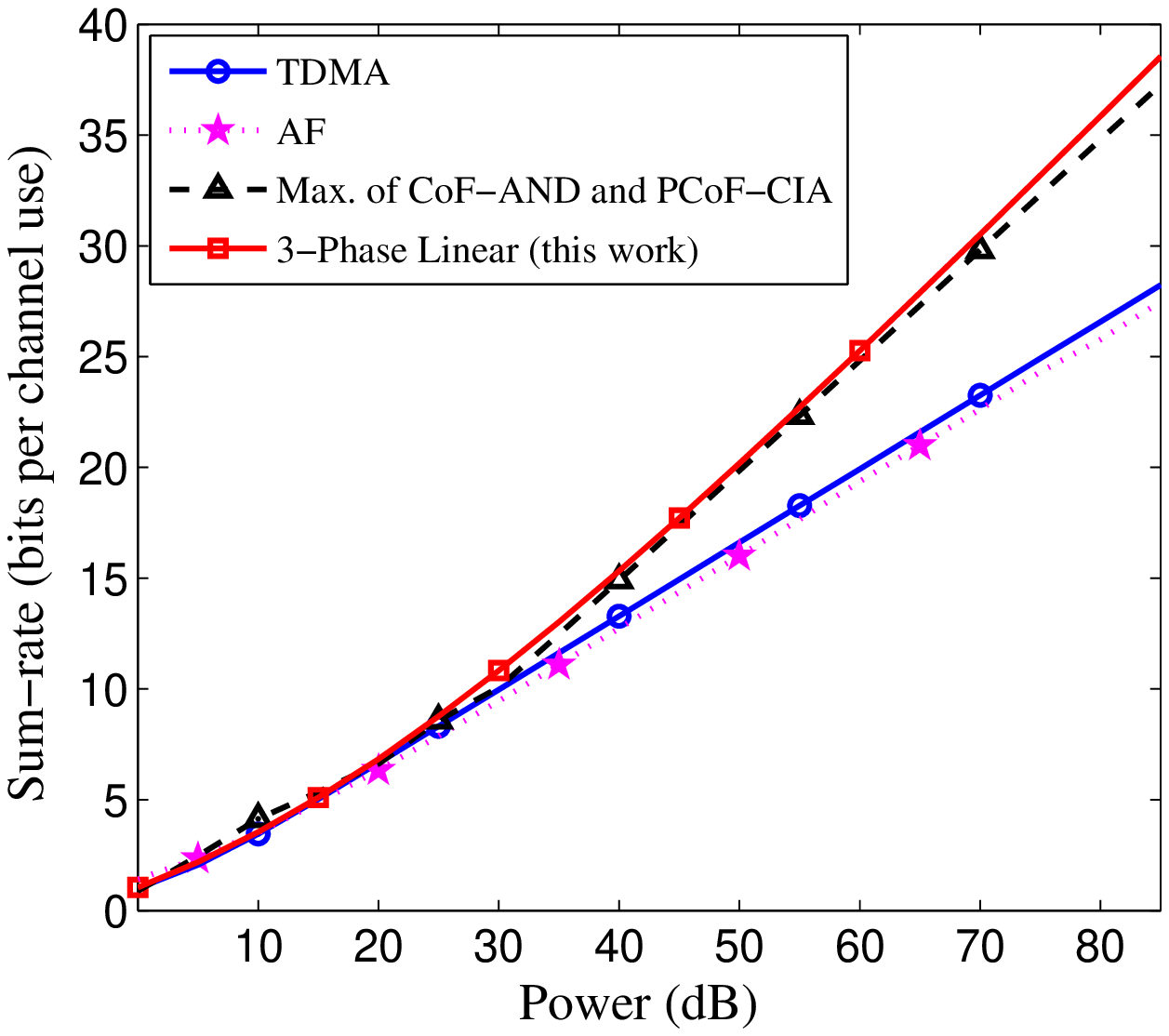}}
\caption{Comparison among TDMA, AF, CoF-AND, PCoF-CIA and our 3-phase linear schemes.}  \label{complexCase}
\end{figure}
We now consider the numerical analysis of the aforementioned schemes for the network of Figure~\ref{twoHopICSimulationComplex} and focus on two values of $r$: $r=1$ and $r=0.5$ corresponding respectively to moderate and high interference regimes. 
In Figure~\ref{complexCase}, we plot for both regimes the average values of $R_{\text{TDMA}}$, $R_{\text{AF}}$, $R_{\text{3-phase}}$ and $\max\{R_{\text{CoF-AND}},R_{\text{PCoF-CIA}}\}$ against power~$P$, where the average is obtained by Monte Carlo simulation which assumes that $\theta_{ab}$'s are i.i.d.\ phases uniformly distributed over $[0, 2\pi]$ (cf.\ Figure~\ref{twoHopICSimulationComplex}). In both cases, we note that our 3-phase linear scheme starts to outperform the other schemes at moderate SNR, particularly at about 15dB for $r=0.5$ and 25dB for $r=1$ as shown in Figures~\ref{complexCase}(a) and (b) respectively. Also, in both cases, the gap between the 3-phase scheme and the other schemes widens as $P$ increases, due to the fact that our scheme has a strictly higher sum-DoF. For the high interference regime (i.e., $r=1$), we note that the performance of 3-phase scheme and the best of CoF-AND and PCoF-CIA schemes are very similar at moderate SNR (before 60dB). However, from the complexity perspective, since the  3-phase scheme only relies on simple linear operations over blocks of size 3, it can be more appealing. 



\section{Conclusion} \label{conclusion}

In this paper, we analyzed the sum-DoF of the two-hop IC with real constant channel gains when relays are restricted to perform vector-linear schemes. We characterized the sum-DoF achievable by such schemes to be 4/3 for almost all values of real channel gains. We then extended the result to the case where each node has $M$ antennas. We showed that the linear sum-DoF in this setup is $2M-2/3$ for almost all values of channel gains. Furthermore, we adapted this result to the case of complex channel gains and $M$-antenna nodes, for which we characterized the sum-DoF to be $2M-1/3$ for almost all values of complex channel gains. Finally, we analytically computed the rates achieved by  our proposed scheme for the single-antenna two-hop IC with complex channel gains for different SNR values, and compared them with achievable rates of state-of-the-art schemes. Simulation results show that the proposed scheme is robust against changes in the interference strength, and outperforms state-of-the-art schemes even at moderate SNR.

\begin{figure}[!htp]
\centering
\includegraphics[scale=0.6]{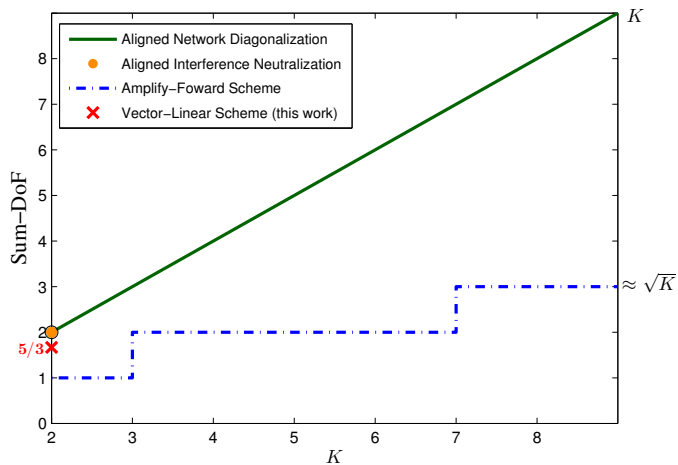}
\caption{Summary of known results of interest for the achievable sum-DoF for general $K \times K \times K$ networks.} \label{current}
\end{figure}

This study can be extended in several directions. One direction could be to investigate the performance of linear schemes in more general two-unicast networks, such as those studied in \cite{Ilan2Unicast,2UnicastConf}. Another interesting direction is the study of linear schemes for general $K \times K \times K$ networks. A summary of current results is shown in Figure~\ref{current}, where the $x$-axis represents the number of users $K$, and the $y$-axis represents the achievable sum-DoF for the corresponding $K \times K \times K$ network (with complex channel gains). We already know, due to the result of~\cite{Jafar}, that the point (2,2) is achievable. More generally, we know that all points of the form $(K,K)$ are achievable due to the result of~\cite{IlanKKK}. If we are restricted to amplify-forward schemes, then we know by the result of~\cite{RankovWittnebenLinear} that for a $K \times N \times K$ network, where $N = K(K-1)+1$, all interference links can be canceled by appropriately choosing the relay coefficients, and thus $K$ sum-DoF is achievable. This implies that for the $K \times K \times K$ network we can achieve a sum-DoF of the order of $\sqrt{K}$, by canceling as many interference links as possible. In this work, we showed that 5/3 sum-DoF is achievable for the $2 \times 2 \times 2$ network. And so we ask: what can linear schemes achieve for general $K \times K \times K$ networks?

\newpage
\begin{appendices}
\section{Proof of Lemma~\ref{lemmaLinearDecomposition}} \label{Lemma3}
We first show that there exist four unique real numbers $\lambda_1^\prime$, $\lambda_2^\prime$, $\mu_1^\prime$ and $\mu_2^\prime$ which are functions of $(\mathbf{H}_1,\mathbf{H}_2)$ such that
\begin{equation}
h_{s_1u}h_{ud_1}x + h_{s_1v}h_{vd_1} y = \lambda_{1}^\prime(h_{s_2u}h_{ud_1}x + h_{s_2v}h_{vd_1}y)+ \lambda_{2}^\prime (h_{s_1u}h_{ud_2}x + h_{s_1v}h_{vd_2}y)
\label{G11Decomposition}
\end{equation}
and
\begin{equation}
h_{s_2u}h_{ud_2}x + h_{s_2v}h_{vd_2}y = \mu_{1}^\prime(h_{s_2u}h_{ud_1}x + h_{s_2v}h_{vd_1} y)+ \mu_{2}^\prime(h_{s_1u}h_{ud_2}x + h_{s_1v}h_{vd_2}y)
\label{G22Decomposition}
\end{equation}
for all $x\in\mathbb{R}$ and all $y\in\mathbb{R}$. The lemma will then follow from \eqref{Gij}, \eqref{G11Decomposition} and \eqref{G22Decomposition} by letting $\lambda_1=\lambda_1^\prime$,  $\lambda_2=\lambda_2^\prime$, $\mu_1=\mu_1^\prime$ and $\mu_2=\mu_2^\prime$. Comparing coefficients of $x$ and $y$ on both sides of \eqref{G11Decomposition}, we obtain the following matrix equation:
\begin{equation}
\left[ \begin{array}{cc}h_{s_2u}h_{ud_1} & h_{s_1u}h_{ud_2} \\ h_{s_2v}h_{vd_1} & h_{s_1v}h_{vd_2} \end{array}  \right]\left[\begin{array}{c}\lambda_{1}^\prime \\ \lambda_{2}^\prime\end{array}\right] = \left[\begin{array}{c}h_{s_1u}h_{ud_1} \\ h_{s_1v}h_{vd_1} \end{array}\right].\label{linearSystem1}
\end{equation}
To facilitate discussion, let $D= \det\left( \left[ \begin{array}{cc}h_{s_2u}h_{ud_1} & h_{s_1u}h_{ud_2} \\ h_{s_2v}h_{vd_1} & h_{s_1v}h_{vd_2} \end{array}  \right]\right)$.
Since $D \ne 0$ by condition (c-3), it follows from Cramer's rule that the unique solution for \eqref{linearSystem1} is
\begin{equation}
\begin{cases} \lambda_1^\prime = \frac{h_{s_1u}h_{s_1v}\det(\mathbf{H}_2)}{D},  \\ \lambda_2^\prime = \frac{-h_{ud_1}h_{vd_1}\det(\mathbf{H}_1)}{D}.  \end{cases} \label{linearSystem1Solution}
\end{equation}
Substituting \eqref{linearSystem1Solution} into \eqref{G11Decomposition} followed by comparing coefficients of $x$ and $y$ on both sides of \eqref{G11Decomposition}, we find that \eqref{G11Decomposition} is satisfied by $\lambda_1^\prime$ and $\lambda_2^\prime$.
 Following similar procedures for obtaining $\lambda_1^\prime$ and $\lambda_2^\prime$ that satisfy \eqref{G11Decomposition}, we obtain $\mu_1^\prime$ and $\mu_2^\prime$ that satisfy \eqref{G22Decomposition}.

\section{Proof of Lemma~\ref{lemmaDifferenceEntropy}} \label{Lemma4}
Let $\mathbf{L}$ be an $n\times n$ matrix. Expanding $h(X^n+Z_1^n,Y^n+Z_2^n)$ in two different ways, we obtain
\begin{align*}
& h(X^n+Z_1^n) +h(Y^n+Z_2^n|X^n+Z_1^n) \\
&\quad  = h(Y^n+Z_2^n)+h(X^n+Z_1^n|Y^n+Z_2^n),
\end{align*}
which then implies that
\begin{align*}
&h(X^n+Z_1^n) - h(Y^n+Z_2^n)\\
 &= h(X^n+Z_1^n|Y^n+Z_2^n)- h(Y^n+Z_2^n|X^n+Z_1^n) \\
&\le h(X^n-\mathbf{L} Y^n+Z_1^n-\mathbf{L} Z_2^n|Y^n+Z_2^n)- h(Y^n+Z_2^n|Y^n,X^n+Z_1^n)\\
&\le h(X^n-\mathbf{L} Y^n+Z_1^n-\mathbf{L} Z_2^n|Y^n+Z_2^n)- h(Z_2^n|Y^n, X^n+Z_1^n)\\
& = h(X^n-\mathbf{L} Y^n+Z_1^n-\mathbf{L} Z_2^n|Y^n+Z_2^n)- h(Z_2^n),
\end{align*}
where the last equality follows from the fact that $Z_1^n$, $Z_2^n$ and $(X^n,Y^n)$ are independent.
\bigskip 

\section{Proof of Lemma~\ref{lemmaDifferentialEntropyBound1}} \label{Lemma5}
We need the following three propositions for proving Lemma~\ref{lemmaDifferentialEntropyBound1}.
\bigskip
\begin{Proposition}\label{lemmaRank}
Let $\mathcal{V}$ be a finite set of $\ell \times \ell$ real matrices and let $\mathbb{M}$ be the set of real matrices. Then, there exist two mappings $\phi_\mathcal{V}: \mathcal{V} \rightarrow \mathbb{M}$ and $\psi_\mathcal{V}: \mathcal{V} \rightarrow  \mathbb{M}$ such that for any $\mathbf{G}\in \mathcal{V}$, $\phi_\mathcal{V}(\mathbf{G})$ is an invertible $\ell\times \ell$ matrix and $\psi_\mathcal{V}(\mathbf{G})$ is a $\rank(\mathbf{G})\times \ell$ matrix that satisfy
\begin{equation*}
\begin{cases}
|\det(\phi_\mathcal{V}(\mathbf{G}))|=1, ~~ \text{and}\\
 \phi_\mathcal{V}(\mathbf{G}) \mathbf{G}= \left[\begin{array}{c} \psi_\mathcal{V}(\mathbf{G}) \\ \mathbf{0}^{(\ell-\rank(\mathbf{G}))\times \ell}\end{array}\right].
 \end{cases}
 \end{equation*}
\end{Proposition}
In addition, there exists a real number~$K$ which is only a function of $\ell$ and $\mathcal{V}$ such that $K$ is an upper bound on the magnitudes of the entries in each $\phi_\mathcal{V}(\mathbf{G})$ and each $\psi_\mathcal{V}(\mathbf{G})$.
\begin{IEEEproof}
Suppose $\mathbf{G}$ is a matrix in $\mathcal{V}$. By linear algebra, there exists an invertible $\ell\times \ell$ matrix denoted by $\mathbf{L}_{\mathbf{G}}$ and a full-rank $\rank(\mathbf{G})\times \ell$ matrix denoted by $\mathbf{G}^*$ such that $|\det(\mathbf{L}_{\mathbf{G}})|=1$ and $\mathbf{L}_{\mathbf{G}} \mathbf{G}= \left[\begin{array}{c} \mathbf{G}^*\\ \mathbf{0}^{(\ell-\rank(\mathbf{G}))\times \ell} \end{array}\right]$. The lemma then follows by letting $\phi_\mathcal{V}(\mathbf{G})=\mathbf{L}_{\mathbf{G}}$ and $\psi_\mathcal{V}(\mathbf{G}) = \mathbf{G}^*$ for each $\mathbf{G}\in \mathcal{V}$ and letting $K$ be the maximum of $\max\{|g| : g \text{ is an entry in some }\phi_\mathcal{V}(\mathbf{G})\}$ and $\max\{|g| : g \text{ is an entry in some }\psi_\mathcal{V}(\mathbf{G})\}$.
\end{IEEEproof}
\bigskip
\begin{Proposition} \label{lemmaBoundedNorm}
For any $x^\ell \in \mathbb{R}^{\ell\times 1}$ and any vector $\vec \phi \in \mathbb{R}^{1\times \ell}$ such that the magnitude of each entry of $\vec \phi$ is less than some $K \ge 0$, $(\vec \phi x^\ell)^2 \le \ell^2 K^2 ||x^\ell||^2$.
\end{Proposition}
\begin{IEEEproof}
Let $x^\ell \in \mathbb{R}^{\ell\times 1}$ and $\vec \phi \in \mathbb{R}^{1\times \ell}$ such that the magnitude of each entry of $\vec \phi$ is less than some $K\ge 0$. Since the magnitude of each entry of $\vec \phi$ is less than $K$, $|\vec \phi x^\ell| \le K(|x_1|+|x_2|+\ldots + |x_\ell|)$ by triangle inequality, which then implies that
\begin{align*}
(\vec \phi x^\ell)^2 & \le K^2(|x_1|+|x_2|+\ldots + |x_\ell|)^2 \notag\\
& \leq K^2 (\ell \max_{i\in\{1,\dots,\ell\}} |x_i| )^2 \notag \\
&\le \ell^2 K^2(|x_1|^2+|x_2|^2+\ldots + |x_\ell|^2) \notag \\
&\le \ell^2 K^2 ||x^\ell||^2.
\end{align*}
\end{IEEEproof}
\bigskip
\begin{Proposition} \label{differentialEntropyUpperBoundGeneric}
Let $\ell$ be a natural number, $\vec u$ be a standard basis vector in $\mathbb{R}^{1\times \ell}$ and $\Lambda_1$, $\Lambda_2$, $\Omega_1$, $\Omega_2$ and $\Omega_3$ be five matrices in $\mathbb{R}^{\ell \times \ell}$. In addition, let $X_1^\ell$, $X_2^\ell$, $Z_1^\ell$, $Z_2^\ell$ and $Z_3^\ell$ be five independent random vectors in $\mathbb{R}^{\ell \times 1}$ such that $Z_i^\ell$ is $\ell$ independent copies of $\mathcal{N}(0,1)$ for each $i\in\{1,2,3\}$. If there exists a real number~$K$ such that $K$ is an upper bound on the magnitudes of the entries in  $\Lambda_1$, $\Lambda_2$, $\Omega_1$, $\Omega_2$ and $\Omega_3$, then there exists a real number~$\kappa$ which is only a function of $K$ and $\ell$ such that
\begin{align}
& h(\vec u (\Lambda_1 X_1^\ell + \Lambda_2 X_2^\ell + \Omega_1 Z_1^\ell + \Omega_2 Z_2^\ell + \Omega_3 Z_3^\ell)) \notag \\
&\quad \le \log_2\sqrt{1+ \E\left[\sum_{m=1}^\ell (X_{1,m}^2+X_{2,m}^2) \right]/\ell} + \kappa \label{propositionBoundGenericSt1}
\end{align}
and
\begin{equation}
h(\vec u (\Omega_1 Z_1^\ell + \Omega_2 Z_2^\ell + \Omega_3 Z_3^\ell)) \le \kappa. \label{propositionBoundGenericSt2}
\end{equation}
\end{Proposition}
\begin{IEEEproof}
Let $K\in \mathbb{R}$ be an upper bound on the magnitudes of the entries in $\Lambda_1$, $\Lambda_2$, $\Omega_1$, $\Omega_2$ and $\Omega_3$. Consider
\begin{align}
& \E[(\vec u (\Lambda_1 X_1^\ell + \Lambda_2 X_2^\ell + \Omega_1 Z_1^\ell + \Omega_2 Z_2^\ell + \Omega_3 Z_3^\ell))^2] \notag\\
& \quad \stackrel{\text{(a)}}{\le} 2\E[(\vec u (\Lambda_1 X_1^\ell + \Lambda_2 X_2^\ell))^2 + (\vec u (\Omega_1 Z_1^\ell + \Omega_2 Z_2^\ell + \Omega_3 Z_3^\ell))^2] \notag\\
&\quad \stackrel{\text{(b)}}{\le} 2 \ell^2 K^2 (\E[||X_1^\ell + X_2^\ell ||^2] + \E[|| Z_1^\ell +  Z_2^\ell +  Z_3^\ell ||^2])\notag\\
&\quad \stackrel{\text{(c)}}{\le} 18\ell^2 K^2 \left(\E\left[\sum_{m=1}^\ell  (X_{1,m}^2+X_{2,m}^2)\right] + \E\left[\sum_{m=1}^\ell (Z_{1,m}^2 + Z_{2,m}^2 + Z_{3,m}^2)\right]\right) \notag\\
&\quad \le   18\ell^2 K^2 \left(\E\left[\sum_{m=1}^\ell  (X_{1,m}^2+X_{2,m}^2)\right] + 3\ell\right) \notag\\
&\quad \le 54\ell^3 K^2 \left(1+ \E\left[\sum_{m=1}^\ell  (X_{1,m}^2+X_{2,m}^2)\right]/\ell \right), \label{appendixDiffBound1GenericProof}
\end{align}
where
\begin{enumerate}
\item[(a)] follows from the fact that $(a+b)^2 \le 2a^2 + 2b^2$ for all real numbers $a$ and $b$.
\item[(b)] follows from Proposition~\ref{lemmaBoundedNorm}.
\item[(c)] follows from the fact that $(a+b+c)^2 \le (3\max\{|a|,|b|,|c|\})^2 \le 9(a^2+b^2+c^2)$ for all real numbers $a$, $b$ and $c$.
\end{enumerate}
Since the differential entropy of a random variable $X$ is upper bounded by $\log_2\sqrt{2\pi e \E[X^2]}$, it follows from \eqref{appendixDiffBound1GenericProof} that  \eqref{propositionBoundGenericSt1} holds by choosing $\kappa=\log_2\sqrt{108\ell^3 K^2 \pi e}$. Following similar procedures for proving \eqref{propositionBoundGenericSt1}, we obtain \eqref{propositionBoundGenericSt2} for the same $\kappa$ chosen above.
\end{IEEEproof}
\bigskip
\begin{IEEEproof}[Proof of Lemma~\ref{lemmaDifferentialEntropyBound1}]
Let
\begin{equation}
\mathcal{G}_{ij}^{\ell\times \ell}=\left\{ h_{s_j u}h_{ud_i}\mathbf{A}   +    h_{s_j v}h_{v d_i}\mathbf{B}\right|\left. \mathbf{A}\text{ and } \mathbf{B}\text{ are in } \mathcal{U}^{\ell  \times \ell }\right\}\label{GijSet}
\end{equation}
be a finite set for each $i,j\in\{1,2\}$.
Since $\mathcal{U}$ is finite, it follows from \eqref{GijSet} that $\mathcal{G}_{12}^{\ell \times \ell}$ is finite, which then implies from Proposition~\ref{lemmaRank} that there exist two mappings denoted by $\phi_{\mathcal{G}_{12}}$ and $\psi_{\mathcal{G}_{12}}$ such that for any $\mathbf{G}_{\mathbf{12}}\in \mathcal{G}_{12}^{\ell\times \ell}$,
\begin{equation}
|\det(\phi_{\mathcal{G}_{12}}(\mathbf{G}_{\mathbf{12}}))| =1,\label{detPhi}
\end{equation}
and
\begin{equation}
\phi_{\mathcal{G}_{12}}(\mathbf{G}_{\mathbf{12}}) \mathbf{G}_{\mathbf{12}}= \left[\begin{array}{c} \psi_{\mathcal{G}_{12}}(\mathbf{G}_{\mathbf{12}}) \\ \mathbf{0}^{(\ell-\rank(\mathbf{G}_{\mathbf{12}}))\times \ell}\end{array}\right].  \label{functionsPhiPsi12}
\end{equation}
In addition, there exists by Proposition~\ref{lemmaRank} a real number $\bar K$ which is only a function of $\ell$ and $\mathcal{U}$ such that $\bar K$ is an upper bound on the magnitudes of the entries in each $\phi_{\mathcal{G}_{12}}(\mathbf{G}_{\mathbf{12}})$ and each $\psi_{\mathcal{G}_{12}}(\mathbf{G}_{\mathbf{12}})$.
For each $k \in \{1, 2, \ldots, n\}$, consider
\begin{align}
  &h(\mathbf{G}_{\mathbf{12},k}(\lambda_1 X_{1, \ell_{k-1}} +  X_{2,\ell_{k-1}}) + Z_{1, \ell_k} - \lambda_2 Z_{2, \ell_k})
\notag \\
  &\stackrel{\text{(a)}}{=}h(\phi_{\mathcal{G}_{12}}(\mathbf{G}_{\mathbf{12},k})(\mathbf{G}_{\mathbf{12},k}(\lambda_1 X_{1, \ell_{k-1}} +  X_{2,\ell_{k-1}}) + Z_{1, \ell_k} - \lambda_2 Z_{2, \ell_k})) \notag \\
  &\stackrel{\text{\eqref{functionsPhiPsi12}}}{=}h\left(\left[\begin{array}{c} \psi_{\mathcal{G}_{12}}(\mathbf{G}_{\mathbf{12},k}) \\ \mathbf{0}^{(\ell-\rank(\mathbf{G}_{\mathbf{12},k})) \times \ell} \end{array}\right](\lambda_1 X_{1, \ell_{k-1}} +  X_{2,\ell_{k-1}}) + \phi_{\mathcal{G}_{12}}(\mathbf{G}_{\mathbf{12},k})( Z_{1, \ell_k} - \lambda_2 Z_{2, \ell_k})\right), \label{thmFirstIneq12*}
\end{align}
where (a) follows from \eqref{detPhi} and the fact that $h(\mathbf{L} X^\ell)=h(X^\ell)+\log_2 |\det(\mathbf{L})|$ for any invertible matrix $\mathbf{L}$.
To facilitate discussion, let
\begin{equation}
\mathbf{I}_k^{\text{left}} =   \left[\begin{array}{cc} \mathbf{I}_{\rank(\mathbf{G}_{\mathbf{12},k})} & \mathbf{0}^{\rank(\mathbf{G}_{\mathbf{12},k}) \times (\ell-\rank(\mathbf{G}_{\mathbf{12},k}))} \end{array}\right] \label{identityLeft}
 \end{equation}
 and
 \begin{equation*}
 \mathbf{I}_k^{\text{right}} =   \left[\begin{array}{cc} \mathbf{0}^{(\ell-\rank(\mathbf{G}_{\mathbf{12},k})) \times \rank(\mathbf{G}_{\mathbf{12},k})} & \mathbf{I}^{(\ell-\rank(\mathbf{G}_{\mathbf{12},k}))} \end{array}\right] \label{identityRight}
 \end{equation*}
 such that $\left[\begin{array}{c} \mathbf{I}_k^{\text{left}} \\ \mathbf{I}_k^{\text{right}} \end{array}\right] = \mathbf{I}_{\ell}$.
Then, it follows from \eqref{thmFirstIneq12*} that for each $k\in\{1, 2, \ldots, n\}$,
\begin{align}
& h\left(\left[\begin{array}{c} \psi_{\mathcal{G}_{12}}(\mathbf{G}_{\mathbf{12},k}) \\ \mathbf{0}^{(\ell-\rank(\mathbf{G}_{\mathbf{12},k})) \times \ell} \end{array}\right](\lambda_1 X_{1, \ell_{k-1}} +  X_{2,\ell_{k-1}}) + \phi_{\mathcal{G}_{12}}(\mathbf{G}_{\mathbf{12},k})( Z_{1, \ell_k} - \lambda_2 Z_{2, \ell_k})\right)\notag \\
 & \quad\le h\left(
  \mathbf{I}_k^{\text{left}}
   \left(\left[\begin{array}{c} \psi_{\mathcal{G}_{12}}(\mathbf{G}_{\mathbf{12},k}) \\ \mathbf{0}^{ \ell \times (\ell-\rank(\mathbf{G}_{\mathbf{12},k}))  \times \ell} \end{array}\right](\lambda_1 X_{1, \ell_{k-1}} +  X_{2,\ell_{k-1}}) + \phi_{\mathcal{G}_{12}}(\mathbf{G}_{\mathbf{12},k})( Z_{1, \ell_k} - \lambda_2 Z_{2, \ell_k})\right)\right) \notag \\
   &\qquad + h\left(
  \mathbf{I}_k^{\text{right}}
  \phi_{\mathcal{G}_{12}}(\mathbf{G}_{\mathbf{12},k})( Z_{1, \ell_k} - \lambda_2 Z_{2, \ell_k})\right)  \notag \\
 &\quad \le \rank(\mathbf{G}_{\mathbf{12},k}) \log_2 \sqrt{1+P_{k-1}/\ell} +\kappa \label{thmFirstIneq**Simp}
 \end{align}
 for some $\kappa$ that does not depend on $n$ and $P$, where the last inequality follows from Proposition~\ref{differentialEntropyUpperBoundGeneric} by setting \linebreak
$\Lambda_1 = \lambda_1 \left[\begin{array}{c} \psi_{\mathcal{G}_{12}}(\mathbf{G}_{\mathbf{12},k}) \\ \mathbf{0}^{ \ell \times (\ell-\rank(\mathbf{G}_{\mathbf{12},k}))  \times \ell} \end{array}\right]$, $\Lambda_2 = \left[\begin{array}{c} \psi_{\mathcal{G}_{12}}(\mathbf{G}_{\mathbf{12},k}) \\ \mathbf{0}^{ \ell \times (\ell-\rank(\mathbf{G}_{\mathbf{12},k}))  \times \ell} \end{array}\right]$, $\Omega_1 = \phi_{\mathcal{G}_{12}}(\mathbf{G}_{\mathbf{12},k})$, $\Omega_2 = -\lambda_2 \phi_{\mathcal{G}_{12}}(\mathbf{G}_{\mathbf{12},k})$, $\Omega_3 = \mathbf{0}^{\ell \times \ell}$ and $K = \max\{\bar K, |\lambda_1|\bar K, |\lambda_2|\bar K\}$.
Following similar procedures for deriving \eqref{thmFirstIneq**Simp}, we obtain that there exists some $\kappa^\prime$ that do not depend on $n$ and $P$ such that
\begin{align*}
& h(\mathbf{G}_{\mathbf{21},k}(X_{1, \ell_{k-1}} + \mu_2X_{2,\ell_{k-1}})  + Z_{2, \ell_k}-\mu_1Z_{1, \ell_k})\notag \\
 &\quad \le \rank(\mathbf{G}_{\mathbf{21},k}) \log_2 \sqrt{1+ P_{k-1}/\ell} +\kappa^\prime
 \end{align*}
for each $k\in\{1, 2, \ldots, n\}$.
\end{IEEEproof}

\section{Proof of Lemma~\ref{lemmaDifferentialEntropyBound2}} \label{Lemma6}
We need the following proposition to prove Lemma~\ref{lemmaDifferentialEntropyBound2}.
\bigskip
\begin{Proposition}\label{lemmaRank*}
Let $\mathcal{V}$ be a finite set of $\ell \times \ell$ real matrices and let $\mathbb{M}$ denote the set of real matrices. Then, there exist three mappings $\rho_\mathcal{V}: \mathcal{V} \rightarrow \mathbb{M}$, $\sigma_\mathcal{V}: \mathcal{V} \rightarrow \mathbb{M}$ and $\tau_\mathcal{V}: \mathcal{V} \rightarrow \mathbb{M}$ such that for any $\mathbf{G}\in \mathcal{V}$, $\rho_\mathcal{V}(\mathbf{G})$ is an $(\ell-\rank(\mathbf{G}))\times \ell$ matrix, $\sigma_\mathcal{V}(\mathbf{G})$ is an $\ell\times \ell$ matrix and $\tau_\mathcal{V}(\mathbf{G})$ is an $\ell\times \ell$ matrix that satisfy
\begin{equation*}
\mathbf{I}_{\ell}= \sigma_\mathcal{V}(\mathbf{G}) \mathbf{G} + \tau_\mathcal{V}(\mathbf{G}) \left[\begin{array}{c} \rho_\mathcal{V}(\mathbf{G}) \\ \mathbf{0}^{\rank(\mathbf{G})\times \ell}\end{array}\right].
\end{equation*}
In addition, there exists a real number~$K$ which is only a function of $\ell$ and $\mathcal{V}$ such that $K$ is an upper bound on the magnitudes of the entries in each $\rho_\mathcal{V}(\mathbf{G})$, each $\sigma_\mathcal{V}(\mathbf{G})$ and each $\tau_\mathcal{V}(\mathbf{G})$.
\end{Proposition}
\begin{IEEEproof}
Suppose $\mathbf{G}$ is a matrix in $\mathcal{V}$. By linear algebra, there exists an $(\ell-\rank(\mathbf{G}))\times \ell$ matrix $\mathbf{G}^\bot$ such that the rows of $\mathbf{G}$ and $\mathbf{G}^\bot$ together span $\mathbb{R}^\ell$. In other words, there exist an $\ell\times \ell$ matrix $\mathbf{\Omega_1}$ and an $\ell \times \ell$ matrix $\mathbf{\Omega_2}$ such that $\mathbf{I}_{\ell}= \mathbf{\Omega_1} \mathbf{G} + \mathbf{\Omega_2} \left[\begin{array}{c} \mathbf{G}^\bot \\ \mathbf{0}^{\rank(\mathbf{G})\times \ell}\end{array}\right]$. The lemma then follows by letting $\rho_\mathcal{V}(\mathbf{G})=\mathbf{G}^\bot$, $\sigma_\mathcal{V}(\mathbf{G})=\mathbf{\Omega_1}$ and $\tau_\mathcal{V}(\mathbf{G})=\mathbf{\Omega_2}$ for each $\mathbf{\mathbf{G}}\in \mathcal{V}$ and letting $K$ be the maximum of $\max\{|g| : g \text{ is an entry in some }\rho_\mathcal{V}(\mathbf{G})\}$, $\max\{|g| : g \text{ is an entry in some }\sigma_\mathcal{V}(\mathbf{G})\}$ and $\max\{|g| : g \text{ is an entry in some }\tau_\mathcal{V}(\mathbf{G})\}$.
\end{IEEEproof}
\bigskip
\begin{IEEEproof}[Proof of Lemma~\ref{lemmaDifferentialEntropyBound2}]
Let $\mathcal{G}_{ij}^{\ell \times \ell}$ be the set defined in \eqref{GijSet} for each $i,j\in\{1,2\}$. Since $\mathcal{U}$ is finite, it follows that $\mathcal{G}_{ij}^{\ell \times \ell}$ is finite for all $i,j\in\{1,2\}$, which implies that there exists a real number $K^*$ which is only a function of $\mathcal{U}$ such that $K^*$ is an upper bound on the magnitudes of the entries in each $\mathbf{G}_{\mathbf{ij}}\in \mathcal{G}_{ij}^{\ell \times \ell}$ for all $i,j\in\{1,2\}$. Then,
\begin{align*}
 h(\tilde Y_{1,\ell_k}) & \stackrel{\eqref{tildeY1}}{=} h(\mathbf{G}_{\mathbf{11},k} X_{1, \ell_k} + \mathbf{G}_{\mathbf{12},k} X_{2, \ell_k} + Z_{1, \ell_k}) \notag \\
 &\le \ell  \log_2 \sqrt{1+P_{k-1}/\ell} +\kappa
\end{align*}
for some $\kappa$ that does not depend on $n$ and $P$, where the inequality follows from Proposition~\ref{differentialEntropyUpperBoundGeneric} by setting $\Lambda_1 =\mathbf{G}_{\mathbf{11},k}$,
$\Lambda_2 =\mathbf{G}_{\mathbf{12},k}$,
$\Omega_1 = \mathbf{I}_{\ell}$,
$\Omega_2 = \Omega_3 = \mathbf{0}^{\ell \times \ell}$ and
$K =\max\{K^*,1\}$.

In addition, it follows from \eqref{GijSet} that $\mathcal{G}_{12}^{\ell \times \ell}$ is finite, which then implies from Proposition~\ref{lemmaRank*} that there exist three mappings denoted by $\rho_{\mathcal{G}_{12}}$, $\sigma_{\mathcal{G}_{12}}$ and $\tau_{\mathcal{G}_{12}}$ respectively such that for any $\mathbf{G}_{\mathbf{12}}\in \mathcal{G}_{12}^{\ell\times \ell}$,
\begin{equation}
\mathbf{I}_{\ell}= \sigma_{\mathcal{G}_{12}}(\mathbf{G}_{\mathbf{12}}) \mathbf{G}_{\mathbf{12}} + \tau_{\mathcal{G}_{12}}(\mathbf{G}_{\mathbf{12}}) \left[\begin{array}{c} \rho_{\mathcal{G}_{12}}(\mathbf{G}_{\mathbf{12}}) \\ \mathbf{0}^{\rank(\mathbf{G}_{\mathbf{12}})\times \ell}\end{array}\right].  \label{functionsSigmaRho12}
\end{equation}
Let $\bar K$ be the real number in Proposition~\ref{lemmaRank*} which is only a function of $\ell$ and $\mathcal{U}$ such that $\bar K$ is an upper bound on the magnitudes of the entries in each $\rho_{\mathcal{G}_{12}}(\mathbf{G}_{\mathbf{12}})$, each $\sigma_{\mathcal{G}_{12}}(\mathbf{G}_{\mathbf{12}})$ and each $\tau_{\mathcal{G}_{12}}(\mathbf{G}_{\mathbf{12}})$.
Let $\bar Z_2^{\ell n}$ be $\ell n$ copies of $\mathcal{N}(0,1)$ such that $\bar Z_2^{\ell n}$, $X_1^{\ell n}$, $X_2^{\ell n}$, $Z_1^{\ell n}$ and $Z_2^{\ell n}$ are independent, and let
\begin{equation}
\vec Y_{2,k}^\prime = \left[\begin{array}{c} \rho_{\mathcal{G}_{12}} (\mathbf{G}_{\mathbf{12},k})  \\ \mathbf{0}^{\rank(\mathbf{G}_{\mathbf{12},k}) \times \ell} \end{array}\right] X_{2, \ell_{k-1}} +  \bar Z_{2,\ell_k}. \label{Y2Prime}
\end{equation}
Using \eqref{functionsSigmaRho12} and \eqref{Y2Prime}, we obtain
\begin{align}
& \sigma_{\mathcal{G}_{12}} (\mathbf{G}_{\mathbf{12},k})(\mathbf{G}_{\mathbf{12},k} X_{2, \ell_{k-1}}+Z_{1, \ell_k}) + \tau_{\mathcal{G}_{12}} (\mathbf{G}_{\mathbf{12},k}) \vec Y_{2,k}^\prime \, \notag \\
&\quad =X_{2, \ell_{k-1}} + \sigma_{\mathcal{G}_{12}}(\mathbf{G}_{\mathbf{12},k})Z_{1, \ell_k}  + \tau_{\mathcal{G}_{12}} (\mathbf{G}_{\mathbf{12},k}) \bar Z_{2, \ell_k}. \label{lemmaDifferentialEntropyBound2ProofTemp}
\end{align}
Then, for each $k\in\{1, 2, \ldots, n\}$,
\begin{align*}
& h(\mathbf{G}_{\mathbf{22},k}X_{2, \ell_{k-1}} + Z_{2, \ell_k}|\mathbf{G}_{\mathbf{12},k} X_{2, \ell_{k-1}}+Z_{1, \ell_k}) \notag \\
 &\quad = h(\mathbf{G}_{\mathbf{22},k}X_{2, \ell_{k-1}} + Z_{2, \ell_k}|\mathbf{G}_{\mathbf{12},k} X_{2, \ell_{k-1}}+Z_{1, \ell_k}, \vec Y_{2,k}^\prime)   + I(\mathbf{G}_{\mathbf{22},k}X_{2, \ell_{k-1}} + Z_{2, \ell_k};\vec Y_{2,k}^\prime|\mathbf{G}_{\mathbf{12},k} X_{2, \ell_{k-1}}+Z_{1, \ell_k}) \notag \\
   &\quad \le h(\mathbf{G}_{\mathbf{22},k}X_{2, \ell_{k-1}} + Z_{2, \ell_k}|\mathbf{G}_{\mathbf{12},k} X_{2, \ell_{k-1}}+Z_{1, \ell_k}, \vec Y_{2,k}^\prime)  + h(\vec Y_{2,k}^\prime) -h(\vec Y_{2,k}^\prime| X_{2, \ell_{k-1}},Z_{1,\ell_k}, Z_{2,\ell_k} ) \notag \\
 &\quad \stackrel{\text{(a)}}{=} h(\mathbf{G}_{\mathbf{22},k}X_{2, \ell_{k-1}} + Z_{2, \ell_k}|\mathbf{G}_{\mathbf{12},k} X_{2, \ell_{k-1}}+Z_{1, \ell_k}, \vec Y_{2,k}^\prime)  + h(\vec Y_{2,k}^\prime) -h(\bar Z_{2, \ell_k}) \notag \\
 &\quad \stackrel{\text{(b)}}{\le} h(\vec Y_{2,k}^\prime) + h(\mathbf{G}_{\mathbf{22},k}X_{2, \ell_{k-1}} + Z_{2, \ell_k}|\mathbf{G}_{\mathbf{12},k} X_{2, \ell_{k-1}}+Z_{1, \ell_k}, \vec Y_{2,k}^\prime)  \notag \\
 &\quad \le  h(\vec Y_{2,k}^\prime) + h(\mathbf{G}_{\mathbf{22},k}X_{2, \ell_{k-1}} + Z_{2, \ell_k}
-\mathbf{G}_{\mathbf{22},k} (\sigma_{\mathcal{G}_{12}}( \mathbf{G}_{\mathbf{12},k}) (\mathbf{G}_{\mathbf{12},k} X_{2,\ell_{k-1}} + Z_{1,\ell_k}) + \tau_{\mathcal{G}_{12}}(\mathbf{G}_{\mathbf{12},k})  \vec{Y}_{2,k}^\prime)) \notag \\
 &\quad \stackrel{\eqref{lemmaDifferentialEntropyBound2ProofTemp}}{\le} h(\vec Y_{2,k}^\prime) + h(Z_{2, \ell_k} - \mathbf{G}_{\mathbf{22},k}\,\sigma_{\mathcal{G}_{12}}(\mathbf{G}_{\mathbf{12},k})Z_{1, \ell_k}  - \mathbf{G}_{\mathbf{22},k}\,\tau_{\mathcal{G}_{12}} (\mathbf{G}_{\mathbf{12},k}) \bar Z_{2, \ell_k}),
\end{align*}
where
\begin{enumerate}
\item[(a)] follows from \eqref{Y2Prime} and the fact that $\bar Z_2^{\ell n}$, $X_2^{\ell n}$, $Z_1^{\ell n}$ and $Z_2^{\ell n}$ are independent.
\item[(b)] follows from the fact that $\{\bar Z_{2,m}\}_{m=1}^{\ell n}$ are independent and the differential entropy of $\mathcal{N}(0,1)$ is positive.
\end{enumerate}
For each $k\in\{1, 2, \ldots, n\}$,
\begin{align*}
 h(\vec Y_{2,k}^\prime) & \stackrel{\eqref{Y2Prime}}{=} h\left(\left[\begin{array}{c} \rho_{\mathcal{G}_{12}} (\mathbf{G}_{\mathbf{12},k})  \\ \mathbf{0}^{\rank(\mathbf{G}_{\mathbf{12},k}) \times \ell} \end{array}\right] X_{2, \ell_{k-1}} +  \bar Z_{2,\ell_k}\right) \notag \\
& \stackrel{\text{(a)}}{\le} h\left( \mathbf{I}_k^{\text{left}}\left( \left[\begin{array}{c} \rho_{\mathcal{G}_{12}} (\mathbf{G}_{\mathbf{12},k})  \\ \mathbf{0}^{\rank(\mathbf{G}_{\mathbf{12},k}) \times \ell} \end{array}\right] X_{2, \ell_{k-1}} +  \bar Z_{2,\ell_k}\right)\right)  + \sum_{m=\ell-\rank(\mathbf{G}_{\mathbf{12}})+1}^{\ell} h(\bar Z_{2, \ell(k-1)+m}) \notag \\
&\stackrel{\text{(b)}}{\le} (\ell-\rank(\mathbf{G}_{\mathbf{12},k})) \log_2 \sqrt{1+P_{k-1}/\ell} +\kappa
\end{align*}
and
\begin{align*}
& h(Z_{2, \ell_k} - \mathbf{G}_{\mathbf{22},k}\,\sigma_{\mathcal{G}_{12}}(\mathbf{G}_{\mathbf{12},k})Z_{1, \ell_k}  - \mathbf{G}_{\mathbf{22},k}\,\tau_{\mathcal{G}_{12}} (\mathbf{G}_{\mathbf{12},k}) \bar Z_{2, \ell_k}) \notag \\
 &\stackrel{\text{(c)}}{\le} \ell  \log_2 \sqrt{1+P_{k-1}/\ell} +\kappa^\prime
\end{align*}
for some $\kappa$ and $\kappa^\prime$ that do not depend on $n$ and $P$, where
\begin{enumerate}
\item[(a)] follows from defining $\mathbf{I}_k^{\text{left}}=   \left[\begin{array}{cc} \mathbf{I}_{\ell -\rank(\mathbf{G}_{\mathbf{12},k})} & \mathbf{0}^{(\ell -\rank(\mathbf{G}_{\mathbf{12},k})) \times \rank(\mathbf{G}_{\mathbf{12},k})} \end{array}\right]$.
\item[(b)] follows from Proposition~\ref{differentialEntropyUpperBoundGeneric} by setting
$\Lambda_2 =\left[\begin{array}{c} \rho_{\mathcal{G}_{12}} (\mathbf{G}_{\mathbf{12},k})  \\ \mathbf{0}^{\rank(\mathbf{G}_{\mathbf{12},k}) \times \ell} \end{array}\right]$,
$\Omega_1 = \mathbf{I}_{\ell}$,
$\Lambda_1=\Omega_2 = \Omega_3 = \mathbf{0}^{\ell \times \ell}$ and
$K =\max\{\bar K,1\}$.
\item[(c)] follows from Proposition~\ref{differentialEntropyUpperBoundGeneric} by setting
$\Omega_1 = \mathbf{I}_{\ell}$,
$\Omega_2 =  - \mathbf{G}_{\mathbf{22},k}\,\sigma_{\mathcal{G}_{12}}(\mathbf{G}_{\mathbf{12},k})$,
$\Omega_3 = \mathbf{G}_{\mathbf{22},k}\,\tau_{\mathcal{G}_{12}} (\mathbf{G}_{\mathbf{12},k})$ and
$K =\max\{K^* \bar K ,1\}$.
\end{enumerate}
\end{IEEEproof} 

\section{Proofs of Lemmas~\ref{lemmaS} and~\ref{lemmaZ} } \label{LemmaS}
The proofs of lemmas~\ref{lemmaS} and~\ref{lemmaZ} are similar. Therefore we present the proof of Lemma~\ref{lemmaS}, and omit that of Lemma~\ref{lemmaZ}.
We will assume that all channel submatrices are invertible since this is true for almost all values of channel gains. So, let
\begin{equation*}
\mathbf{A^S} = - H_{ud_1}^{-1}H_{vd_1} \mathbf{B^S} H_{s_2v}H_{s_2u}^{-1}.
\end{equation*}
Then we get
\begin{align*}
\mathbf{G}_{\mathbf{12}}^{\mathbf{(A,B)}} & = \mathbf{0}^{M\times M}, \\
\mathbf{G}_{\mathbf{11}}^{\mathbf{(A,B)}} & =H_{vd_1} \mathbf{B^S} ( H_{s_1v} - H_{s_2v}H_{s_2u}^{-1} H_{s_1u}), \\
\mathbf{G}_{\mathbf{22}}^{\mathbf{(A,B)}} & = (-H_{ud_2} H_{ud_1}^{-1}H_{vd_1} + H_{vd_2}) \mathbf{B^S}  H_{s_2v}, \\
\mathbf{G}_{\mathbf{21}}^{\mathbf{(A,B)}} & = (-H_{ud_2}H_{ud_1}^{-1}H_{vd_1} \mathbf{B^S} H_{s_2v}H_{s_2u}^{-1}H_{s_1u} + H_{vd_2} \mathbf{B^S} H_{s_1v}).
\end{align*}

\noindent Since $(H_{s_1v} - H_{s_2v}H_{s_2u}^{-1} H_{s_1u})$ and $(-H_{ud_2} H_{ud_1}^{-1}H_{vd_1} + H_{vd_2})$ are invertible for almost all values of channel gains, then it sufficient (and necessary) to choose an invertible $\mathbf{B^S}$ to get $\text{rank}(\mathbf{G}_{\mathbf{11}}^{\mathbf{(A,B)}}) =\text{rank}(\mathbf{G}_{\mathbf{22}}^{\mathbf{(A,B)}})=M$.
So the problem is reduced to finding an invertible $\mathbf{B^S}$ such that
\begin{equation*} \label{tosolve}
-H_{ud_2}H_{ud_1}^{-1}H_{vd_1} \mathbf{B^S} H_{s_2v}H_{s_2u}^{-1}H_{s_1u} + H_{vd_2} \mathbf{B^S} H_{s_1v} = \begin{bmatrix}
  \mathbf{0}^{M \times M-1} & c
 \end{bmatrix}.
\end{equation*}
We can rewrite the equation as
\begin{equation} \label{Sylvester}
 \mathbf{B^S} \underbrace{H_{s_1v}H_{s_1u}^{-1} H_{s_2u}H_{s_2v}^{-1}}_{\triangleq A} -\underbrace{H_{vd_2}^{-1}H_{ud_2}H_{ud_1}^{-1}H_{vd_1}}_{\triangleq B} \mathbf{B^S}  = \underbrace{H_{vd_2}^{-1} \begin{bmatrix}
 \mathbf{0}^{M \times M-1} & c
 \end{bmatrix} H_{s_1u}^{-1} H_{s_2u}H_{s_2v}^{-1}}_{\triangleq C}.
\end{equation}
If we denote $A=H_{s_1v}H_{s_1u}^{-1} H_{s_2u}H_{s_2v}^{-1}$, $B=H_{vd_2}^{-1}H_{ud_2}H_{ud_1}^{-1}H_{vd_1}$, and $C=H_{vd_2}^{-1} \begin{bmatrix}
 \mathbf{0}^{M \times M-1} & c
 \end{bmatrix} H_{s_1u}^{-1} H_{s_2u}H_{s_2v}^{-1}$, then $\mathbf{B^S}$ will be the solution to $X$ in the following equation, which is known as the Sylvester equation.
 \begin{equation*}
 XA-BX=C. \label{Sylvestershort}
 \end{equation*}
Since $c=\begin{bmatrix}
1 & 0^{1 \times M-1}
\end{bmatrix}^T$, then $C=qp^T$, where $q$ denotes the first column vector of $H_{vd_2}^{-1}$, and $p^T$ denotes the last row vector of $H_{s_1u}^{-1} H_{s_2u}H_{s_2v}^{-1}$. Now consider the following proposition (proof found in \cite{sylvesterEquation} and \cite{Luenberger}):
\begin{Proposition} \label{InvertibleSylvestre}
Let $A$, $B$, and $C$ be three $M \times M$ matrices. Suppose the eigenvalues of $A$ are distinct from the eigenvalues of $B$. Then, the equation $XA-BX = C$ has a unique solution. In addition, if $C=qp^T$ for some column vectors $q$ and $p$, then the unique solution is invertible if and only if
\begin{enumerate}
\item the (row vectors) $p^T, p^T A, \dots, p^T A^{M-1} $ are linearly independent, and
\item the (column vectors) $q, Bq, \dots, B^{M-1} q$ are linearly independent. \\
\end{enumerate}
\end{Proposition}

\noindent Note that  $p^T, p^T A, \dots, p^T A^{M-1} $ are linearly independent $\iff \text{rank} \left(\begin{bmatrix} p & A^Tp & \dots & (A^T)^{M-1}p \end{bmatrix} \right) = M$.
 So it remains to show that the following conditions hold for almost all values of channel gains:
\begin{enumerate}
\item[] (C-1) The eigenvalues of $A$ are distinct from the eigenvalues of $B$. \vspace{0.04 in}
\item[] (C-2) $\text{rank}\left( \begin{bmatrix}
p & A^Tp & \dots & (A^T)^{M-1}p
\end{bmatrix} \right)=M$. \vspace{0.04 in}
\item[] (C-3) $\text{rank}\left( \begin{bmatrix}
q & Bq & \dots & B^{M-1}q
\end{bmatrix} \right)=M$.\\
\end{enumerate}
Thus, the proof of the lemma is concluded with the following proposition, the proof of which is given in Appendix~\ref{proplemmaS}.
\bigskip
\begin{Proposition} \label{conditionsprop}
Conditions (C-1), (C-2), and (C-3) hold for almost all values of channel gains.
\end{Proposition}

\section{Proof of Proposition~\ref{conditionsprop}} \label{proplemmaS}
First, consider the following proposition.\\
\begin{Proposition} \label{eigenvalue}
Let $\lambda \in \mathbb{C}$ be given, and let $\mathbb{M}$ be the set of $M \times M$ real matrices. \\
Let $\mathbb{M}_\lambda = \{ \mathbf{L} \in \mathbb{M}$ s.t. $\lambda$ is an eigenvalue of $\mathbf{L} \}$. Then $\mathbb{M}_\lambda$ has Lebesgue measure zero.\\
\end{Proposition}

\begin{IEEEproof}
Write $\mathbf{L} = \begin{bmatrix}
\ell_{11} & \cdots & \ell _{1M} \\
\vdots & \ddots & \vdots\\
\ell_{M1} & \cdots & \ell_{MM}
\end{bmatrix}$. Since
$\mathbf{L} \in \mathbb{M}_\lambda \iff \lambda$ is an eigenvalue of $\mathbf{L} \iff \det(\lambda \mathbf{I}_M - \mathbf{L})=0$, $\det(\lambda \mathbf{I}_M - \mathbf{L} )$ is a non-zero multivariate polynomial in the variables $(\ell_{11}, \ldots, \ell_{MM})$, which then implies that the set of roots has Lebesgue measure zero (see~\cite{Federer} for proof). Therefore
$\mathbb{M}_\lambda$ has Lebesgue measure zero.
\end{IEEEproof}
\bigskip

Now, consider condition (C-1): The eigenvalues of $H_{s_1v}H_{s_1u}^{-1} H_{s_2u}H_{s_2v}^{-1}$ are distinct from the eigenvalues of $H_{vd_2}^{-1}H_{ud_2}H_{ud_1}^{-1}H_{vd_1}$.\\
Fix $H_{ud_1}$, $H_{vd_1}$, $H_{vd_2}$, and $H_{ud_2}$. This gives fixed eigenvalues for $H_{vd_2}^{-1}H_{ud_2}H_{ud_1}^{-1}H_{vd_1}$; call them $\lambda_1$, $\lambda_2, \dots, \lambda_M$. Also fix $H_{s_2u}$, $H_{s_2v}$ and $H_{s_1u}$. Assume they are fixed to invertible matrices (we can make this assumption since it's true for almost all values of channel gains). Now, define the set $S_{\text{(C-1)}}$ as
\begin{equation*}
S_{\text{(C-1)}} =\{ A \in \mathbb{M} ~s.t.~ A H_{s_1u}^{-1} H_{s_2u}H_{s_2v}^{-1} \text{ and } H_{vd_2}^{-1}H_{ud_2}H_{ud_1}^{-1}H_{vd_1} \text{ have at least one common eigenvalue} \},
\end{equation*}
where $\mathbb{M}$ is the set of $l \times l$ real matrices.  Then, we get
\begin{align*}
A \in S_{\text{(C-1)}} & \iff AH_{s_1u}^{-1} H_{s_2u}H_{s_2v}^{-1} \in \bigcup_{i=1}^l \mathbb{M}_{\lambda_i} \\
& \iff A \in \bigcup_{i=1}^l \mathbb{M}_{\lambda_i} H_{s_2v}H_{s_2u}^{-1} H_{s_1u},
\end{align*}
where, for $i \in \{1,\dots,l\}$, $\mathbb{M}_{\lambda_i}$ is as defined in Proposition~\ref{eigenvalue}, and $\mathbb{M}_{\lambda_i} H_{s_2v}H_{s_2u}^{-1} H_{s_1u}$ is defined as
\begin{equation*}
\mathbb{M}_{\lambda_i} H_{s_2v}H_{s_2u}^{-1} H_{s_1u} = \{ B H_{s_2v}H_{s_2u}^{-1} H_{s_1u} \text{ where } B \in \mathbb{M}_{\lambda_i} \}.
\end{equation*}
We can easily see that $\mathbb{M}_{\lambda_i}$ and $\mathbb{M}_{\lambda_i} H_{s_2v}H_{s_2u}^{-1} H_{s_1u}$ have the same cardinality, which yields that $\mathbb{M}_{\lambda_i} H_{s_2v}H_{s_2u}^{-1} H_{s_1u}$ has measure zero by Proposition~\ref{eigenvalue}. Therefore $S_{\text{(C-1)}}$ has measure zero. Finally, $H_{s_1v}H_{s_1u}^{-1} H_{s_2u}H_{s_2v}^{-1}$ and $H_{s_1v}H_{s_1u}^{-1} H_{s_2u}H_{s_2v}^{-1}$ have common eigenvalues only if $H_{s_1v} \in S_{\text{(C-1)}}$. Thus, (C-1) holds for almost all values of channel gains.

 It remains to prove that conditions (C-2) and (C-3) hold for almost all values of channel gains. The proofs of (C-2) and (C-3) are similar, so we will focus on (C-2) only.
First, note that $\det\left( \begin{bmatrix}
p & A^Tp & \dots & (A^T)^{M-1}p
\end{bmatrix} \right)$ is a ratio of polynomials in the channel gains. Since the roots of any non-identically zero multivariate polynomial have Lebesgue measure zero~\cite{Federer}, it suffices to show that the numerator and denominator are not identically zero. For that end, it suffices to find \emph{one} realization of channel gains such that $\det\left( \begin{bmatrix}
p & A^Tp & \dots & (A^T)^{M-1}p
\end{bmatrix} \right) \neq 0$ and $\det\left( \begin{bmatrix}
q & Bq & \dots & B^{M-1}q
\end{bmatrix} \right)\neq 0$ in order to prove our claim.
So, let
\begin{equation*}
H_{s_1u}=H_{s_2u}=H_{s_2v}=H_{vd_2}=H_{ud_2}=H_{ud_1}=\mathbf{I}_M,
\end{equation*}
and let
\begin{equation*}
H_{s_1v}=H_{vd_1}=
\begin{bmatrix}
0 & 1 & 0 & \cdots & \cdots & 0\\
 0 & 0 & 1 &0 & \cdots & 0 \\
 \vdots & & \ddots & \ddots & & \vdots\\
 0& \cdots &\cdots& 0& 1 & 0\\
 0 & \cdots & \cdots & \cdots & 0 &  1 \\
  -1 & 0 & \cdots & \cdots & \cdots & 0
\end{bmatrix}
\end{equation*}
be a matrix with non-zero entries at only the upper diagonal and the bottom left corner.
Let $\boldsymbol{\Pi}$ denote the above matrix. Then we get $A=B=\boldsymbol{\Pi}$, $p = [\ 0 \ \ldots \ 0 \ 1\ ]^T$, and $q=[\ 1 \ 0 \ \ldots \ 0\ ]^T$. We get
\begin{equation} \label{ATpControllable}
\begin{bmatrix}
p & A^Tp & \dots & (A^T)^{M-1}p
\end{bmatrix} =
\begin{bmatrix}
0 & -1 & 0 & \cdots & \cdots & 0\\
 0 & 0 & -1 &0 & \cdots & 0 \\
 \vdots & & \ddots & \ddots & & \vdots\\
 0& \cdots &\cdots& 0& -1 & 0\\
 0 & \cdots & \cdots & \cdots & 0 &  -1 \\
  1 & 0 & \cdots & \cdots & \cdots & 0
\end{bmatrix},
\end{equation}
and
\begin{equation} \label{BqControllable}
\begin{bmatrix}
q & Bq & \dots & B^{M-1}q
\end{bmatrix} =
\begin{bmatrix}
1 & 0 & \cdots & \cdots & \cdots & 0\\
0 & \cdots & \cdots & \cdots & 0 & -1\\
 0 & \cdots & \cdots &  0 & -1 & 0 \\
 \vdots & & \iddots & \iddots & & \vdots\\
 0& 0 & -1 & 0 & \cdots & 0 \\
 0& -1 &0 & \cdots & \cdots & 0 
\end{bmatrix}.
\end{equation}
We can easily see that the above two matrices are invertible, and thus (C-2) and (C-3) hold for almost all values of channel gains. 

\section{Proof of Lemma~\ref{lemmaLinearDecompositionMIMO}} \label{proofOfLemmaLinearDecompositionMIMO}
Recall
\begin{equation}
\mathbf{G_{ij}^{(\mathbf{A},\mathbf{B})}}= H_{ud_i} \mathbf{A} H_{s_j u}  +   H_{vd_i}\mathbf{B} H_{s_j v} \label{GijMIMOProof}
\end{equation}
for each $(\mathbf{A},\mathbf{B})\in \mathcal{U}^{M\times M} \times \mathcal{U}^{M\times M}$. We need the following proposition to prove Lemma~\ref{lemmaLinearDecompositionMIMO}.
\bigskip
\begin{Proposition} \label{propositionNonzeroMIMO}
If at least one of $\mathbf{A}$ and $\mathbf{B}$ is not the zero matrix, then at least one of $\mathbf{G_{12}^{(\mathbf{A},\mathbf{B})}}$ and $\mathbf{G_{21}^{(\mathbf{A},\mathbf{B})}}$ is not the zero matrix.
\end{Proposition}
\begin{IEEEproof}
Assume the contrary holds, i.e.,
\begin{equation}
\mathbf{G_{12}^{(\mathbf{A},\mathbf{B})}}=\mathbf{G_{21}^{(\mathbf{A},\mathbf{B})}} = \mathbf{0}^{M\times M}. \label{assumptionZero}
\end{equation}
We will show that \eqref{assumptionZero} implies $\mathbf{A}=\mathbf{B}=\mathbf{0}^{M\times M}$.
Using \eqref{assumptionZero} and \eqref{GijMIMOProof}, we obtain
\begin{equation} \label{AfuncB}
-\mathbf{A} =H_{ud_1}^{-1} H_{vd_1} \mathbf{B} H_{s_2v}H_{s_2u}^{-1} = H_{ud_2}^{-1}H_{vd_2}\mathbf{B} H_{s_1 v}H_{s_1 u}^{-1},
\end{equation}
which then implies that
\begin{equation}
\mathbf{B} H_{s_1 v}H_{s_1 u}^{-1}H_{s_2u}H_{s_2v}^{-1} -H_{vd_2}^{-1}H_{ud_2}H_{ud_1}^{-1} H_{vd_1} \mathbf{B} =0. \label{sylvesterEquation*}
\end{equation}
Since $H_{s_1 v}H_{s_1 u}^{-1}H_{s_2u}H_{s_2v}^{-1}$ and $H_{vd_2}^{-1}H_{ud_2}H_{ud_1}^{-1} H_{vd_1}$ do not have a common eigenvalue by Condition~(C-1) (cf.\ Proposition~\ref{conditionsprop} in Appendix~\ref{LemmaS}), it follows from \eqref{sylvesterEquation*} and Proposition~\ref{InvertibleSylvestre} that $\mathbf{B}=\mathbf{0}^{M\times M}$, which then implies from \eqref{AfuncB} that $\mathbf{A}=\mathbf{0}^{M\times M}$.
\end{IEEEproof}
\bigskip
\begin{IEEEproof}[Proof of Lemma~\ref{lemmaLinearDecompositionMIMO}]
Since $\mathcal{U}$ is finite, it suffices to show that for each $(\mathbf{A},\mathbf{B})\in \mathcal{U}^{M\times M} \times \mathcal{U}^{M\times M}$, there exist six matrices in $\mathbb{R}^{M\times M}$, denoted by $\boldsymbol{\Lambda_{\mathbf{1}}^{(\mathbf{A},\mathbf{B})}}$, $\boldsymbol{\Lambda}_{\mathbf{2}}^{(\mathbf{A},\mathbf{B})}$, $\boldsymbol{\Lambda}_{\mathbf{3}}^{(\mathbf{A},\mathbf{B})}$, $\boldsymbol{\Omega}_{\mathbf{1}}^{(\mathbf{A},\mathbf{B})}$, $\boldsymbol{\Omega}_{\mathbf{2}}^{(\mathbf{A},\mathbf{B})}$ and $\boldsymbol{\Omega}_{\mathbf{3}}^{(\mathbf{A},\mathbf{B})}$ respectively, and two matrices in $\mathbb{R}^{(M-1)\times M}$, denoted by $\boldsymbol{\Gamma}_{\mathbf{1}}^{(\mathbf{A},\mathbf{B})}$ and $\boldsymbol{\Gamma}_{\mathbf{2}}^{(\mathbf{A},\mathbf{B})}$ respectively, such that
\begin{equation}
 \mathbf{G}_{\mathbf{11}}^{(\mathbf{A},\mathbf{B})} = \mathbf{G}_{\mathbf{12}}^{(\mathbf{A},\mathbf{B})}\boldsymbol{\Lambda}_{\mathbf{1}}^{(\mathbf{A},\mathbf{B})} + \boldsymbol{\Lambda}_{\mathbf{2}}^{(\mathbf{A},\mathbf{B})}\mathbf{G}_{\mathbf{21}}^{(\mathbf{A},\mathbf{B})} + \boldsymbol{\Lambda}_{\mathbf{3}}^{(\mathbf{A},\mathbf{B})}\left[ \begin{array}{c} \boldsymbol{\Gamma}_{\mathbf{1}}^{(\mathbf{A},\mathbf{B})} \\ \mathbf{0}^{1\times M}\end{array}\right]\label{G11ThmStatementSumRankMIMOProof}
 \end{equation}
 and
\begin{equation}
\mathbf{G}_{\mathbf{22}}^{(\mathbf{A},\mathbf{B})} = \boldsymbol{\Omega}_{\mathbf{1}}^{(\mathbf{A},\mathbf{B})}\mathbf{G}_{\mathbf{12}}^{(\mathbf{A},\mathbf{B})} + \mathbf{G}_{\mathbf{21}}^{(\mathbf{A},\mathbf{B})}\boldsymbol{\Omega}_{\mathbf{2}} + \boldsymbol{\Omega}_{\mathbf{3}}^{(\mathbf{A},\mathbf{B})} \left[\begin{array}{c} \boldsymbol{\Gamma}_{\mathbf{2}}^{(\mathbf{A},\mathbf{B})}  \\ \mathbf{0}^{1\times M}\end{array}\right]. \label{G22ThmStatementSumRankMIMOProof}
 \end{equation}
 The lemma will then follow from \eqref{GijMIMOProof} by letting
 \[
 K_{M, \mathcal{U}}= \max_{(\mathbf{A},\mathbf{B})\in \mathcal{U}^{M\times M} \times \mathcal{U}^{M\times M}}\left\{ |a| \left|\: \parbox[c]{2.8 in}{
 $a$ is an entry of $\boldsymbol{\Lambda_1^{(\mathbf{A},\mathbf{B})}}$, $\boldsymbol{\Lambda_2^{(\mathbf{A},\mathbf{B})}}$, $\boldsymbol{\Lambda_3^{(\mathbf{A},\mathbf{B})}}$, $\boldsymbol{\Omega_1^{(\mathbf{A},\mathbf{B})}}$, $\boldsymbol{\Omega_2^{(\mathbf{A},\mathbf{B})}}$, $\boldsymbol{\Omega_3^{(\mathbf{A},\mathbf{B})}}$, $\boldsymbol{\Gamma_{\mathbf{1}}^{(\mathbf{A},\mathbf{B})}}$ or $\boldsymbol{\Gamma_{\mathbf{2}}^{(\mathbf{A},\mathbf{B})}}$}\right\}\right. .
 \]
If $\mathbf{A}=\mathbf{B}=\mathbf{0}^{M\times M}$, then \eqref{G11ThmStatementSumRankMIMOProof} and \eqref{G22ThmStatementSumRankMIMOProof} follow trivially from \eqref{GijMIMOProof}. Therefore, we assume in the rest of the proof that at least one of $\mathbf{A}$ and $\mathbf{B}$ is not the zero matrix, which implies from Proposition~\ref{propositionNonzeroMIMO} that
\begin{equation}
\rank(\mathbf{G_{12}^{(\mathbf{A},\mathbf{B})}})+ \rank(\mathbf{G_{21}^{(\mathbf{A},\mathbf{B})}}) \ge 1. \label{sumRankAtLeastOne}
\end{equation}
Consider the following two cases: \vspace{0.04 in} \\
\textbf{Case $\rank(\mathbf{G_{12}^{(\mathbf{A},\mathbf{B})}})\ge 1$}:  \vspace{0.04 in} \\
By linear algebra, there exist two matrices denoted by $\mathbf{\tilde L}$ and $\mathbf{\tilde L^*}$ such that $\rank(\mathbf{\tilde L^*})\le M-1$ and
\[
\mathbf{G_{11}^{(\mathbf{A},\mathbf{B})}} = \mathbf{G_{12}^{(\mathbf{A},\mathbf{B})}} \mathbf{\tilde L} + \mathbf{\tilde L^*}.
\]
\textbf{Case }otherwise:  \vspace{0.04 in} \\
It follows from \eqref{sumRankAtLeastOne} that $\rank(\mathbf{G_{21}^{(\mathbf{A},\mathbf{B})}})\ge 1$. Then, there exist by linear algebra two matrices denoted by $\mathbf{\bar{L}}$ and $\mathbf{\bar L^*}$ respectively such that $\rank(\mathbf{\bar L^*})\le M-1$ and
\[
\mathbf{G_{11}^{(\mathbf{A},\mathbf{B})}} = \mathbf{\bar L} \mathbf{G_{21}^{(\mathbf{A},\mathbf{B})}}  + \mathbf{\bar L^*}.
\]
\vspace{0.04 in} \\
Combining the two cases, there exist three matrices denoted by $\mathbf{L_1}$, $\mathbf{L_2}$ and $\mathbf{L_3}$ respectively such that $\rank(\mathbf{L_3})\le M-1$ and
\[
\mathbf{G_{11}^{(\mathbf{A},\mathbf{B})}} = \mathbf{G_{12}^{(\mathbf{A},\mathbf{B})}} \mathbf{L_1}+\mathbf{L_2} \mathbf{G_{21}^{(\mathbf{A},\mathbf{B})}}  + \mathbf{L_3},
\]
which then implies \eqref{G11ThmStatementSumRankMIMOProof}. Similarly, there exist three matrices denoted by $\mathbf{\hat L_1}$, $\mathbf{\hat L_2}$ and $\mathbf{\hat L_3}$ respectively such that $\rank(\mathbf{L_3})\le M-1$ and
\[
\mathbf{G_{22}^{(\mathbf{A},\mathbf{B})}} = \mathbf{\hat L_1}\mathbf{G_{12}^{(\mathbf{A},\mathbf{B})}} + \mathbf{G_{21}^{(\mathbf{A},\mathbf{B})}}\mathbf{\hat L_2}  + \mathbf{\hat L_3},
\]
which then implies \eqref{G22ThmStatementSumRankMIMOProof}.
\end{IEEEproof}

\section{Proof of Lemma~\ref{lemmaDifferentialEntropyBound1MIMO}} \label{appendixLemmaDifferentialEntropyBound1}
Let
\begin{equation}
\mathcal{G}_{ij}^{M\times M}=\left\{ H_{ud_i}\mathbf{A}H_{s_j u}   +    H_{v d_i}\mathbf{B}H_{s_j v}\right|\left. \mathbf{A}\text{ and } \mathbf{B}\text{ are in } \mathcal{U}^{M  \times M }\right\}\label{GijSetMIMO}
\end{equation}
be a finite set for each $i,j\in\{1,2\}$. Since $\mathcal{U}$ is finite, it follows from \eqref{GijSetMIMO} that $\mathcal{G}_{12}^{M \times M}$ is finite, which then implies from Proposition~\ref{lemmaRank} that there exist two mappings denoted by $\phi_{\mathcal{G}_{12}}$ and $\psi_{\mathcal{G}_{12}}$ respectively such that for any $\mathbf{G}_{\mathbf{12}}\in \mathcal{G}_{12}^{M\times M}$,
\begin{equation}
|\det(\phi_{\mathcal{G}_{12}}(\mathbf{G}_{\mathbf{12}}))|=1\label{detPhiMIMO}
\end{equation}
and
\begin{equation}
\phi_{\mathcal{G}_{12}}(\mathbf{G}_{\mathbf{12}}) \mathbf{G}_{\mathbf{12}}= \left[\begin{array}{c} \psi_{\mathcal{G}_{12}}(\mathbf{G}_{\mathbf{12}}) \\ \mathbf{0}^{(M-\rank(\mathbf{G}_{\mathbf{12}}))\times M}\end{array}\right].  \label{functionsPhiPsi12MIMO}
\end{equation}
In addition, there exists by Proposition~\ref{lemmaRank} a real number $\bar K$ which is only a function of $M$ and $\mathcal{U}$ such that $\bar K$ is an upper bound on the magnitudes of the entries in each $\phi_{\mathcal{G}_{12}}(\mathbf{G}_{\mathbf{12}})$ and each $\psi_{\mathcal{G}_{12}}(\mathbf{G}_{\mathbf{12}})$.
For each $k\in\{1, 2, \ldots, n\}$, consider
\begin{align}
  &h(\mathbf{G}_{\mathbf{12},k}(\boldsymbol{\Lambda}_{\mathbf{1},k} X_{1, M_{k-1}} +  X_{2,M_{k-1}}) + Z_{1, M_k}  - \boldsymbol{\Lambda}_{\mathbf{3},k}\hat Z_{1, M_k} - \boldsymbol{\Lambda}_{\mathbf{2},k} Z_{2, M_k})
\notag \\
  &\stackrel{\text{(a)}}{=}h(\phi_{\mathcal{G}_{12}}(\mathbf{G}_{\mathbf{12},k})(\mathbf{G}_{\mathbf{12},k}(\boldsymbol{\Lambda}_{\mathbf{1},k} X_{1, M_{k-1}} +  X_{2,M_{k-1}}) + Z_{1, M_k} - \boldsymbol{\Lambda}_{\mathbf{3},k}\hat Z_{1, M_k} - \boldsymbol{\Lambda}_{\mathbf{2},k} Z_{2, M_k})) \notag \\
  &\stackrel{\text{\eqref{functionsPhiPsi12MIMO}}}{=}h\left(\left[\begin{array}{c} \psi_{\mathcal{G}_{12}}(\mathbf{G}_{\mathbf{12},k}) \\ \mathbf{0}^{(M-\rank(\mathbf{G}_{\mathbf{12},k})) \times M} \end{array}\right](\boldsymbol{\Lambda}_{\mathbf{1},k} X_{1, M_{k-1}} +  X_{2,M_{k-1}}) + \phi_{\mathcal{G}_{12}}(\mathbf{G}_{\mathbf{12},k})( Z_{1, M_k} - \boldsymbol{\Lambda}_{\mathbf{3},k}\hat Z_{1, M_k}- \boldsymbol{\Lambda}_{\mathbf{2},k} Z_{2, M_k})\right)\label{thmFirstIneq12*MIMO}
\end{align}
where (a) follows from \eqref{detPhiMIMO} and the fact that $h(\mathbf{L} X^M)=h(X^M)+\log_2 |\det(\mathbf{L})|$ for any invertible matrix $\mathbf{L}$.
Since $K_{M, \mathcal{U}}$ is an upper bound on the magnitudes of the entries in $\boldsymbol{\Lambda}_{\mathbf{1},k}$, $\boldsymbol{\Lambda}_{\mathbf{2},k}$ and $\boldsymbol{\Lambda}_{\mathbf{3},k}$ (cf.\ \eqref{Y1MkMIMOTemp*} and \eqref{Y2MkMIMOTemp*}), it follows from \eqref{thmFirstIneq12*MIMO} that for each $k\in\{1, 2, \ldots, n\}$,
\begin{align}
&h(\mathbf{G}_{\mathbf{12},k}(\boldsymbol{\Lambda}_{\mathbf{1},k} X_{1, M_{k-1}} +  X_{2,M_{k-1}}) + Z_{1, M_k}  - \boldsymbol{\Lambda}_{\mathbf{3},k}\hat Z_{1, M_k} - \boldsymbol{\Lambda}_{\mathbf{2},k} Z_{2, M_k})\notag \\
 & \quad \le \rank(\mathbf{G}_{\mathbf{12},k}) \log_2 \sqrt{1+P_{k-1}} +\kappa_1. \label{thmFirstIneq**SimpMIMO}
 \end{align}
 for some $\kappa_1$ that does not depend on $n$ and $P$, where the last inequality follows from Proposition~\ref{differentialEntropyUpperBoundGeneric} by setting \linebreak
$\Lambda_1 = \left[\begin{array}{c} \psi_{\mathcal{G}_{12}}(\mathbf{G}_{\mathbf{12},k}) \\ \mathbf{0}^{(M-\rank(\mathbf{G}_{\mathbf{12},k})) \times M} \end{array}\right]\boldsymbol{\Lambda}_{\mathbf{1},k}$, $\Lambda_2 = \left[\begin{array}{c} \psi_{\mathcal{G}_{12}}(\mathbf{G}_{\mathbf{12},k}) \\ \mathbf{0}^{(M-\rank(\mathbf{G}_{\mathbf{12},k})) \times M} \end{array}\right]$, $\Omega_1 = \phi_{\mathcal{G}_{12}}(\mathbf{G}_{\mathbf{12},k})$ $\Omega_2 = -\phi_{\mathcal{G}_{12}}(\mathbf{G}_{\mathbf{12},k}) \boldsymbol{\Lambda}_{\mathbf{3},k} $, \vspace{0.04 in} \linebreak $\Omega_3 = -\phi_{\mathcal{G}_{12}}(\mathbf{G}_{\mathbf{12},k}) \boldsymbol{\Lambda}_{\mathbf{2},k} $ and $K= \max\{\bar K,\bar K K_{M, \mathcal{U}}\}$.
 Following similar procedures for proving \eqref{thmFirstIneq**SimpMIMO}, we obtain that there exists some $\kappa_2$ that do not depend on $n$ and $P$ such that for each $k\in\{1, 2, \ldots, n\}$,
\begin{align*}
& h(\mathbf{G}_{\mathbf{21},k}(X_{1, M_{k-1}} + \boldsymbol{\Omega}_{\mathbf{2},k}X_{2,M_{k-1}})  + Z_{2, M_k}- \boldsymbol{\Omega}_{\mathbf{3},k}\hat Z_{2, M_k}-\boldsymbol{\Omega}_{\mathbf{1},k}Z_{1, M_k})\notag \\
 &\quad \le \rank(\mathbf{G}_{\mathbf{21},k}) \log_2 \sqrt{1+P_{k-1}} +\kappa_2.
 \end{align*}
It remains to upper bound $h(\vec Y_{1,k}^*)$ and $h(\vec Y_{2,k}^*)$ defined in \eqref{Y1*MIMO} and \eqref{Y2*MIMO} respectively. Since $K_{M, \mathcal{U}}$ is an upper bound on the magnitudes of the entries in $\boldsymbol{\Gamma}_{k}$ (cf.\ \eqref{Y1*MIMO} and \eqref{Y2*MIMO}) for each $k\in\{1, 2, \ldots, n\}$ and $K_{M, \mathcal{U}}$ does not depend on $n$ and $P$, we can obtain from \eqref{Y1*MIMO} and \eqref{Y2*MIMO} by following similar procedures for proving \eqref{thmFirstIneq**SimpMIMO} that there exist some $\kappa_i^*$ that does not depend on $n$ and $P$ such that for each $k\in\{1, 2, \ldots, n\}$,
\begin{equation*}
 h(\vec Y_{i,k}^*) \le (M-1) \log_2 \sqrt{1+P_{k-1}} + \kappa_i^*
\end{equation*}
for each $i\in\{1,2\}$.

\section{MIMO Complex Channel} \label{CondComplexMIMO}

For achievability purposes, we only need to show that the conditions for Lemmas~\ref{lemmaS},~\ref{lemmaZ}, and~\ref{lemmaX} hold for almost values of the augmented channel gains (since it can be easily checked that all the steps in the proof of Lemma~\ref{lemmaS} for the case of real channel gains in Appendix~\ref{LemmaS} hold for the case of the augmented channel gains up to replacing $H_{ij}$'s by $\bar{H}_{ij}$'s). We will first prove the conditions for Lemma~\ref{lemmaS}. The proof for Lemma~\ref{lemmaZ} is similar and thus omitted.
Conditions~(C-1),~(C-2),~(C-3), and the condition of the invertibility of channel submatrices are reformulated as follows.

First, let $\bar q$ denote the first column of $\bar H_{vd_2}^{-1}$, $\bar p^T$ denote the last row of $\bar H_{s_1u}^{-1}\bar H_{s_2 u}\bar H_{s_2v}^{-1}$, $\bar A$ denote $\bar H_{s_1v}\bar H_{s_1u}^{-1}\bar H_{s_2u}\bar H_{s_2v}^{-1}$, and $\bar B$ denote $\bar H_{vd_2}^{-1}\bar H_{ud_2}\bar H_{ud_1}^{-1}\bar H_{vd_1}$. The conditions can be restated as follows:
\begin{enumerate}
\item[] (C-I) $\det(\bar H_{ij})\ne 0$ for each $(i,j)\in \{s_1, s_2\}\times \{u,v\}$, and $\det(\bar H_{k,l})\ne 0$ for each $(k,l)\in \{u, v\}\times \{d_1,d_2\}$.
\item[] (C-II) There does not exist a $\lambda\in \mathbb{C}$ that satisfy both $\det(\bar{A}-\lambda \mathbf{I}_{2M})=0$ and $\det(\bar{B}-\lambda \mathbf{I}_{2M})=0$.
\item[] (C-III) $\det([\begin{array}{cccc} \bar p & \bar A^T \bar p & \ldots &(\bar A^T)^{2M-1}\bar p \end{array}])\ne 0$ and $\det([\begin{array}{cccc} \bar q & \bar B \bar q & \ldots &(\bar B)^{2M-1}\bar q \end{array}])\ne 0$.
\end{enumerate}

Also note that all the converse steps in Section~\ref{upperBound} still hold for the case of augmented channel gains. The only condition needed is (C-1), which is rewritten above as condition (C-II). So proving the above three conditions plus the conditions for Lemma~\ref{lemmaX} (stated later) is sufficient for both achievability and converse.

It easy to check that condition (C-I) holds for almost all values of channel gains. Condition (C-II) can be shown to be true by the same argument used in Appendix~\ref{proplemmaS} (since $\det(\bar{A}-\lambda \mathbf{I}_{2M})$ and $\det(\bar{B}-\lambda \mathbf{I}_{2M})$ are non-zero polynomials). It remains to shown that condition (C-III) holds. Similarly to the proofs of conditions (C-2) and (C-3), it suffices to demonstrate a particular choice of $\left(\left[\begin{array}{cc}\bar H_{s_1 u} & \bar H_{s_2 u}\\ \bar H_{s_1 v} & \bar H_{s_2 v}\end{array}\right], \left[\begin{array}{cc}\bar H_{u d_1} & \bar H_{v d_1}\\ \bar H_{u d_2} & \bar H_{v d_2}\end{array}\right]\right)$ that satisfies Condition (C-III), which will imply that $\det([\begin{array}{cccc} \bar p & \bar A^T \bar p & \ldots &(\bar A^T)^{M-1}\bar p \end{array}])$ and $\det([\begin{array}{cccc} \bar q & \bar B \bar q & \ldots &(\bar B)^{M-1}\bar q \end{array}])$ are non-zero polynomials in terms of the entries of  $\Re\{\mathbf{H}_{\boldsymbol{1}}\}$, $\Im\{\mathbf{H}_{\boldsymbol{1}}\}$, $\Re\{\mathbf{H}_{\boldsymbol{2}}\}$ and $\Im\{\mathbf{H}_{\boldsymbol{2}}\}$ (cf.\ \eqref{H1Complex}, \eqref{H2Complex} and \eqref{barHij}), and then (C-III) follows for almost all channel gains. Consider two $M\times M$ real matrices denoted by $\boldsymbol{\Pi}=[\pi_{ij}]_{1\le i,j\le M}$ and $\boldsymbol{\Omega}=[\omega_{ij}]_{1\le i,j\le M}$ respectively such that
\begin{equation*}
\pi_{ij}= \begin{cases} 1 & \text{if $j=i+1$,}\\ 0 & \text{otherwise,} \end{cases} \label{piIJProof}
 \end{equation*}
and
 \begin{equation*}
 \omega_{ij}= \begin{cases} 1 & \text{if $(i,j)=(M,1)$,}\\ 0 & \text{otherwise.} \end{cases} \label{omegaIJProof}
 \end{equation*}
 Letting
 \begin{equation}
\bar H_{s_1u}=\bar H_{s_2 u}=\bar H_{s_2v}= \bar H_{vd_2}=\bar H_{ud_2}=\bar H_{ud_1}=\mathbf{I}_{2M} \label{specificBarHProof1}
 \end{equation}
 and
 \begin{equation}
 \bar H_{s_1v}=\bar H_{vd_1} = \left[\begin{array}{cc} \boldsymbol{\Pi}&\boldsymbol{\Omega} \\ -\boldsymbol{\Omega}& \boldsymbol{\Pi}\end{array}\right] =  \begin{bmatrix}
0 & 1 & 0 & \cdots & \cdots & 0\\
 0 & 0 & 1 &0 & \cdots & 0 \\
 \vdots & & \ddots & \ddots & \ddots & \vdots\\
 0& \cdots &\cdots& 0& 1 & 0\\
 0 & \cdots & \cdots & \cdots & 0 &  1 \\
  -1 & 0 & \cdots & \cdots & \cdots & 0
\end{bmatrix}
 \label{specificBarHProof2}.
 \end{equation}
Note that the assignment of matrices in~\eqref{specificBarHProof1} and~\eqref{specificBarHProof2} satisfies the structure dictated by~\eqref{barHij}. So
 we get
$
 \bar p = [0\ \ldots \ 0 \ 1]^T$,
$
 \bar q = [1\ 0\ \ldots \ 0]^T$,
$
 \bar A^T = \left[\begin{array}{cc} \boldsymbol{\Pi}&\boldsymbol{\Omega} \\ -\boldsymbol{\Omega}& \boldsymbol{\Pi}\end{array}\right] ^T
$
 and
$
 \bar B = \left[\begin{array}{cc} \boldsymbol{\Pi} & \boldsymbol{\Omega} \\ -\boldsymbol{\Omega} & \boldsymbol{\Pi} \end{array} \right] $.
It easy to check that $\bar{p}$, $\bar{A}^T$, $\bar{q}$, and $\bar{B}$ satisfy condition \vspace{0.04 in} \linebreak (C-III) (cf.\ \eqref{ATpControllable} and \eqref{BqControllable}). Therefore (C-III) holds for almost all values of augmented channel gains.

We still need to check that the conditions for Lemma~\ref{lemmaX} hold as well. Note that we need to parse out the proof for the conditions for Lemma~\ref{lemmaX} in the case of complex channel gains because when we define the modified sources (as done in~\eqref{modifieds1} and~\eqref{modifieds2}), the corresponding channel matrices between the modified sources and the relays lose the structure imposed by~\eqref{barHij}. Therefore, the proof is different from the one given above.

We will define the modified sources differently from~\eqref{modifieds1} and~\eqref{modifieds2}. In particular, for the case of real channel gains, the modified sources essentially correspond to ``flipping'' the last antenna of $s_1$ with the first antenna of $s_2$. In this case, we will flip the first antenna of $s_1$ with the first antenna of $s_2$ in the \emph{augmented} channel. Note that this is not fundamental to the proof (as noted in Remark~\ref{LemmaAspect}), but it makes it easier. More specifically, let
\begin{align*}
\tilde{X}_{1,M_k} & = [\Re\{X_{2,M(k-1)+1}\}, \Re\{X_{1,M(k-1)+2}\}, \dots, \Re\{X_{1,Mk}\},\Im\{X_{1,M(k-1)+1}\}, \Im\{X_{1,M(k-1)+2}\}, \dots, \Im\{X_{1,Mk}\}]^T
\\
\intertext{and}
\tilde{X}_{2,M_k} & = [\Re\{X_{1,M(k-1)+1}\}, \Re\{X_{2,M(k-1)+2}\}, \dots, \Re\{X_{2,Mk}\},\Im\{X_{2,M(k-1)+1}\}, \Im\{X_{2,M(k-1)+2}\}, \dots, \Im\{X_{2,Mk}\}]^T.
\end{align*}
As before, define $\tilde{H}_{s_i,r}$ to be the channel submatrix between the modified source
$\tilde{s}_i$ ($i \in \{1,2\}$) and relay $r$ ($r \in \{u,v\}$).
Note that the matrix $\tilde{H}_{s_i,r}$  is obtained by taking the matrix $\bar{H}_{s_i,r}$ and replacing its first column by the first column of $\bar{H}_{s_{\bar{i}},r}$, where $\bar{i}=3-i$. Now, similarly to~\eqref{Sylvester}, we need to find an invertible $\mathbf{B^X}$ such that
\begin{equation}
\mathbf{B^X} \tilde{A} - \tilde{B} \mathbf{B^X} = \bar{H}_{vd_2}^{-1} \begin{bmatrix}
c & \mathbf{0}^{M \times M-1} \end{bmatrix}  \tilde{H}_{s_1u}^{-1} \tilde{H}_{s_2u} \tilde{H}_{s_2v}^{-1}, \label{Sylvestercomplex}
\end{equation}
where $\tilde{A} = \tilde{H}_{s_1v} \tilde{H}_{s_1u}^{-1} \tilde{H}_{s_2u} \tilde{H}_{s_2v}^{-1}$, and $\tilde{B} =  \bar{H}_{vd_2}^{-1} \bar{H}_{ud_2} \bar{H}_{ud_1}^{-1} \bar{H}_{vd_1}$. Note that the change in position of $c$ in the RHS (as compared to~\eqref{Sylvester}) is due to the fact that we placed the first antenna of $s_2$ as the \emph{first} antenna of the modified source $\tilde{s}_1$ (while it was the \emph{last} antenna in the previous formulation). Finally, let $\tilde{q}$ be the first column of $\bar{H}_{vd_2}^{-1}$, and $\tilde{p}^T$ be the first row of $\tilde{H}_{s_1u}^{-1} \tilde{H}_{s_2u} \tilde{H}_{s_2v}^{-1}$, we get RHS of~\eqref{Sylvestercomplex} equal to $\tilde{q}\tilde{p}^T$. Then the condition needed is the following:
\begin{itemize}
\item[] (C-A) $\det([\begin{array}{cccc} \tilde p & \tilde A^T \tilde p & \ldots &(\tilde A^T)^{2M-1}\tilde p \end{array}])\ne 0$ and $\det([\begin{array}{cccc} \tilde q & \tilde B \tilde q & \ldots &(\tilde B)^{2M-1}\tilde q \end{array}])\ne 0$.
\end{itemize}
The conditions equivalent to (C-I) and (C-II) are dropped since the proofs are similar. Similarly to the proof of (C-III), we only need to find one realization of the channel gains such that (C-A) holds to conclude the proof.

Consider the following assignment of channel gains. Let $\bar{H}_{vd_2}=\bar{H}_{ud_2}=\bar{H}_{ud_1}=\mathbf{I}_{2M}$, and let $\bar H_{vd_1}$ be as defined in~\eqref{specificBarHProof2}. Then $\tilde q = \begin{bmatrix}
1 & 0 & \ldots & 0
\end{bmatrix}^T$, and $\tilde{B} = \bar H_{vd_2}$. This is similar to the case above in the proof of (C-III) and thus satisfies (C-A). Furthermore, let
\begin{equation*}
\bar H_{s_1u} = \bar H_{s_2u} = \bar H_{s_2v} = \mathbf{I}_{2M}, \label{complexXch1}
\end{equation*}
and let
\begin{equation*}
\bar H_{s_1v} = 
\begin{bmatrix}
1 & 1 & 0 & \cdots & \cdots & 0 \\
0 & 1 & 1 & 0 & \cdots & 0 \\
\vdots & \ddots & \ddots & \ddots & \ddots & \vdots \\
 0& \cdots & 0 & 1 & 1 & 0 \\
  0& \cdots & 0 & 0 & 1 & 1 \\
 -1 & 0 & \cdots & \cdots & 0 & 1
\end{bmatrix}. 
\label{complexXch2}
\end{equation*}
be a matrix with non-zero entries at only the main diagonal, the upper diagonal and the bottom left corner.
Now we get
\begin{equation*}
\tilde{H}_{s_1u} = \tilde{H}_{s_2u} = \mathbf{I}_{2M}, \label{tildech1}
\end{equation*}
\begin{equation*}
\tilde{H}_{s_2v} =  
\begin{bmatrix}
1  & 0 & \cdots & \cdots & 0  \\
0 &  1 & 0 &\cdots & 0 \\
\vdots  & \ddots & \ddots & \ddots &\vdots \\
0 &  \cdots & \cdots &1 & 0 \\
 -1 & 0 & \cdots & 0 & 1
\end{bmatrix},
~~\text{ and }~~
\tilde{H}_{s_1v} = 
\begin{bmatrix}
1 & 1 & 0 & \cdots & \cdots & 0 \\
0 & 1 & 1 & 0 & \cdots & 0 \\
\vdots & \ddots & \ddots & \ddots & \ddots & \vdots \\
 0& \cdots & 0 & 1 & 1 & 0 \\
   0& \cdots & 0 & 0 & 1 & 1 \\
 0 & \cdots & \cdots & \cdots & 0 & 1
\end{bmatrix}. \label{tildech2}
\end{equation*}
Now, it is easy to verify that
\begin{equation*}
\tilde{H}_{s_2v}^{-1} =  
\begin{bmatrix}
1  & 0 & \cdots & \cdots & 0  \\
0 &  1 & 0 &\cdots & 0 \\
\vdots  & \ddots & \ddots & \ddots &\vdots \\
0 &  \cdots & \cdots &1 & 0 \\
 1 & 0 & \cdots & 0 & 1
\end{bmatrix},
~~\text{ and consequently }~~
\tilde{A} = \tilde{H}_{s_1v} \tilde{H}_{s_2v}^{-1} =  
\begin{bmatrix}
1 & 1 & 0 & \cdots & \cdots & 0 \\
0 & 1 & 1 & 0 & \cdots & 0 \\
\vdots & \ddots & \ddots & \ddots & \ddots & \vdots \\
 0& \cdots & 0 & 1 & 1 & 0 \\
   1& 0 & \cdots & 0 & 1 & 1 \\
 1 & 0 & \cdots & \cdots & 0 & 1
\end{bmatrix}.
\end{equation*}
Recall $\tilde{p}^T$ is the first row of $\tilde{H}_{s_1u}^{-1} \tilde{H}_{s_2u}^{-1} \tilde{H}_{s_2v}^{-1} $, so $\tilde{p}^T=[1\ 0\ \ldots \ 0]^T$. We need to verify that $\tilde{p}$ and $\tilde{A}$ satisfy condition (C-A). Note that (C-A) says (by definition) that the pair $(\tilde{A}^T,\tilde p)$ is controllable (see~\cite[Definition 4.1.1]{BertsekasControl}). But by~\cite[Theorem 6.8]{Controllability}, we know that $(\tilde A^T,\tilde p)$ is controllable iff $\tilde p$ is not orthogonal to any left eigenvector of $\tilde A^T$, i.e. any eigenvector of $\tilde A$. So, we need to show that, given $\mathbf{v} = [v_1\ v_2\ \ldots \ v_{2M}]^T \in \mathbb{R}^{2M}$, if $\exists \lambda \in \mathbb{C}$ such that
\begin{equation*}
\begin{cases}
\tilde{p}^T \mathbf{v} = 0, \\
\tilde{A} \mathbf{v} = \lambda \mathbf{v},
\end{cases} \label{controllability}
\end{equation*}
then $\mathbf{v}= \mathbf{0}^{2M \times 1}$. Consider $\mathbf{v}$ such that $\tilde{p}^T \mathbf{v} =0$, then $v_1 = 0$. So we get
\begin{equation*}
\tilde{A} \mathbf{v} = 
\begin{bmatrix}1 & 1 & 0 & \cdots & \cdots & 0 \\
0 & 1 & 1 & 0 & \cdots & 0 \\
\vdots & \ddots & \ddots & \ddots & \ddots & \vdots \\
 0& \cdots & 0 & 1 & 1 & 0 \\
   1& 0 & \cdots & 0 & 1 & 1 \\
 1 & 0 & \cdots & \cdots & 0 & 1
 \end{bmatrix}
 \begin{bmatrix}
v_1 \\ v_2 \\ \vdots \\ v_{2M-2} \\ v_{2M-1} \\ v_{2M}
\end{bmatrix} = \begin{bmatrix}
v_1 + v_2 \\ v_2 + v_3 \\ \vdots \\ v_{2M-2} + v_{2M-1} \\ v_1 + v_{2M-1} + v_{2M} \\ v_1 + v_{2M}
\end{bmatrix} =
\begin{bmatrix}
v_2 \\ v_2 + v_3 \\ \vdots \\ v_{2M-2} + v_{2M-1} \\ v_{2M-1} + v_{2M} \\ v_{2M}
\end{bmatrix}, \label{eigenvector1}
\end{equation*}
where the last equality follows from $v_1 =0$. Now equating $\tilde A \mathbf{v} = \lambda \mathbf{v}$, we get
\begin{equation}
\begin{bmatrix}
v_2 \\ v_2 + v_3 \\ \vdots \\ v_{2M-2} + v_{2M-1} \\ v_{2M-1} + v_{2M} \\ v_{2M}
\end{bmatrix} = \lambda \begin{bmatrix}
0 \\ v_2 \\ \vdots \\ v_{2M-2} \\ v_{2M-1} \\ v_{2M}
\end{bmatrix}.
\end{equation}
This implies that $v_2=0$, which in turn implies that $v_2+v_3=v_3=0$ (from the second row), and the rest of the entries follow similarly, i.e. $\mathbf{v}=\mathbf{0}^{2M\times 1}$. Therefore, the pair $(\tilde A^T, \tilde p)$ is controllable, and thus satisfies (C-A).
\hfill $\blacksquare$ 

\end{appendices}

\section*{Acknowledgment}
The authors would like to thank Song-Nam Hong and Giuseppe Caire for providing the codes for evaluating the numerical results of their CoF-AND and PCoF-CIA schemes in \cite{CaireFiniteField}. This work is supported in part by NSF Grants CAREER-0953117, CCF-1161720 and ECCS-1247915, Samsung Advanced Institute of Technology (SAIT), AFOSR YIP award, and ONR award N000141310094.

\bibliographystyle{IEEEtran}
\bibliography{IEEEabrv,database}

\end{document}